\documentclass[10pt,twocolumn,secnumarabic, nobibnotes, aps, prd,
floatfix]{revtex4-2}
\usepackage{eurosym}
\usepackage{xcolor,soul}
\usepackage{amsmath}
\usepackage{wrapfig}
\usepackage[utf8]{inputenc}
\usepackage{lipsum}
\usepackage{subfigure}
\usepackage{wrapfig}
\usepackage{xparse,xcoffins}
\usepackage{calligra}
\usepackage{graphicx}

\sethlcolor{red}

\DeclareMathAlphabet{\mathcalligra}{T1}{calligra}{m}{n}
\DeclareFontShape{T1}{calligra}{m}{n}{<->s*[2.2]callig15}{}

\ExplSyntaxOn
\NewCoffin\imagecoffin
\NewCoffin\labelcoffin
\keys_define:nn { miguel/label }
 {
  label   .tl_set:N = \l_miguel_label_tl,
  labelbox .bool_set:N = \l_miguel_label_box_bool,
  labelbox .default:n = true,
  fontsize .tl_set:N = \l_miguel_label_size_tl,
  fontsize .initial:n = \footnotesize,
  pos .choice:,
  pos/nw .code:n = \tl_set:Nn \l_miguel_label_pos_tl { left,up },
  pos/ne .code:n = \tl_set:Nn \l_miguel_label_pos_tl { right,up },
  pos/sw .code:n = \tl_set:Nn \l_miguel_label_pos_tl { left,down },
  pos/se .code:n = \tl_set:Nn \l_miguel_label_pos_tl { right,down },
  pos/n .code:n = \tl_set:Nn \l_miguel_label_pos_tl { hc,up },
  pos/w .code:n = \tl_set:Nn \l_miguel_label_pos_tl { left,vc },
  pos/s .code:n = \tl_set:Nn \l_miguel_label_pos_tl { hc,down },
  pos/e .code:n = \tl_set:Nn \l_miguel_label_pos_tl { right,vc },
  pos .initial:n = nw,
  xpos .tl_set:N = \l_miguel_label_x_tl,
  ypos .tl_set:N = \l_miguel_label_y_tl,
  xpos .initial:n = ,
  ypos .initial:n = ,
  unknown .code:n   = \clist_put_right:Nx \l_miguel_label_clist
                       { \l_keys_key_tl = \exp_not:n { #1 } }
 }
\clist_new:N \l_miguel_label_clist
\box_new:N \l_miguel_label_box
\box_new:N \l_miguel_label_image_box
\NewDocumentCommand{\xincludegraphics}{O{}m}
 {
  \group_begin:
  \tl_clear:N \l_miguel_label_tl
  \clist_clear:N \l_miguel_label_clist
  \keys_set:nn { miguel/label } { #1 }
  \tl_if_empty:NTF \l_miguel_label_tl
   {
    \miguel_includegraphics:Vn \l_miguel_label_clist { #2 }
   }
   {
    \SetHorizontalCoffin\imagecoffin
     {
      \miguel_includegraphics:Vn \l_miguel_label_clist { #2 }
     }
    \SetHorizontalCoffin\labelcoffin
     {
      \raisebox{\depth}
       {
        \bool_if:NTF \l_miguel_label_box_bool
         { \fcolorbox{white}{white}{\l_miguel_label_size_tl\l_miguel_label_tl} }
         { \l_miguel_label_size_tl\l_miguel_label_tl }
       }
     }
    \SetVerticalPole\imagecoffin{left}{3pt+\CoffinWidth\labelcoffin/2}
    \SetVerticalPole\imagecoffin{right}{\Width-3pt-\CoffinWidth\labelcoffin/2}
    \SetHorizontalPole\imagecoffin{up}{\Height-3pt-\CoffinHeight\labelcoffin/2}
    \SetHorizontalPole\imagecoffin{down}{3pt+\CoffinHeight\labelcoffin/2}
    \use:x{\JoinCoffins\imagecoffin[\l_miguel_label_pos_tl]\labelcoffin[vc,hc]}
    \TypesetCoffin\imagecoffin
   }
   \group_end:
 }
\NewDocumentCommand{\setlabel}{m}
 {
  \keys_set:nn { miguel/label } { #1 }
 }
\cs_new_protected:Nn \miguel_includegraphics:nn
 {
  \includegraphics[#1]{#2}
 }
\cs_generate_variant:Nn \miguel_includegraphics:nn { V }
\ExplSyntaxOff
\makeatletter
\newcommand{\twocolumncaption}{\@dblarg\@twocolumncaption}
\def\@twocolumncaption[#1]#2{  \renewcommand{\@makecaption}[2]{    \par\vskip\abovecaptionskip\begingroup\small\rmfamily
    \splittopskip=0pt
    \setbox\@tempboxa=\vbox{
      \@arrayparboxrestore \let \\\@normalcr
      \hsize=.5\hsize \advance\hsize-1em
      \let\\\heading@cr
      \noindent ##1\ ##2\par    }    \vbadness=10000
    \setbox\z@=\vsplit\@tempboxa to .55\ht\@tempboxa
    \setbox\z@=\vtop{\hrule height 0pt \unvbox\z@}
    \setbox\tw@=\vtop{\hrule height 0pt \unvbox\@tempboxa}
    \noindent\box\z@\hfill\box\tw@\par
    \endgroup\vskip \belowcaptionskip
  }  \setlength{\abovecaptionskip}{4ex}  \caption[#1]{#2}}

\begin{document}

\title{Three-dimensional multipole states of the Bose-Einstein condensate of
polar molecules in a shallow two-dimensional trapping potential}
\author{S. Sanjay$^{1,2}$}
\email{s\textunderscore sanjay@cb.students.amrita.edu}
\author{S. Saravana Veni$^{1,2}$}
\email{s\textunderscore saravanaveni@cb.amrita.edu.in}
\author{Boris A. Malomed$^{3,4}$}
\email{malomed@tauex.tau.ac.il}
\affiliation{$^{1}$Department of Physics, Amrita School of Physical Sciences Coimbatore,
Amrita Vishwa Vidyapeetham, Coimbatore 641112, India }

\affiliation{$^{2}$Quantum Complexity and Nonlinear Dynamics Research Laboratory, \\
Amrita School of Engineering Coimbatore, Amrita Vishwa Vidyapeetham,  India}

\affiliation{$^{3}$Department of Physical Electronics, School of Electrical Engineering, Faculty of Engineering,
Tel Aviv University, Tel Aviv 69978, Israel}
\affiliation{$^{4}$Instituto de Alta Investigaci\'{o}n, Universidad de
Tarapac\'{a}, Casilla 7D, Arica, Chile}

\begin{abstract}
We investigate the formation, stability, and dynamics of three-dimensional
(3D) elongated multipole quantum droplets (QDs), in a Bose-Einstein
condensate (BEC) of {polar molecules with long-range interactions between
molecular electric dipoles. The condensate is confined} by a 2D trapping
potential, which is a combination of the harmonic-oscillator potential and a
central Gaussian potential hill. The system is modeled by an extended
Gross-Pitaevskii (GP) equation, incorporating the dipole-dipole interactions
and beyond-mean-field Lee-Huang-Yang (LHY) correction, which play a crucial
role in stabilizing self-bound QD structures. {In regimes of weak and strong
confinement, stable multipole states with the number of lobes $n=2S$ are
created, by means of the imaginary-time-propagation (ITP) method, from
inputs with vorticity $S\leq 2$. The so created multipole QDs keep no
angular momentum. Stable higher-order multipole states can be produced by
the ITP method from the input with $S\geq 3$ up to $S=6$ (i.e., $2S=12$
lobes) only under the action of strong confinement, and with the molecule
number $N$ exceeding a threshold value.} We analyze the dependence of the
chemical potential, energy per particle, and peak density on $N$ and $S$ {in
the weak- and strong-confinement regimes}. A noteworthy fact is that, below
another threshold value of $N$, the minimum-energy (ground) state {is
created by the input with} $S=1$, rather than $S=0$. Simulations of
perturbed evolution reveal that, despite satisfying energetic stability
criteria, only a subset of the multipole QDs are truly stable, {with
higher-order ones tending to develop instabilities or relax toward
lower-order states}. These results highlight the essential roles of the
trapping potential, {anisotropy of the dipole-dipole interactions,} and
LHY-induced repulsion in the formation and structural stability of multipole
QDs.
\end{abstract}

\maketitle


\section{Introduction}

The theoretical and experimental work on quantum matter has led to
substantial progress in studies of macroscopic quantum phenomena, such as
quantum droplets (QDs) in Bose-Einstein condensates (BECs) \cite%
{Bulgac,luo2021new, guo2021new}, supersolidity \cite%
{boninsegni2012colloquium,recati2023supersolidity,ilzhofer2021phase}, and
vortices \cite{verhelst2017vortex, klaus2022observation,
chaika2023making,molignini2025beyond}. Specifically, the investigation of
QDs reveals diverse experimentally accessible phenomenology \cite%
{neely2010observation,weiler2008spontaneous}, sustained by the equilibrium
between mean-field (MF) and beyond-MF effects. The latter ones, known as
Lee-Huang-Yang (LHY) corrections, are caused by quantum fluctuations around
the MF states \cite{petrov, Astrakharchik}. In the experiment, QDs were
observed in BECs of dipolar gases \cite%
{Pfau2,Pfau1,kadau2016observing,bottcher2020new}. A great deal of interest
has been also drawn to the observation of QDs in homonuclear \cite%
{cheiney2018bright,semeghini2018self} and heteronuclear \cite%
{d2019observation,burchianti2020dual} binary non-dipolar BECs. The formation
of stable QDs in the binary condensates is achieved by the balance between
the MF and LHY intrinsic interactions, that may be properly adjusted by
means of the Feshbach resonance \cite{d2007feshbach}. The work in this
direction has also addressed collisions between QDs in one-dimensional (1D)
\cite{katsimiga2023interactions}, 2D \cite{hu2022collisional}, and 3D \cite%
{ferioli2019collisions} geometries, along with the experimentally observed
1D scattering of QDs on localized potentials \cite{debnath2023interaction}.

A fundamentally important extension is the study of QDs with embedded
vorticity, producing vortex and multipole states. The introduction of the
vorticity into QDs not only gives rise to their internal topology but also
creates complex self-organized modes, such as dipole, quadrupole, and
higher-order multipole droplets. Stable vortex QDs (VQDs) with vorticity up
to $S=5$ and up to $S=2$ were predicted in 2D \cite{li2018two} and 3D \cite%
{kartashov2018three} geometries, respectively. In the rotating reference
frame, beyond a critical rotation frequency \cite{dong2021rotating}, the QDs
can be shaped into VQDs in 2D or vortex tori in 3D. In particular, the
transformation of 2D dipole and quadrupole QDs into vortex VQDs with the
increasing rotation frequency has been demonstrated \cite{liu2024rotating}.
Stable VQDs with higher vorticity values can be produced using suitable
trapping potentials. Under a radial lattice potential \cite{huang2023stable}%
, 2D VQDs remain stable up to $S=10$. Stable 3D vortex and multipole QDs
with vorticity up to $S=12$, were predicted by employing a toroidal trapping
potential \cite{dong2024three}. Apart from ring-shaped traps, these states
also tend to be stable in a weak 2D anharmonic potential \cite%
{dong2022bistable}. Multipole states can be observed as well in photonic
lattices, such as $\mathcal{PT}$-symmetric ones \cite%
{wang2015two,wang2018evolution}, Bessel lattices \cite%
{kartashov2005multipole,dong2009bessel,hong2015surface}, circular waveguides
\cite{kartashov2009stabilization}, and media with competing nonlinearities
\cite{dong2023multipole}. Multipole structures can also emerge in dipolar
QDs formed by magnetic atoms \cite{huang2018excited}.

The work with BECs of dipolar atoms with long-range anisotropic interactions
has enabled the realization of stable anisotropic QD configurations \cite%
{wachtler2016ground,xi2016droplet,Boronat}. Notably, Refs. \cite%
{young2022supersolid,young2022supersolidbox} investigated crystalline
structures composed of QDs in BEC of $^{164}$Dy atoms, supported by both 3D
harmonic-oscillator (HO) traps and box-shaped potentials. Multi-stripe
arrangements, referred to as superstripe states, featuring ground and
metastable states in 3D HO potentials, were reported too \cite{young2023mini}%
. A challenging possibility is the incorporation of vorticity into dipolar
self-bound QDs. Usually, the self-interaction causes azimuthal instability,
which tends to split the 3D vortex into fragments \cite%
{mihalache1,book,mihalache2}. However, this problem can be resolved by
incorporating the LHY correction balancing the MF attraction, akin to the
stabilization of vortex solitons induced by the cubic-quintic nonlinearity
in optics \cite{QTM,Pego,mihalache2002stable,reyna2017high}. The integration
of LHY correction within the BEC framework enhances the stabilization of
self-bound states, and these characteristics align well with the outcomes
generated by the Monte Carlo method applied to the respective many-body
problems \cite%
{cikojevic2019universality,cikojevic2020finite,parisi2019liquid,cinti2017superfluid,macia2016droplets,cikojevic2018ultradilute,molignini2025stability,molignini2025interaction,bilinskaya2024realizing,hughes2023accuracy}%
. Similar to the fundamental QDs, the incorporation of the vorticity in
dipolar QDs leads to the formation of vortex and multipole QD modes.
Isotropic dipolar vortex QDs with the vortex axis parallel to the
polarization axis of the atoms result in an unstable configuration \cite%
{cidrim2018vortices}. On the other hand, stable anisotropic dipolar vortex
QDs were predicted with the vortex axis oriented \emph{perpendicular} to the
polarization axis in 2D \cite{li2024two} and 3D \cite{li2024strongly}
dipolar systems. Stable higher-order dipolar VQDs up to vorticity $S=11$
under annular potential were reported too \cite{huang2025anisotropic}.

Despite the significant progress in the studies of vortex QDs, higher-order
multipole QDs in dipolar BECs have not yet been addressed in a systematic
way. Recently, multipole states in the condensate of dipolar $^{164}$Dy
atoms \cite{sanjay2026toroidal} under toroidal confinement were investigated
for phase-imprinted inputs with vorticity up to $S=12$, demonstrating that
the fundamental state ($S=0$) is stable, while higher-order states are only
transiently stable or unstable. These results highlight the instability of
the higher-order multipole configurations, motivating the search for
alternative stabilization mechanisms. In this context, the
microwave-shielding technique \cite{quemener2023electroassociation} has
enabled the creation of the strongly dipolar BEC of NaCs molecules \cite%
{zhang2026observation}, with the effective electric dipole-dipole
interaction (DDI) scattering length $a_{\mathrm{DDI}}$ ranging from $%
1000a_{0}$ to $25000a_{0}$ ($a_{0}$ is the Bohr radius) \cite%
{bigagli2024observation}, far exceeding the respective magnetic-DDI value
for $^{164}$Dy atoms, $a_{\mathrm{DDI}}=130.8a_{0}$. These advances rely on
techniques protected against collisional effects, including static
electric-field shielding \cite%
{gonzalez2017adimensional,mukherjee2023shielding,mukherjee2024controlling},
microwave shielding\cite%
{karman2018microwave,karman2019microwave,karman2020microwave,
karman2022resonant,deng2023effective} and double-microwave shielding \cite%
{karman2025double,deng2025two} schemes, which suppress inelastic losses,
while preserving strong long-range DDIs. Recent theoretical studies have
further examined settings with microwave-shielded polar molecules, using
variational wave functions subject to the Jastrow correlations \cite%
{b8y9-yvz9} and the Monte-Carlo method \cite{langen2025dipolar}. These
settings include the long-range attractive DDI and a short-range repulsive
shielding potential $\sim 1/r^{6}$, addressing the regime with a negative
scattering length. More recently, Ref.~\cite{arnone2026exploring} developed
an extended Gross--Pitaevskii (GP) model that incorporates the LHY
correction for single-microwave-shielded molecules. It has demonstrated
quantitative agreement with the path-integral Monte-Carlo calculations in
the regime of a positive scattering length. Using the extended GP equation,
Ref. \cite{adhikari2025dynamically} reported the formation of elongated
metastable droplets in strongly dipolar NaCs condensates, confined by a
harmonic trap combined with a repulsive Gaussian barrier. That work employed
interaction parameters which represent strongly dipolar molecular systems,
without explicitly modeling the microwave-shielding mechanism. Subsequently,
the same framework was used to demonstrate metastable hollow
cylindrically-shaped droplets under the action of a strong Gaussian
confinement \cite{2025hollow}. The existence of elongated self-bound
droplets under the action of the such confinement suggests the question if
distinct multipole states can be generated from different phase-imprinted
initial configurations, and how their structure and stability are affected
by the interplay between the trapping potential, DDI, and LHY correction.
These issues are the objective of the present work. We aim to systematically
investigate the structural characteristics and stability of elongated
multipole QDs in a strongly dipolar Bose gas, confined by the combination of
a harmonic-oscillator trap and a central repulsive-Gaussian potential. The
multipole states are generated from initial configurations characterized by
an integer phase-imprinting index $S$. Although the initial states are phase
imprinted with the winding number $S$, angular momentum is not conserved
during the subsequent imaginary-time-propagation (ITP). The 
interaction parameters considered here represent a strongly dipolar regime motivated by
recent theoretical studies of polar molecular condensates. However,
microwave shielding effects are not explicitly incorporated into the model,
and the rigorous applicability of the adopted extended GP equation to the
condensate of dipolar molecules with a positive scattering length remains to
be established. Considering, nevertheless, this theoretical framework as
relevant (at least, at the phenomenological level), our results represent
the respective solutions and dynamics. Based on the trapping parameters, we
analyze the formation and stability of multipole states under the action of
the HO confining potential with a Gaussian potential hill inserted at the
center. Our numerical analysis combines the ITP and real-time-propagation
methods to obtain both stationary and dynamically evolving solutions,
focusing on their physically relevant characteristics, such as the chemical
potential, energy per particle, peak density, and the radial and axial
widths, for different values of the phase-imprinting index $S$ used to
initiate the ITP simulations. The paper is organized as follows: in Section~%
\ref{model} we present the model for the dipolar BEC, which includes the MF
and LHY interactions. The numerical results presented in Section~\ref%
{results} are presented in subsections \ref{deg} and \ref{strong} for weak
and strong confinement regimes, respectively, determined by the central
Gaussian potential hill. In each regime, we address stationary states of
multipole QDs, and their dynamical behavior. The paper is concluded by
Section IV.

\section{The model}

\label{model} The dipolar BEC composed of $N$ polar molecules with mass $m$,
whose electric dipole moments are oriented along the $z$-axis, is modeled by
the LHY-amended Gross-Pitaevskii (GP) equation, which includes the contact
interactions and DDI \cite{young2023dipole,adhikari2024quasi}:
\begin{eqnarray}
i\hbar \frac{\partial \psi (\mathbf{r},t)}{\partial t} &=&\bigg[-\frac{\hbar
^{2}}{2m}\nabla ^{2}+V(x,y)+\frac{4\pi \hbar ^{2}}{m}aN|\psi (\mathbf{r}%
,t)|^{2}  \notag \\
&&+\frac{3\hbar ^{2}}{m}a_{\mathrm{dd}}N\int V_{\mathrm{dd}}(\mathbf{R}%
)|\psi (\mathbf{r}^{\prime })|^{2}d\mathbf{r}^{\prime }  \notag \\
&&+\frac{\gamma _{\mathrm{LHY}}\hbar ^{2}N^{3/2}}{m}|\psi (\mathbf{r},t)|^{3}%
\bigg]\psi (\mathbf{r},t),  \label{gpe}
\end{eqnarray}%
\begin{equation}
V(x,y)\equiv \frac{m\omega ^{2}}{2}\big(x^{2}+y^{2}\big)+A_{0}\exp \left( -%
\dfrac{x^{2}+y^{2}}{\alpha ^{2}}\right)  \label{elongated_trap}
\end{equation}%
\begin{equation}
V_{\mathrm{dd}}(\mathbf{R})=\frac{1-3\cos ^{2}\theta }{|\mathbf{R}|^{3}},~%
\mathbf{R}=\mathbf{r}-\mathbf{r}^{\prime },~a_{\mathrm{dd}}=\frac{md_{%
\mathrm{polar}}^{2}}{12\pi \hbar ^{2}\varepsilon _{0}},  \label{add}
\end{equation}%
where $\nabla ^{2}$ is the 3D Laplacian, and $V(x,y)$ represents the 2D
axially symmetric external potential, which includes the HO term with
trapping frequency $\omega $, combined with the central repulsive Gaussian
term with strength $A_{0}$ and width $\alpha=\sqrt{2} $. Coefficients $a$
and $a_{\mathrm{dd}}$ denote the scattering lengths of the contact
interaction and DDI, respectively, with $d_{\mathrm{polar}}$ being the
electric dipole moment of the polar molecule (in particular, NaCs). Kernel $%
V_{\mathrm{dd}}$ determines the long-range interaction between two polar
molecules positioned at $\mathbf{r}=\{x,y,z\}$ and $\mathbf{r}^{\prime
}=\{x^{\prime },y^{\prime },z^{\prime }\}$, where $\theta $ is the angle
between vector $\mathbf{R}$ and the polarization axis $z$. In the numerical
calculations, the singularity in $V_{\mathrm{dd}}$ is taken care of by
evaluating it in the momentum space ($\mathbf{k}$), where the respective
kernel is regularized using the spherical truncation method and setting the $%
k=0$ component equal to zero, thus avoiding the singularity while preserving
the long-range anisotropic character of the DDI \cite{young2023openmp}. The
wavefunction is normalized by $\int |\psi (\mathbf{r},t)|^{2}d\mathbf{r}=1$.
In the scaled form, the effective DDI length is $\epsilon _{\mathrm{dd}%
}\equiv a_{\mathrm{dd}}/a$. For the formation of QDs, we consider the
condensate of strongly dipolar molecules, with $a_{\mathrm{dd}}\gg a$. {%
Taking the value of $a_{\mathrm{dd}}=1000a_{0}$ for NaCs molecules \cite%
{adhikari2025dynamically}, we choose $a=100a_{0}$ as the contact-interaction
scattering length, which thus corresponds to $\epsilon_{\mathrm{dd}}=10$.
Although the choice of the parameters is motivated by strongly dipolar NaCs
molecular condensates, the present effective GP model does not explicitly
include the microwave shielding and molecular losses, which may be a subject
for a separate investigation.} The coefficient of the LHY term \cite%
{young2023spontaneous} is
\begin{equation}
\gamma _{\mathrm{LHY}}=\frac{128}{3}\sqrt{\pi a^{5}}Q_{l}(\epsilon _{\mathrm{%
dd}}),
\end{equation}%
with the auxiliary function \cite{young2024expansion}
\begin{figure}[tbp]
\xincludegraphics[width=1.0\linewidth]{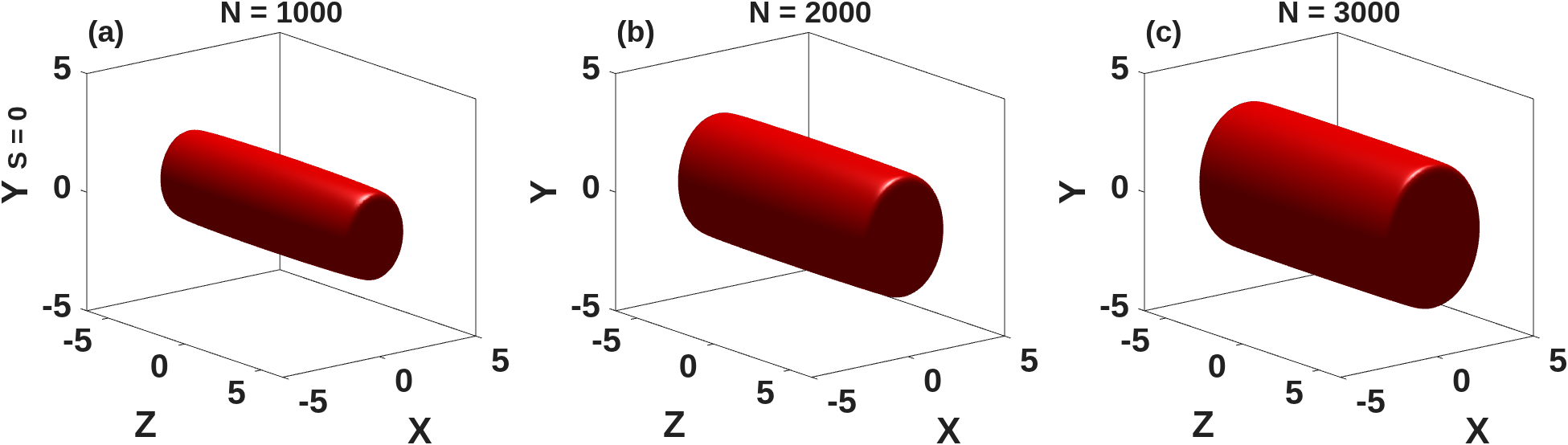} %
\xincludegraphics[width=1.0\linewidth]{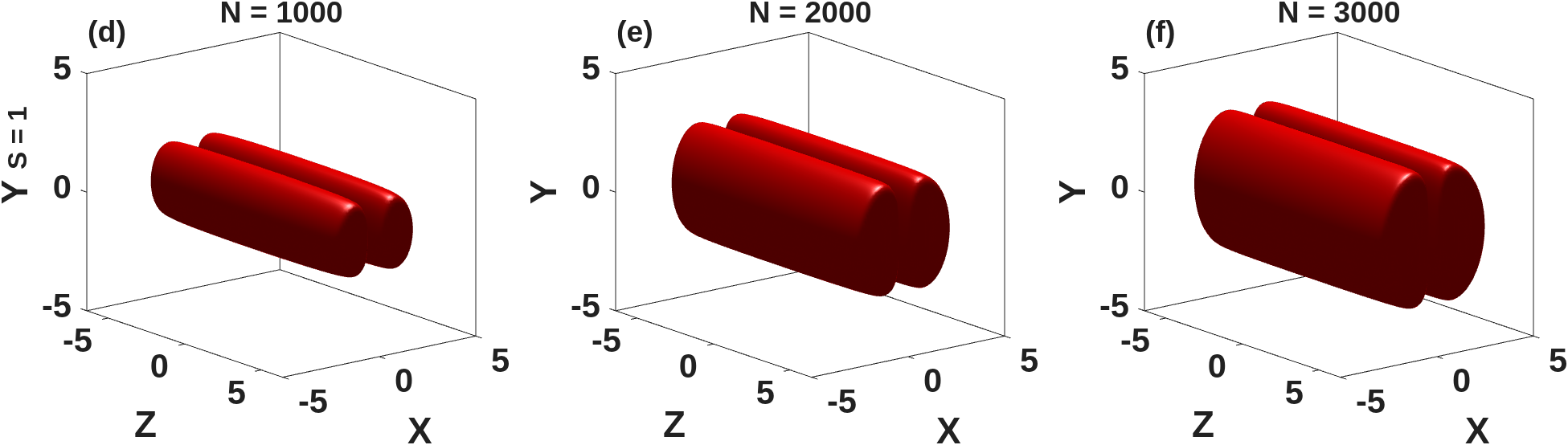} %
\xincludegraphics[width=1.0\linewidth]{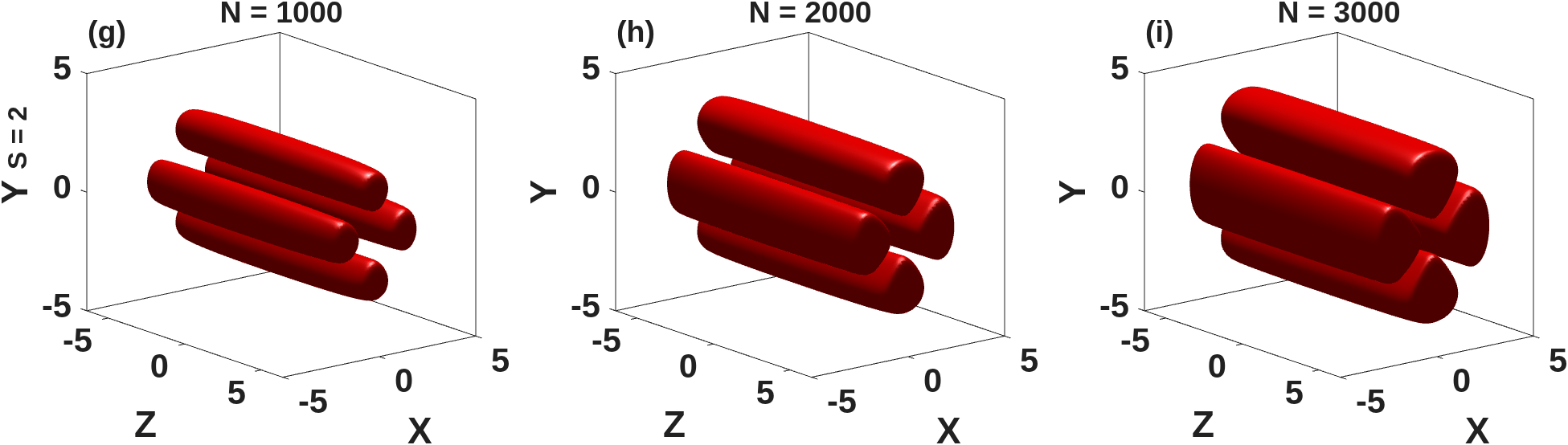} %
\xincludegraphics[width=1.0\linewidth]{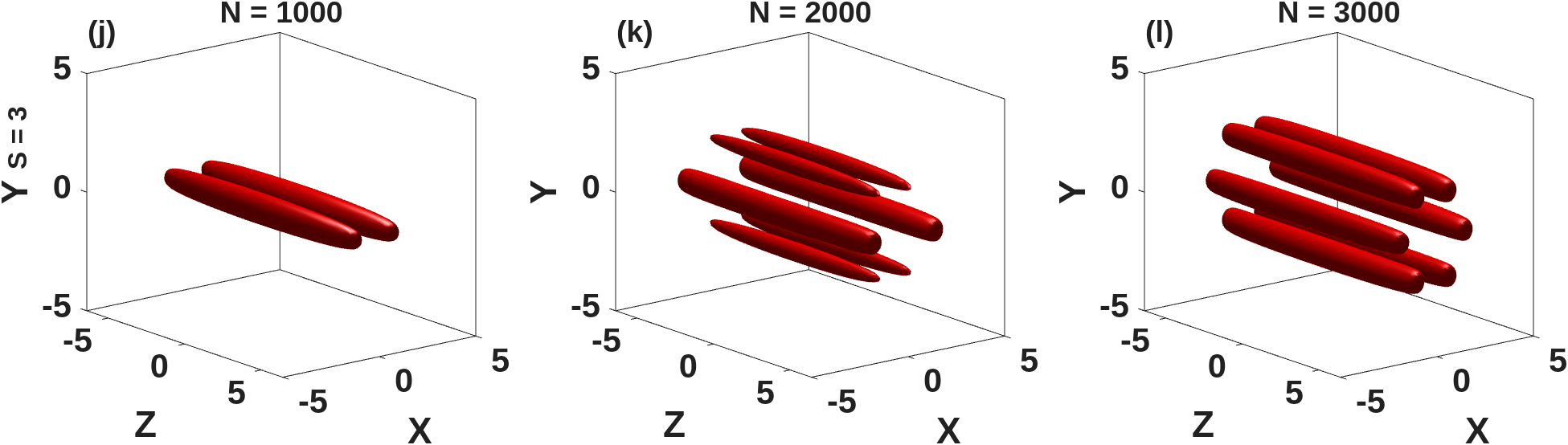}
\caption{ {\ Isosurface plots of stable multipole QD density profiles, as
obtained from different initial phase-imprinting configurations,
characterized by integer indices $S$ and molecule number $N$, with DDI and
contact-interaction scattering lengths $a_{\mathrm{dd}}=1000a_{0}$ and $%
a=100a_{0}$, the LHY strength $\protect\gamma_{\mathrm{LHY}%
}=1.6792\times10^{9}a_{0}^{5/2}$, and the confining potential (\protect\ref%
{elongated_trap}) with $\protect\omega=1$ and $A_{0}=1$. The rows correspond
to the phase-imprinting indices $S=0$, $1$, $2$, and $3$, from top to
bottom, and the columns correspond to $N=1000$, $2000$, and $3000$, from
left to right. All stationary states were produced by means of the
imaginary-time propagation (ITP).}}
\label{el_nacs}
\end{figure}
\begin{equation}
Q_{l}(\epsilon _{\mathrm{dd}})=(1-\epsilon _{\mathrm{dd}})^{l/2}{_{2}}F_{1}%
\bigg(-\frac{l}{2},\frac{1}{2};\frac{3}{2};\frac{3\epsilon _{\mathrm{dd}}}{%
\epsilon _{\mathrm{dd}}-1}\bigg)  \label{hypergeo}
\end{equation}%
where ${_{2}}F_{1}\bigg(\dfrac{-l}{2},\dfrac{1}{2};\dfrac{3}{2};\dfrac{%
3\epsilon _{\mathrm{dd}}}{\epsilon _{\mathrm{dd}}-1}\bigg)$ is the
hypergeometric function. Here, $l$ is an integer which represents the order
of the momentum-space integral in the Bogoliubov treatment of the LHY
correction. Following Ref. \cite{lima2011quantum}, the most significant
value is $l=5$, which is used in the evaluation of $Q_{l}(\epsilon _{\mathrm{%
dd}})$ in our calculations, given by \cite{bisset2016ground}
\begin{eqnarray}
{Q_5(\varepsilon_{\mathrm{dd}})}&{=}& {\frac{(3\varepsilon_{\mathrm{dd}%
})^{5/2}}{48} \Re \left[ \right. (8 + 26\eta + 33\eta^2)\sqrt{1+\eta}}
\notag \\
&&{+ \left. 15\eta^3 \ln \left( \frac{1 + \sqrt{1+\eta}}{\sqrt{\eta}}
\right) \right], \eta = \frac{1 - \varepsilon_{\mathrm{dd}}}{3\varepsilon_{%
\mathrm{dd}}}},  \label{Q5}
\end{eqnarray}
{where $\Re$ denotes the real part.} {We use the evaluated expression $%
Q_{5}(\epsilon _{\mathrm{dd}})$, given by Eq. (\ref{Q5}), in numerical
simulations, for which $\gamma _{\mathrm{LHY}}=1.6792\times
10^{9}a_{0}^{5/2} $}. Equation (\ref{gpe}) is normalized by measuring
lengths in units of $L_{0}=\sqrt{\hbar /m\omega _{0}}$, and time, energy and
angular frequency in units of $T_{0}=\omega _{0}^{-1}$, $\hbar \omega _{0}$,
and $\omega _{0}$ respectively, where $\omega _{0}$ is the reference angular
frequency (not the trapping frequency $\omega $ of the HO potential term).
Further, $A_{0}$ and density $|\psi |^{2}$ are measured in units of $\hbar
\omega _{0}$ and $L_{0}^{-3}$, Thus, the scaled form of Eq. (\ref{gpe}) is
\begin{eqnarray}
i\frac{\partial \psi (\mathbf{r},t)}{\partial t} &=&\bigg[-\frac{1}{2}\nabla
^{2}+V(\rho )+4\pi aN|\psi (\mathbf{r},t)|^{2}  \notag \\
&&+3a_{\mathrm{dd}}N\int \frac{1-3\cos ^{2}\theta }{|\mathbf{R}|^{3}}|\psi (%
\mathbf{r}^{\prime },t)|^{2}d\mathbf{r}^{\prime }  \notag \\
&&+\gamma _{\mathrm{LHY}}N^{3/2}|\psi (\mathbf{r},t)|^{3}\bigg]\psi (\mathbf{%
r},t),  \label{dgpe}
\end{eqnarray}%
\begin{equation}
V(\rho )=\dfrac{\omega ^{2}}{2}\rho ^{2}+A_{0}\exp \left( -\dfrac{\rho ^{2}}{%
\alpha ^{2}}\right) ,\rho ^{2}\equiv x^{2}+y^{2},  \label{scal_pot}
\end{equation}%
where the scaled variables and potential are designated by the same symbols
as in Eqs. (\ref{gpe}) and (\ref{elongated_trap}), as this will not cause
any confusion. Stationary states with real chemical potential are looked for
in the usual form,
\begin{equation}
\psi (\mathbf{r},t)=\exp \left( -i\mu t\right) \phi (\mathbf{r}),
\label{psiphi}
\end{equation}%
where the stationary wave function $\phi (\mathbf{r})$ satisfies the equation%
\begin{eqnarray}
\mu \phi &=&\bigg[-\frac{1}{2}\nabla ^{2}+V(\rho )+4\pi aN|\phi (\mathbf{r}%
)|^{2}  \notag \\
&&+3a_{\mathrm{dd}}N\int \frac{1-3\cos ^{2}\theta }{|\mathbf{R}|^{3}}|\phi (%
\mathbf{r}^{\prime })|^{2}d\mathbf{r}^{\prime }  \notag \\
&&+\gamma _{\mathrm{LHY}}N^{3/2}|\phi (\mathbf{r})|^{3}\bigg]\phi (\mathbf{r}%
),  \label{ph}
\end{eqnarray}

\begin{figure}[tbp]
\xincludegraphics[width=1.0\linewidth]{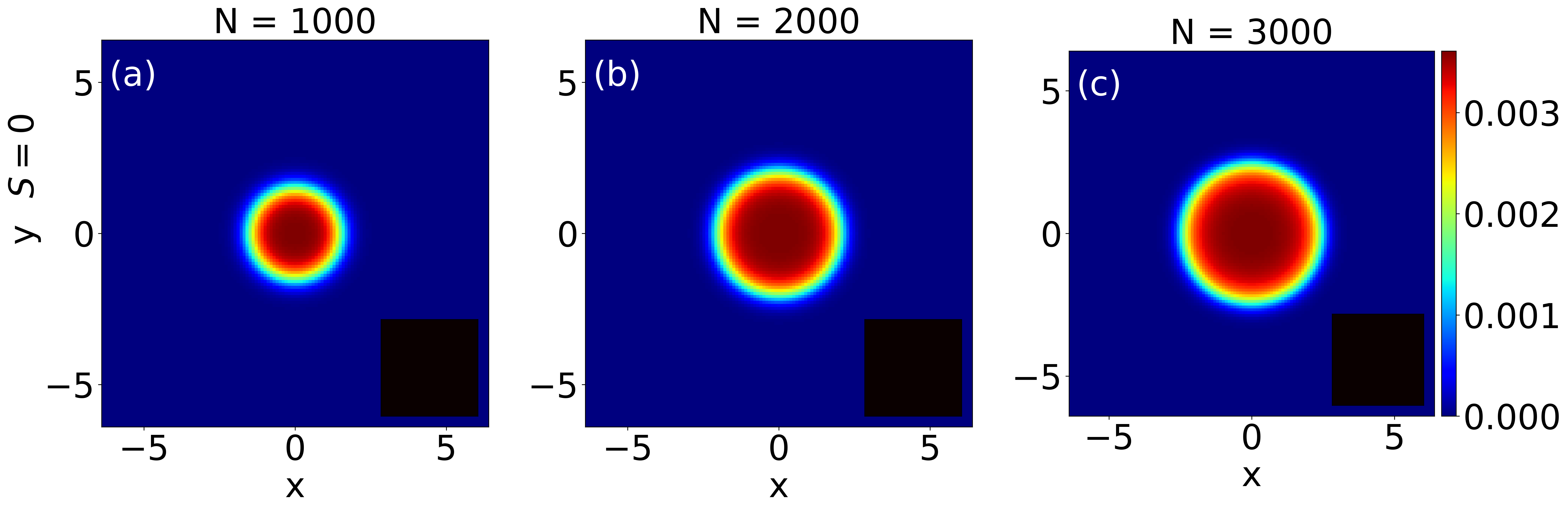} %
\xincludegraphics[width=1.0\linewidth]{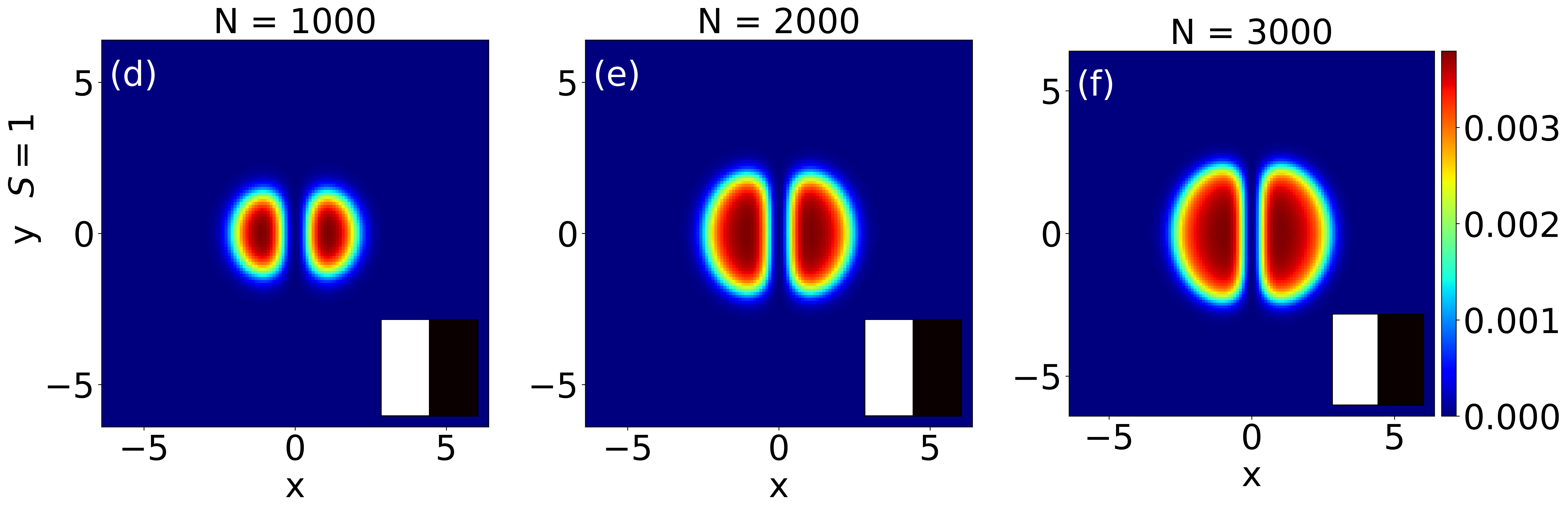} %
\xincludegraphics[width=1.0\linewidth,]{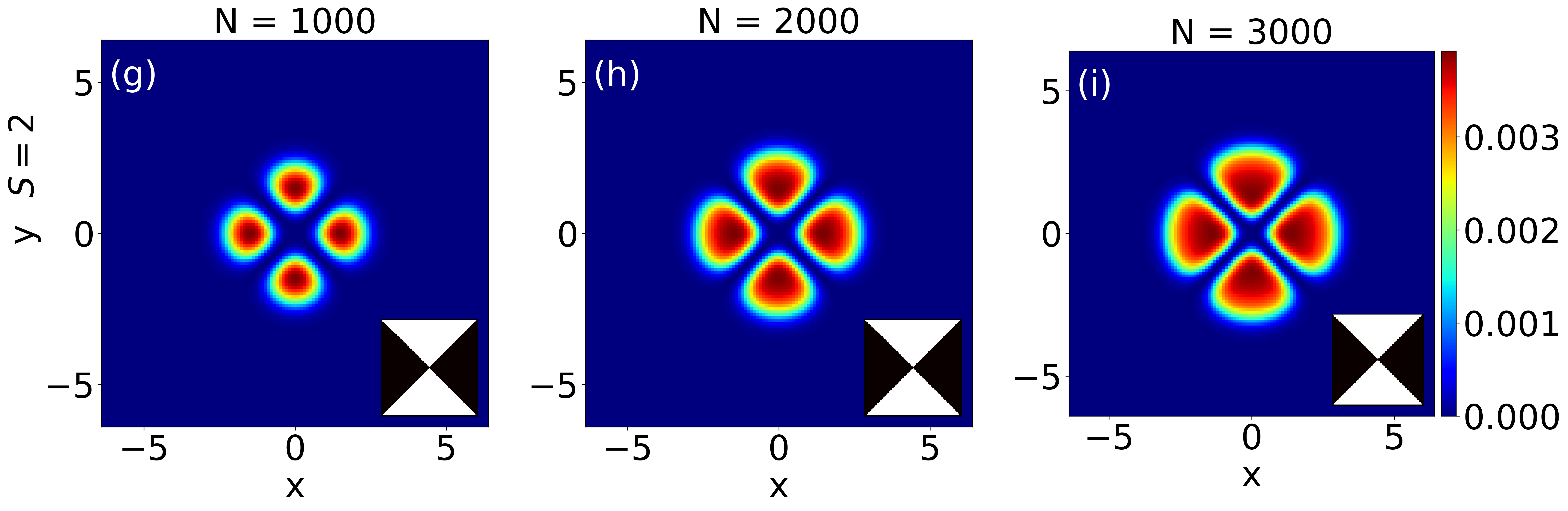}
\caption{Cross-section of the density distribution in the $(x,y)$ plane,
corresponding to the stable stationary isosurface profiles shown in Fig.~%
\protect\ref{el_nacs}(a - i).}
\label{2d}
\end{figure}
In the case of a sufficiently strong central Gaussian potential hill,
\textit{viz}.,
\begin{figure}[tbp]
\xincludegraphics[width=0.32\linewidth,
label={a)},pos={n}]{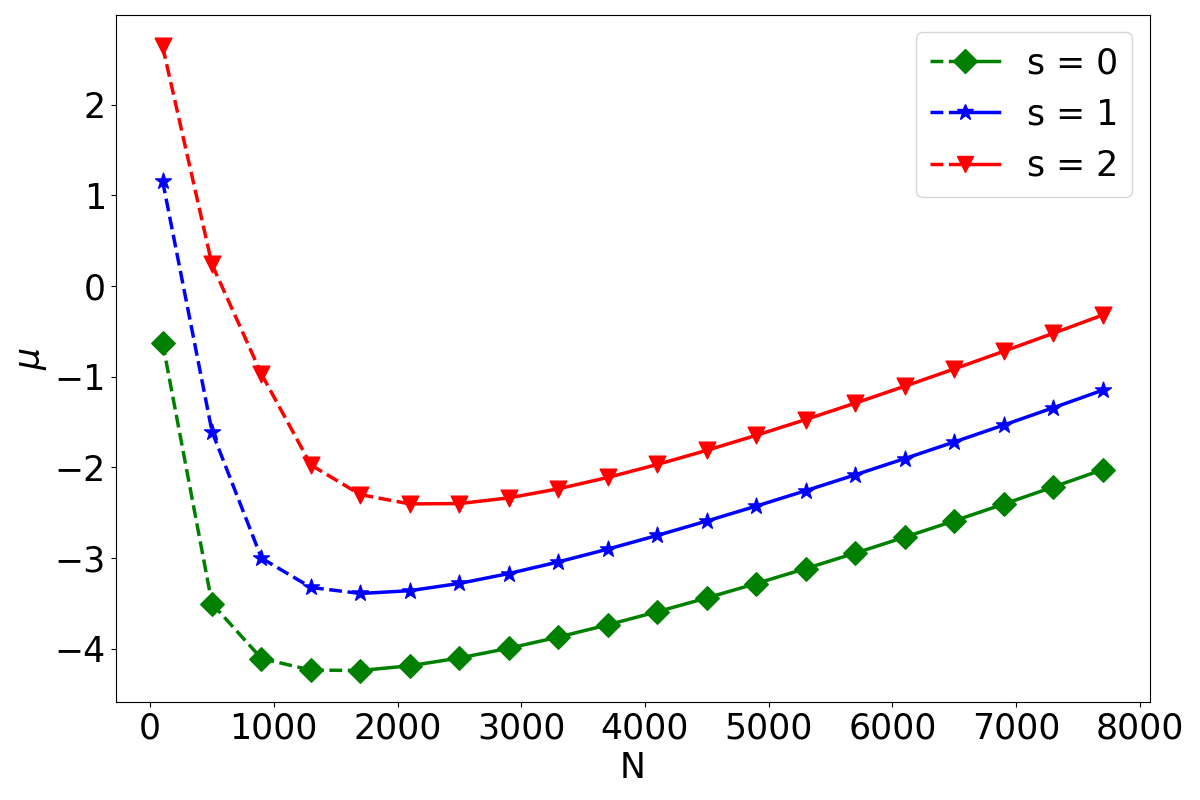}
\xincludegraphics[width=0.32\linewidth,
label={b)},pos={n}]{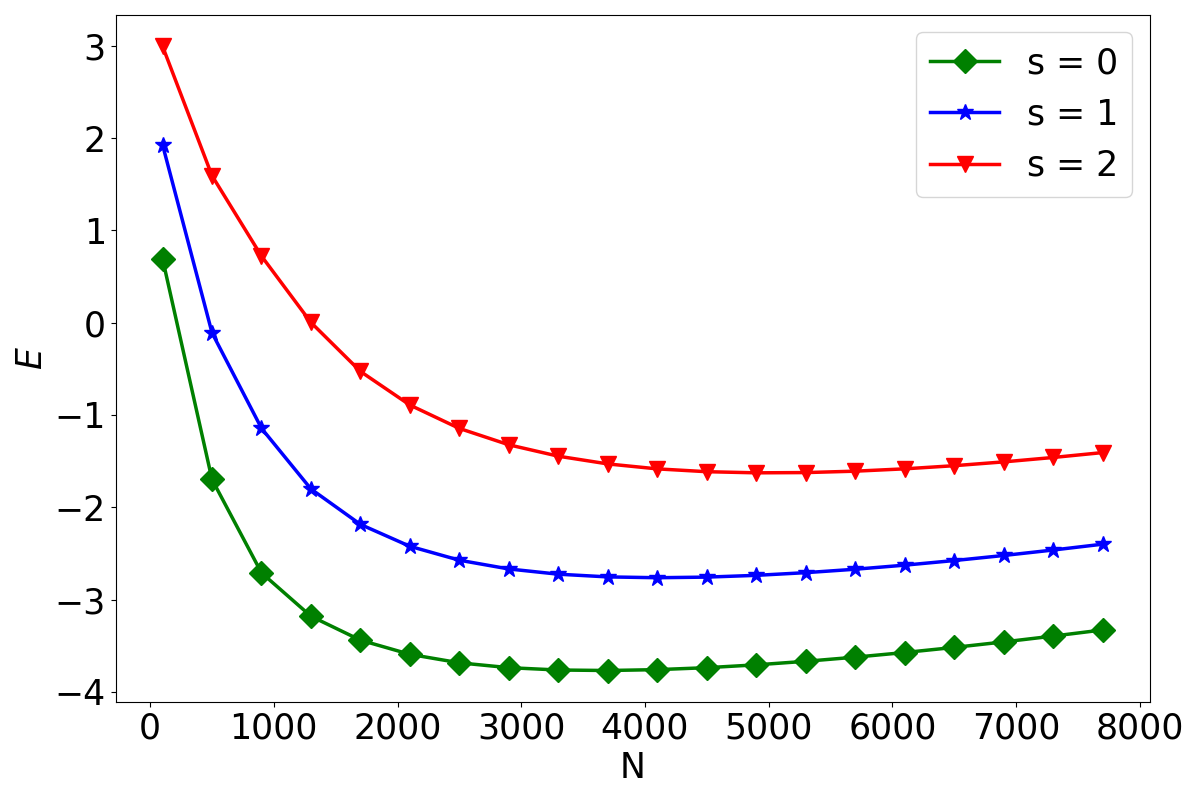}
\xincludegraphics[width=0.32\linewidth,
label={c)},pos={n}]{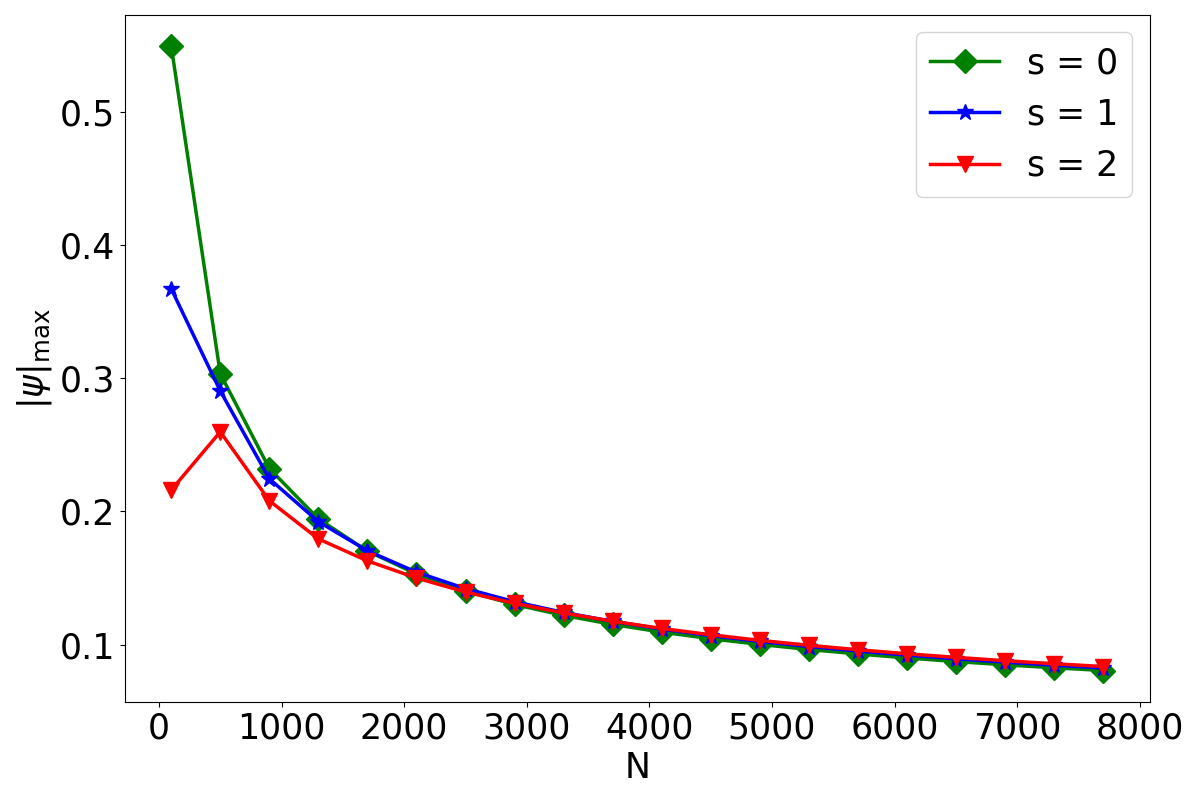}
\caption{{a) Chemical potential $\protect\mu $, b) energy per particle $E$,
and c) amplitude $\protect\psi _{\max }$ vs. the number of molecules $N$ in
stationary QD\ states with the DDI and contact-interaction scattering
lengths $a_{\mathrm{dd}}=1000a_{0}$ and $a=100a_{0}$, and LHY strength $%
\protect\gamma _{\mathrm{LHY}}=1.6792\times 10^{9}a_{0}^{5/2}$, for the the
ITP input, with the phase-imprinting indices $S=0,1,2$, under the action of
the confining potential (\protect\ref{scal_pot}). All the configurations
presented here are stationary solutions obtained by means of the ITP method.
Solid and dashed lines represent families with $d\protect\mu/dN<0$ and $d%
\protect\mu/dN>0$, corresponding to the VK and anti-VK criteria,
respectively.} }
\label{n_mole}
\end{figure}
\begin{equation}
A_{0}>\left( \alpha \omega \right) ^{2}/2,  \label{>}
\end{equation}%
the potential (\ref{scal_pot}) determines an annular potential trough in the
$\left( x,y\right) $ plane, with a potential minimum located at
\begin{equation}
\rho _{0}^{2}=\alpha ^{2}\ln \left( \frac{2A_{0}}{\left( \alpha \omega
\right) ^{2}}\right) .  \label{rho}
\end{equation}%
which is real for $A_{0}>(\alpha \omega )^{2}/2$. It is straightforward to
see that this minimum is a global one, as the value $V\left( \rho
_{0}\right) $ is smaller than $V\left( \rho =0\right) $. The potential
trough provides radial pinning and favours the formation of necklace-shaped
multipole states.

If inequality (\ref{>}) degenerates into the exact equality, Eq. (\ref{rho})
yields a nearly flat potential minimum at the origin, $\rho _{0}=0$. In the
case of $A_{0}<(\omega \alpha )^{2}/2$, the HO term plays the dominant role,
with the entire trapping profile reducing to a single-well potential with
the minimum at $\rho =0$. In this case, the HO potential imposes strong
spatial constraint, compressing the QD toward the center and suppressing
multipole configurations, therefore we do not consider it below.

We begin the analysis with the degenerate limit, which marks the boundary
between the HO-dominated and annular regimes. Addressing this limit allows
us to identify the maximum orders of multipole modes and assess the
stability of the states in the absence of an annular trough. Building on
this baseline, we then proceed to the strong-trough regime, where an annular
minimum develops, and the multipole QDs acquire radial support, leading to
enhanced stability and higher-order multipole configurations. {In both
confinement regimes, we fix the angular frequency as $\omega=1$, while the
amplitude of the Gaussian barrier $A_{0}$ varies.} As concerns other
parameters, for the mass of the NaCs molecule, $m=156\times 1.66054\times
10^{-27}$ kg, and the reference angular frequency $\omega _{0}=2\pi \times
180$ Hz, the characteristic length and time in physical units are $L_{0}=%
\sqrt{\hbar /\left( m\omega _{0}\right) }\equiv 0.6~\mathrm{\mu }$m and $%
T_{0}\equiv \omega _{0}^{-1}=0.884$ ms. {The study of multipole QDs was
carried out by varying the molecule number $N$ and phase-imprinting index $S$
(alias the winding number) of the ITP input (recall that the ITP does not
conserve vorticity, resulting in multipolar zero-vorticity output states),
while fixing the DDI scattering length as $a_{\mathrm{dd}}=1000a_{0}$, the
scattering length of the contact interaction as $a=100a_{0}$, and the LHY
coefficient as $\gamma _{\mathrm{LHY}}=1.6792\times 10^{9}a_{0}^{5/2}$.} The
scaled equation (\ref{dgpe}) keeps the above-mentioned unitary
normalization, i.e.,
\begin{equation}
\int |\psi (\mathbf{r})|^{2}d\mathbf{r}=1.  \label{1}
\end{equation}

Finally, Eq. (\ref{dgpe}) can be presented in the Hamiltonian form,
\begin{equation}
i\frac{\partial \psi }{\partial t}=\frac{\delta E}{\delta \psi ^{\ast }}.
\end{equation}%
Here $\delta /\delta \psi ^{\ast }$ denotes the variational derivative, $%
\psi ^{\ast }$ is the complex conjugate of $\psi $, and $E$ is the energy
per particle,
\begin{eqnarray}
E &=&\int d\mathbf{r}\bigg[\frac{1}{2}|\nabla \psi (\mathbf{r}%
)|^{2}+V(\rho)|\psi (\mathbf{r})|^{2}+2\pi Na|\psi (\mathbf{r})|^{4}  \notag
\\
&&+\frac{3}{2}a_{\mathrm{dd}}N|\psi (\mathbf{r})|^{2}\int \frac{1-3\cos
^{2}\theta }{|\mathbf{R}|^{3}}|\psi (\mathbf{r}^{\prime })|^{2}d\mathbf{r}%
^{\prime }  \notag \\
&&+\frac{2\gamma _{\mathrm{LHY}}}{5}N^{3/2}|\psi (\mathbf{r})|^{5}\bigg].
\label{E}
\end{eqnarray}

{In the framework of the ITP method, the norm of the wave function is
conserved (set equal to unity), whereas, as mentioned above, the angular
momentum is not. Therefore, the converged stationary multipole
configurations are not vortex states carrying angular momentum. Instead,
they are composed as alternating out-of-phase density lobes, whose
contributions to the angular momentum cancel, yielding a vanishing
expectation value.}
\begin{equation}
{M=i\iiint \psi^{} \left(y\frac{\partial}{\partial x} -x\frac{\partial}{%
\partial y}\right) \psi\,d\mathbf{r}=0.}
\end{equation}%
{Thus, throughout this work, the parameter $S$ is regarded as a label
characterizing the initial phase-imprinted configuration, rather than the
angular momentum of the established stationary state.}

\section{Numerical results}

\label{results}
\begin{figure}[tbp]
\xincludegraphics[width=1.0\linewidth]{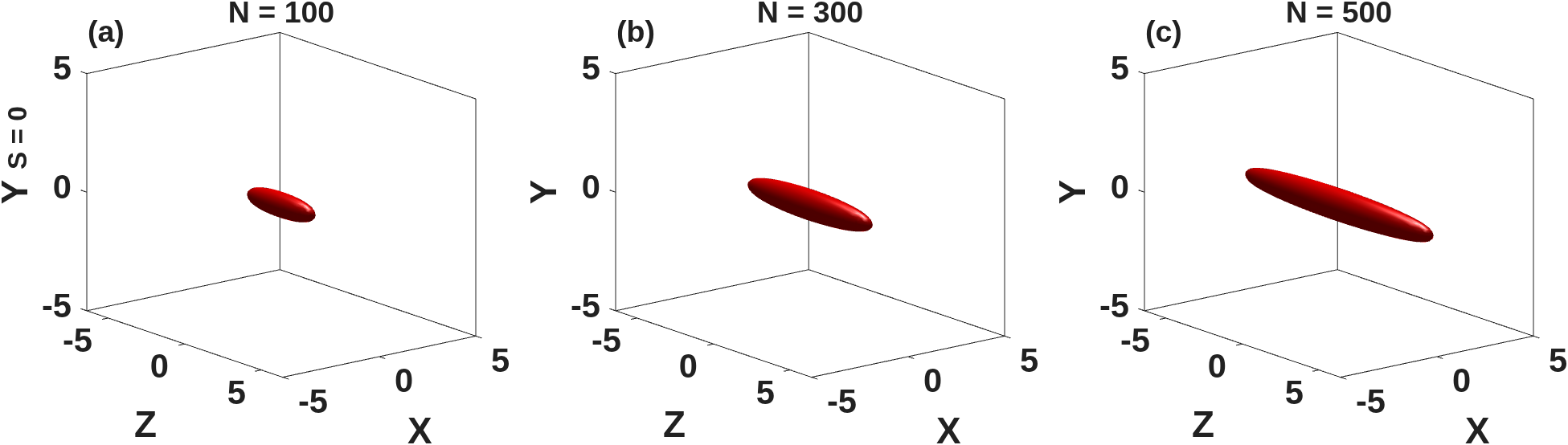} %
\xincludegraphics[width=1.0\linewidth]{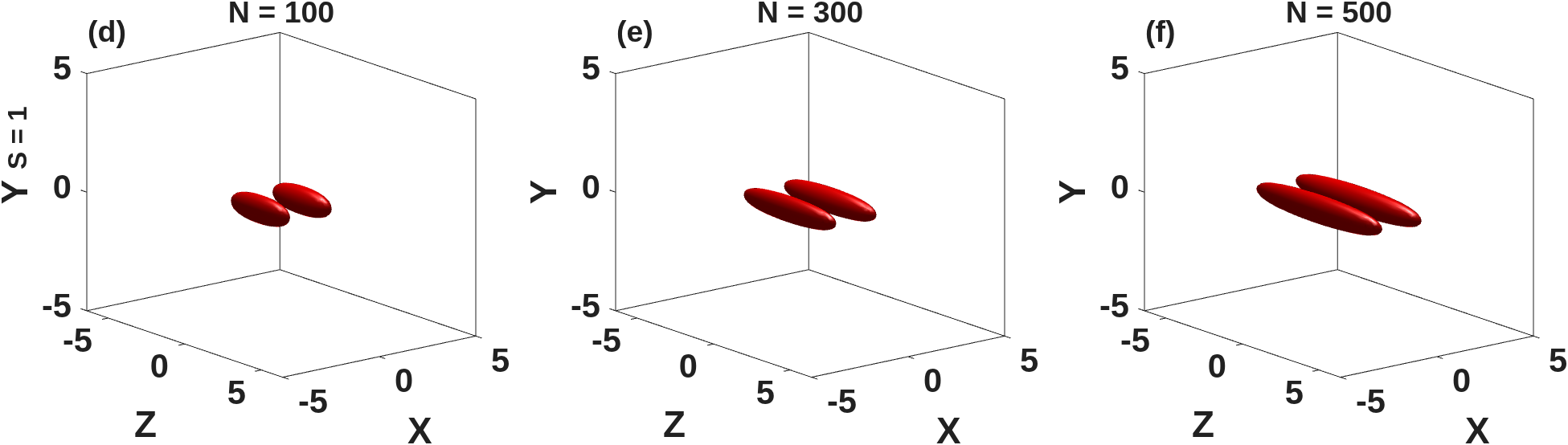}
\caption{ {Isosurface plots of quasi-1D stable multipole density profiles in
the DDI-attractive regime for vorticities $S=0$ and $1$ of the ITP input in
the top and bottom rows, and values of the molecule number $N=100$, $300$,
and $500$ in the columns (from left to right). The DDI and
contact-interaction scattering lengths are $a_{\mathrm{dd}}=1000a_{0}$ and $%
a=100a_{0}$, respectively. All the stationary states were produced by the
ITP method.} }
\label{datt}
\end{figure}
The integrodifferential GP equation (\ref{dgpe}) was solved numerically,
using the split-time-step Crank-Nicolson algorithm and adapting the
FORTRAN/C programs \cite{kumar2015fortran} or their open-multiprocessing
versions \cite{lonvcar2016openmp}. In particular, stationary solutions were
obtained by means of the ITP method with space discretization steps $dx=0.1$%
, $dy=0.1$, and $dz=0.125$. Such solutions are identified by the convergent
outcome of the ITP, with the chemical potential, energy per particle, and
density profile remaining unchanged upon further imaginary-time iterations,
within the available numerical accuracy. The relative changes in the
chemical potential and energy per particle are typically of the order of $%
10^{-4}$ or smaller. The numerical treatment of the dipolar term in Eq.~(\ref%
{dgpe}) in the configuration space is challenging due to the presence of the
singular factor $1/|\mathbf{R}|^{3}$ in the DDI kernel. As mentioned above,
to fix the issue we evaluated the dipolar potential in the momentum space,
using the convolution theorem, which is numerically advantageous because the
Fourier transform of the kernel is smooth and analytic. To regularize the
singularity at the origin, a spherical truncation procedure \cite%
{young2023openmp} was employed, ensuring that the kernel remains finite in
the simulations.

{To enable the numerical investigation of the multipole states, we use the
phase-imprinted Gaussian state as the initial ansatz:} \cite{young2025giant}%
:
\begin{equation}
\psi (x,y,z)\propto \left( x+i\,y\right) ^{S}\exp \left( -\frac{\rho ^{2}}{%
\delta _{r}^{2}}-\frac{z^{2}}{\delta _{z}^{2}}\right) ,\qquad  \label{input}
\end{equation}%
where $\delta _{r}$ and $\delta _{z}$ are the radial and vertical width
parameters. The numerical computations were performed for $\delta
_{r}=\delta _{z}\equiv \sqrt{2}$. {Here, $S$ is the winding number of the
phase-imprinted initial state used to initiate the ITP. It characterizes the
input, and does not represent the angular momentum of the resulting
stationary multipole state.} The chemical potential $\mu $ is evaluated from
the converged stationary spatial mode $\phi (\mathbf{r})$, produced by ITP,
by means of the expression following from Eq. (\ref{ph})
\begin{eqnarray}
\mu =\int &&d\mathbf{r}\Bigg[\frac{1}{2}|\nabla \phi (\mathbf{r}%
)|^{2}+V(r)|\phi (\mathbf{r})|^{2}+4\pi Na|\phi (\mathbf{r})|^{4}  \notag \\
&&+3a_{\mathrm{dd}}N|\phi (\mathbf{r})|^{2}\int \frac{1-3\cos ^{2}\theta }{|%
\mathbf{R}|^{3}}|\phi (\mathbf{r}^{\prime })|^{2}\,d\mathbf{r}^{\prime }
\notag \\
&&+\gamma _{\mathrm{LHY}}N^{3/2}|\phi (\mathbf{r})|^{5}\Bigg].
\end{eqnarray}

\subsection{The weak central Gaussian potential hill, \textbf{$A_{0}=(%
\protect\alpha \protect\omega )^{2}/2$}}

\label{deg}

We first address multipole QDs solutions in the degenerate case, $%
A_{0}=(\alpha \omega )^{2}/2$, which corresponds, as said above (see Eq. (%
\ref{rho})), to a flat (\textit{prolate}) potential minimum at $\rho =0$,
with different integer values of vorticity $S$.

\subsubsection{Stationary states of multipole droplets}

\label{stat_ste} Here, we address stable multipole QDs under the prolate
geometry defined by potential (\ref{scal_pot}). Naturally, multipole QDs
tend to form with an even number of lobes, reflecting the alternating phase
structure of the multipole configurations. {Typical stationary density
profiles of elongated multipole QDs with different numbers of molecules $N$
and phase-imprinting indices $S$ of the ITP inputs are presented in Fig. \ref%
{el_nacs}}. For $S=0$, a monopole (cylindrical-shaped QD) appears in Fig. %
\ref{el_nacs}(a). {As we use frequency $\omega =1$, representing a weak HO
trap, the increase of the number of molecules from $N=1000$ to $2000$ and $%
3000$ results in an increase of the QD radial size, as seen in Figs. \ref%
{el_nacs}(b) and \ref{el_nacs}(c).} For a stronger HO trap, the bound states
elongate along the $z$ direction, resulting in QDs with a quasi-1D shape.
\begin{figure}[tbp]
\xincludegraphics[width=0.49\linewidth, label={a)},pos={nw}]{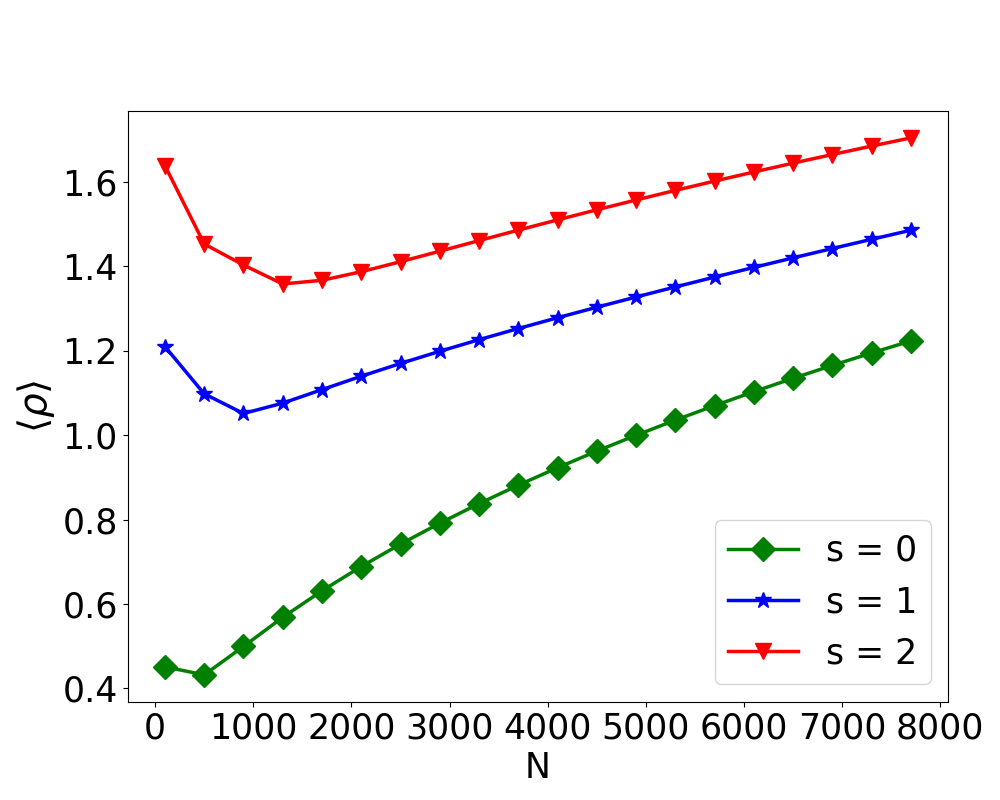} %
\xincludegraphics[width=0.49\linewidth, label={b)},pos={nw}]{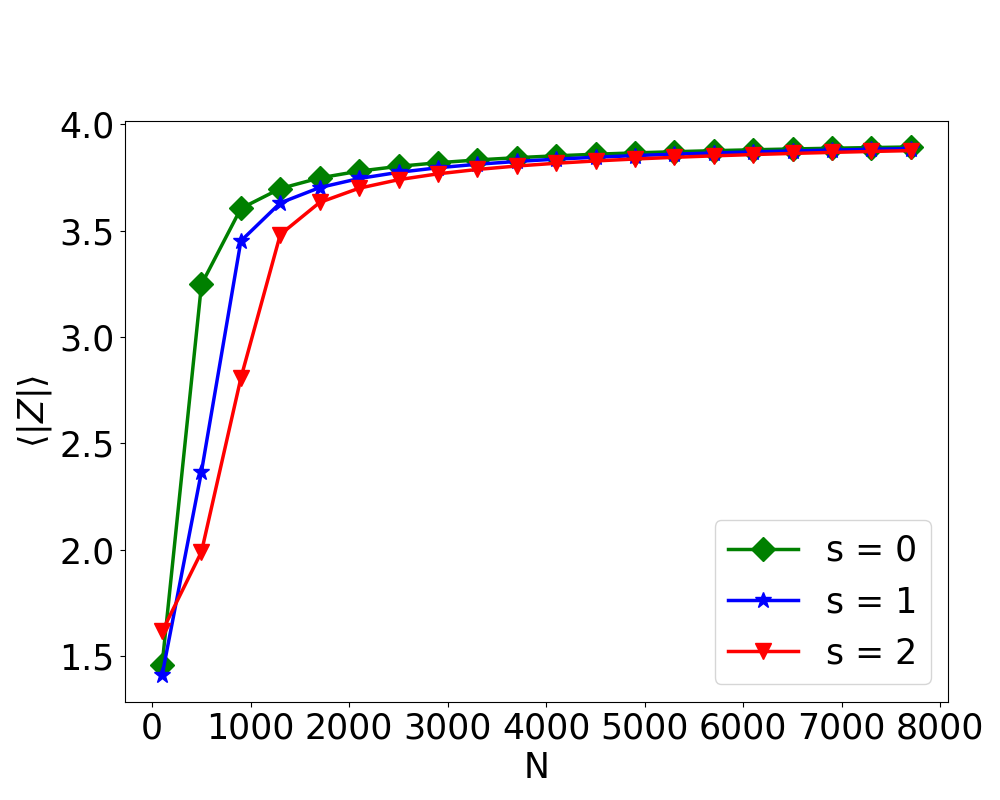}
\caption{{(a) The radial extension $\langle\protect\rho\rangle$ and (b) the
axial size $\langle |z|\rangle$ vs. the molecule number $N$, for winding
numbers vorticities $S=0,1,2$ of the ITP input, calculated according to
definitions (\protect\ref{radial_width}) and (\protect\ref{axial_width}),
for the DDI and contact-interaction scattering lengths $a_{\mathrm{dd}%
}=1000a_{0}$ and $a=100a_{0}$, and $\protect\gamma_{\mathrm{LHY}%
}=1.6792\times 10^{9}a_{0}^{5/2}$. All the configurations are stationary
solutions produced by the ITP method.}}
\label{thickness}
\end{figure}
\begin{figure}[tbp]
\xincludegraphics[width=0.49\linewidth, label={a)}]{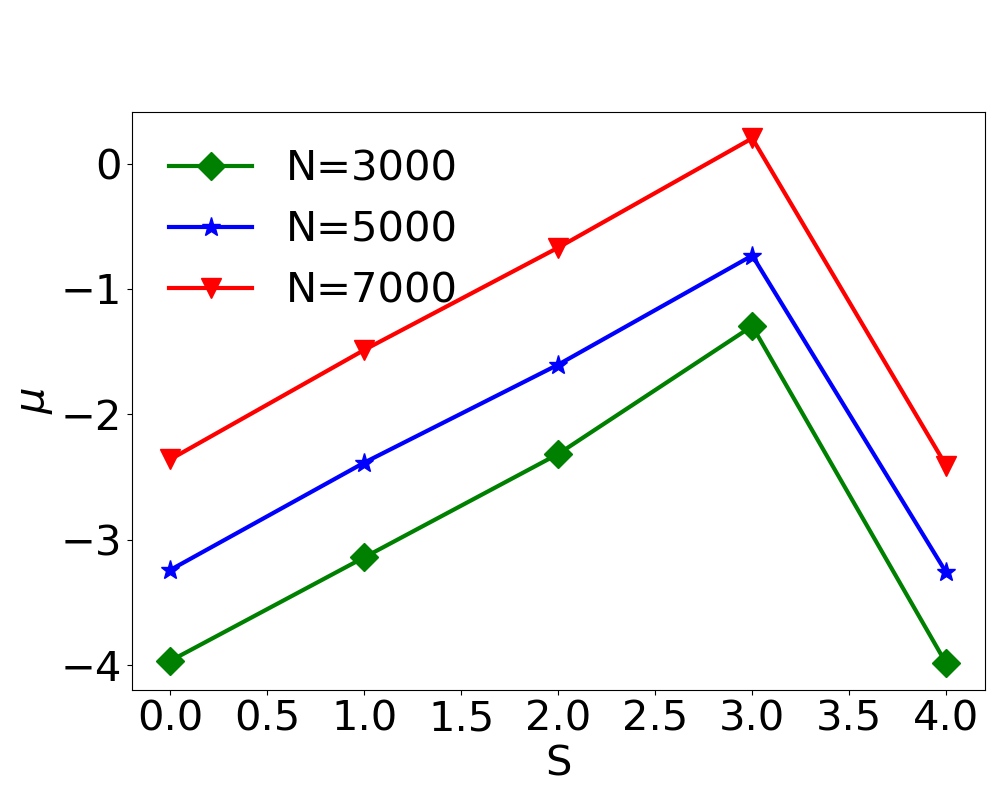} %
\xincludegraphics[width=0.49\linewidth, label={b)}]{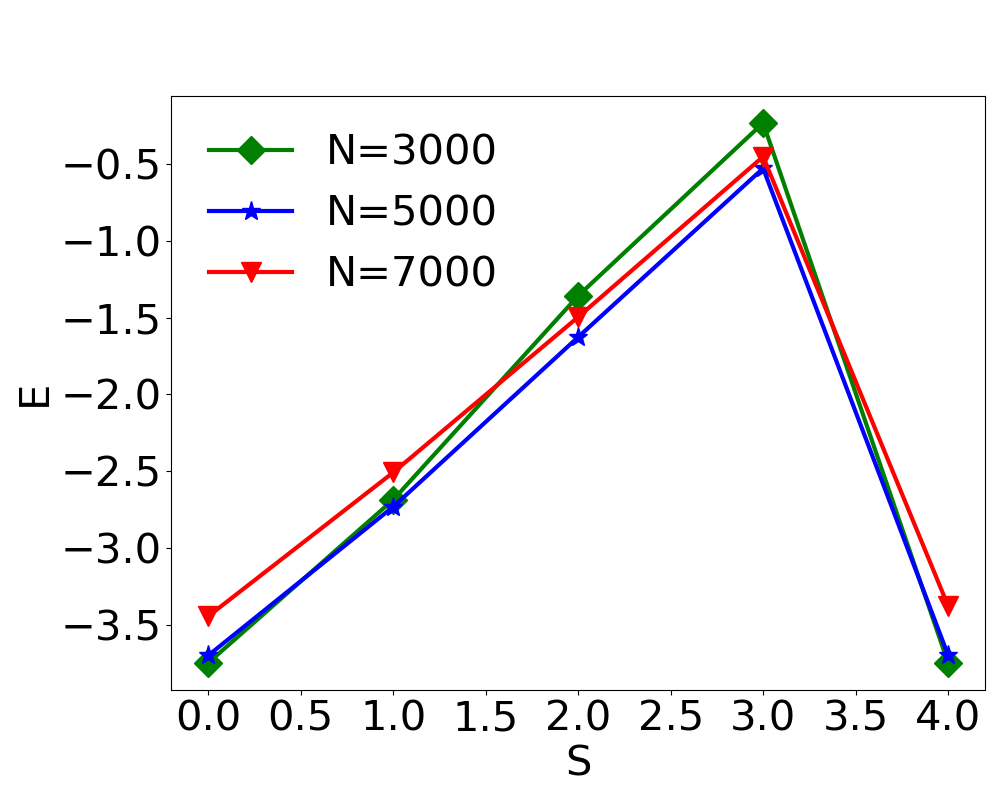}
\caption{{The variation of a) the chemical potential $\protect\mu$ and b)
energy-per-particle $E$ (defined as per Eq. (\protect\ref{E})) with respect
to the phase-imprinting index (winding number) $S$ of the ITP input, as
produced by the stationary solutions for different molecule numbers $N$,
with the DDI and contact-interaction scattering lengths $a_{\mathrm{dd}%
}=1000a_{0}$ and $a=100a_{0}$, and $\protect\gamma _{\mathrm{LHY}%
}=1.6792\times 10^{9}a_{0}^{5/2}$. The ground-state (minimum-energy)
configuration corresponds to $S=0$.}}
\label{s_vs_E}
\end{figure}
\begin{figure}[tbp]
\xincludegraphics[width=0.45\linewidth,label={a)$\quad S=0$}]{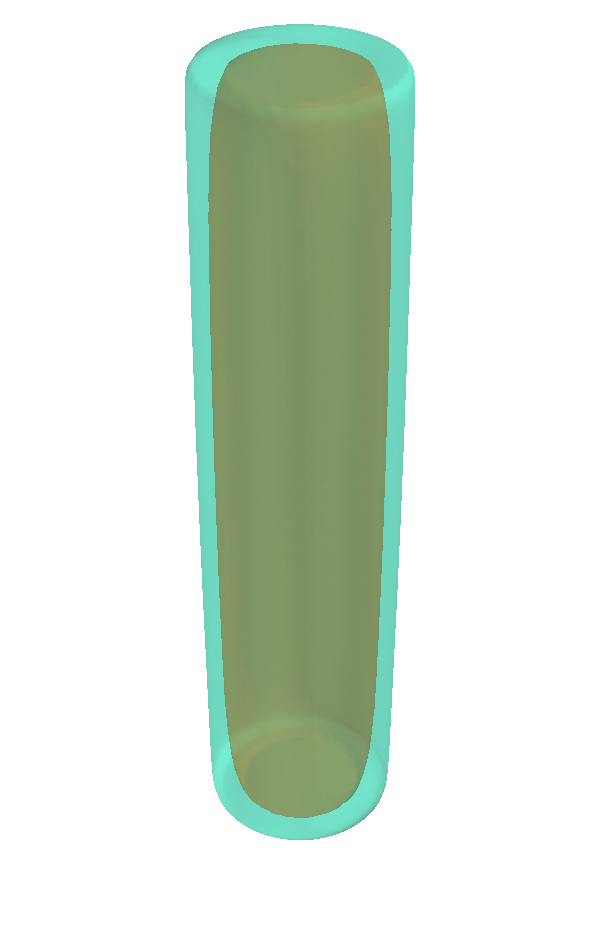} %
\xincludegraphics[width=0.45\linewidth,label={b)$\quad S=1$}]{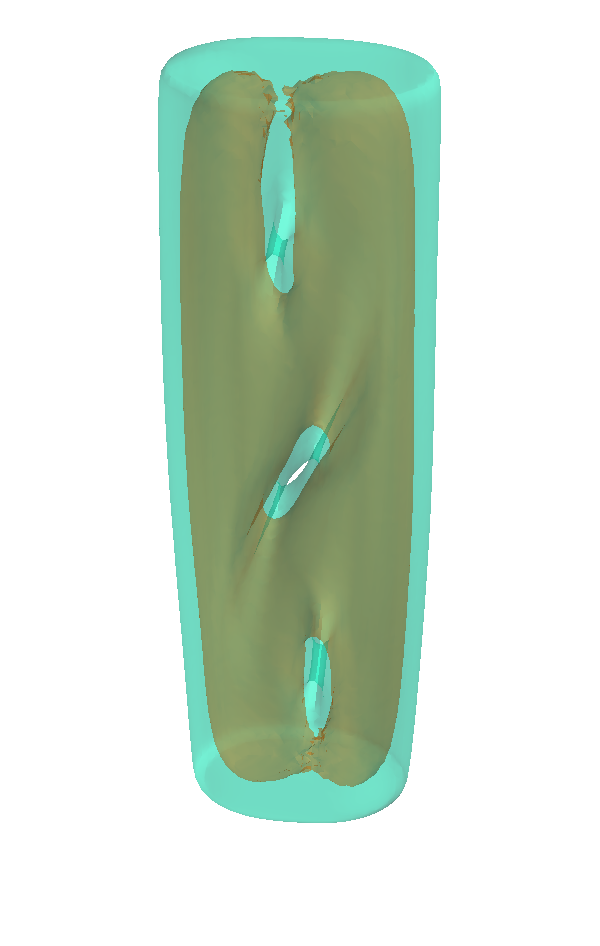} %
\xincludegraphics[width=0.45\linewidth,label={c)$\quad S=2$}]{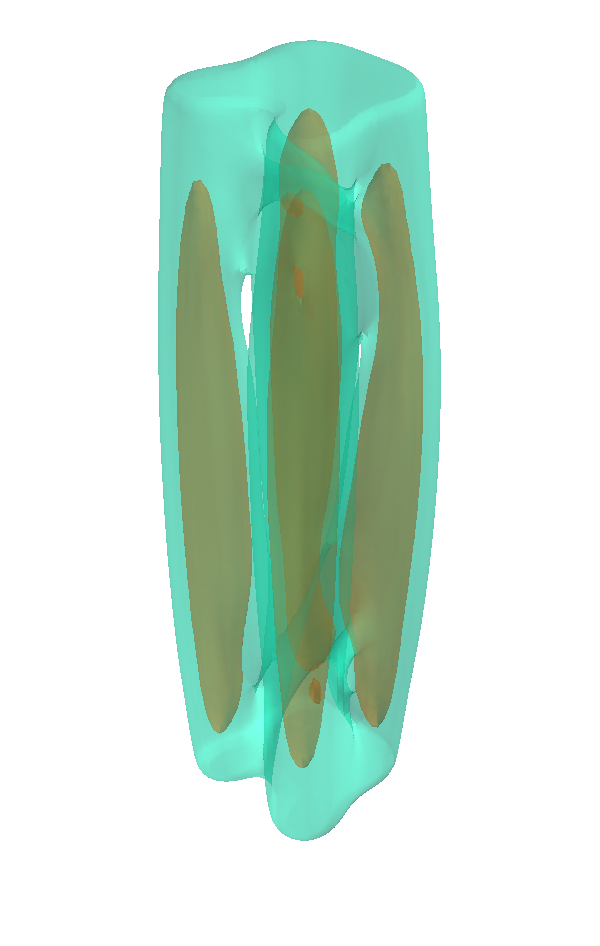} %
\xincludegraphics[width=0.45\linewidth,label={d)$\quad S=3$}]{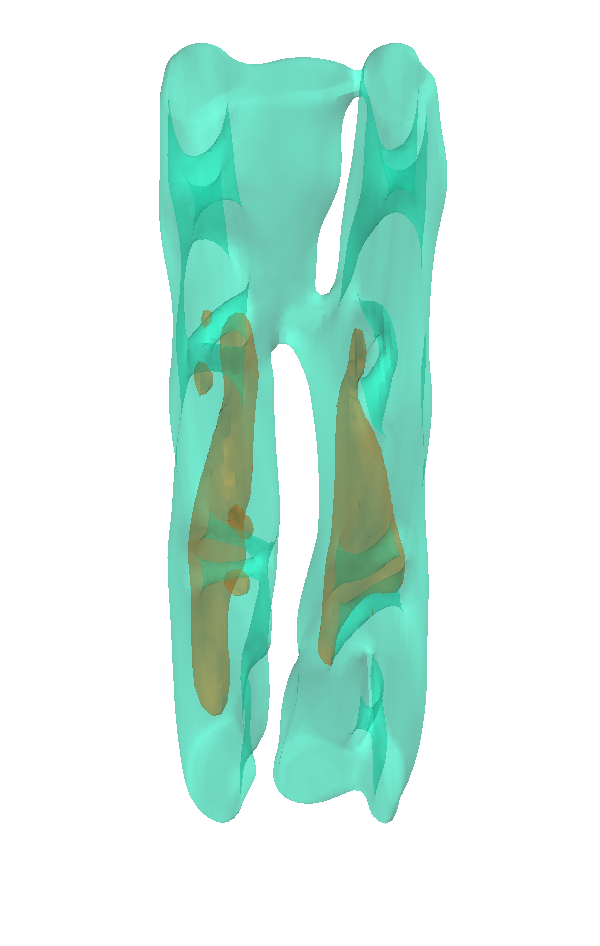}
\caption{{The results produced, at time $t=4.6415T_{0}\approx 4.1031$ ms
(with $T_{0}=0.884$ ms) by real-time simulations of the evolution of QDs,
initially perturbed by random noise at the $1\%$ level. The phase-imprinting
index (winding number) of the ITP input is (a) $S=0$, (b) $S=1$, (c) $S=2$,
and (d) $S=3$, showing that solely the QD with $S=0$ is stable. The results
are produced for $N=5000$, with the DDI and contact-interaction scattering
lengths $a_{\mathrm{dd}}=1000a_{0}$ and $a=100a_{0}$.}}
\label{rl_tme}
\end{figure}
When the phase-imprinting index is changed to $S=1$, QDs exhibits break the
rotational symmetry, and, due to the strong DDI, the droplets align along
the polarization direction, resulting in an elongated dipole-like
configuration, as shown in Fig. \ref{el_nacs}(d). Figures \ref{el_nacs}(e,f)
display the QDs of the same type, but for the larger numbers of molecules, $%
N=2000$ and $N=3000$, resulting in a larger QD radial size, as expected. A
similar study was performed for $S=2$, resulting in a quadrupole structure
of the droplets, as seen in Fig. \ref{el_nacs}(g-i). It is observed in Figs. %
\ref{el_nacs}(e,h) and (f,i) that, for the same value of $N$, the radial
width of each dipole QD is greater than that of its quadrupole counterpart,
because each lobe in the dipole configuration contains the double number of
molecules in comparison to the quadrupole. {The phase-imprinted input with $%
S=3$ and $N=1000$, shown in Fig. \ref{el_nacs} (j-l), produces a
dipole-shaped QD, while $N=2000$ leads to a sextupole state with an uneven
density distribution, which can be resolved by further increasing the
molecule number to $N=3000$, see panels (k) and (l).} Further, Fig. \ref%
{el_nacs} demonstrates that the number ($n$) of the density lobes (poles)
increase with $S$. In the simplest case of $S\leq 2$, the trend follows the
simple relation, $n=2S$, for any $N$. However, this correspondence is not a
universal rule and tends to deviate for $S=3$, where the interplay between
nonlinear interactions, and confinement geometry leads to a more specific
evolution, with the jump from $n=2$ to $n=6$ between panels (j) and (k). {%
Similar behavior has been reported for dipolar $^{164}$Dy QDs under the
action of the toroidal confinement~\cite{sanjay2026toroidal}, where the
multipole structures follow the relation $n=2S$ up to $S=6$.}

{The spatial extension of these QDs can be studied by analyzing 2D density
profiles (cross sections) in the $\left( x,y\right) $ plane. In Fig. \ref{2d}%
, they are plotted for the states displayed in Figs. \ref{el_nacs}(a)--(i),
for different values of $S$ and $N$. As $N$ increases in the monopole
configuration plotted in Fig. \ref{2d} for $N=1000$ in (a), $N=2000$ in (b),
and $N=3000$ in (c), the cross-section expands in the $\left( x,y\right) $
plane.} The same is true for the dipole and quadrupole states, as seen in
Figs. \ref{2d}(d-f) and \ref{2d}(g-i), respectively. The phase profiles of
these multipole states exhibit out-of-phase behavior (phase difference $\pi $%
) between adjacent lobes, which implies repulsion between them. The
interplay of the repulsive interactions with the overall trapping potential
helps to stabilize the multipole structure. Because the configurations
presented in Figs. \ref{el_nacs} are stationary solutions obtained by means
of the ITP method, they may be expected to represent the ground state (hence
they may be stable) for $S=0$. However, for $S\geq 1$ \cite{yrast1,yrast2},
the ITP procedure generates distinct stationary multipole configurations
from different phase-imprinted initial states, but these configurations are
not necessarily the ground state. Their dynamical stability therefore needs
to be examined independently by real-time perturbed propagation.
\begin{figure}[tbp]
\xincludegraphics[width=0.32\linewidth,label={a)$\quad t=3.61$
ms}]{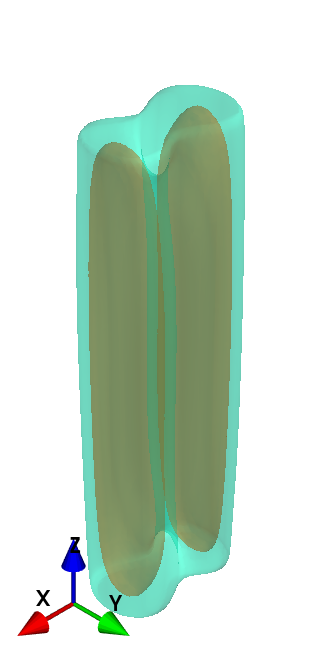}
\xincludegraphics[width=0.32\linewidth,label={b)$\quad
t=6.15$} ms]{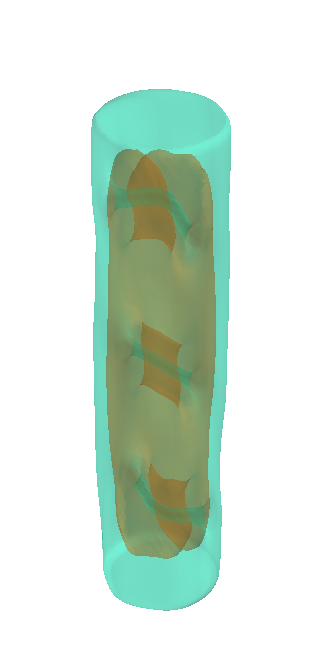} \xincludegraphics[width=0.32\linewidth,label={c)$%
\quad t=9.02$ ms}]{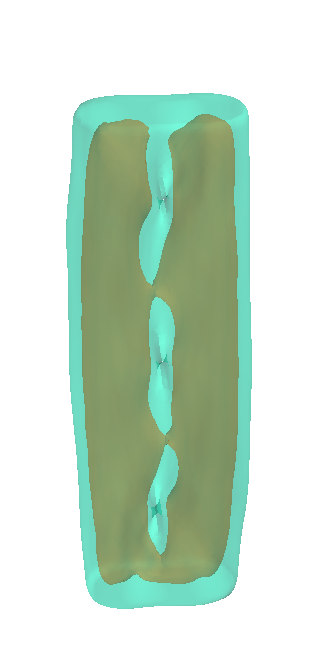} \xincludegraphics[width=0.32%
\linewidth,label={d)}]{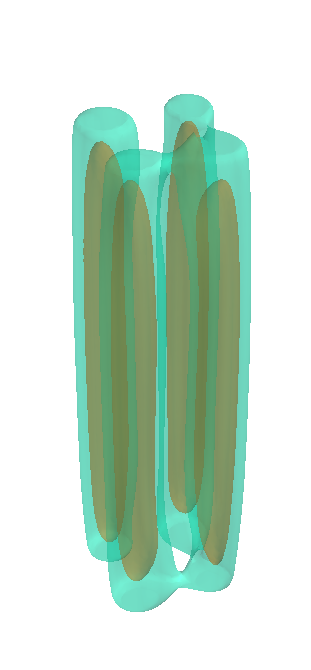} \xincludegraphics[width=0.32%
\linewidth,label={e)}]{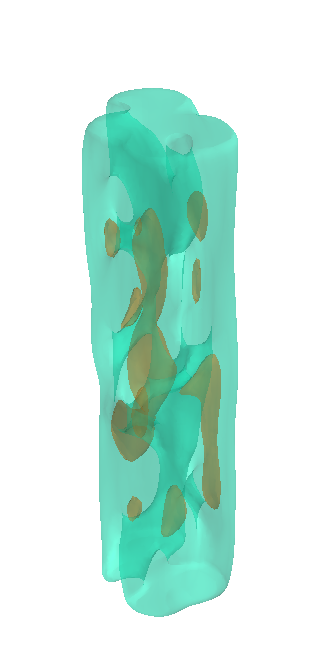} \xincludegraphics[width=0.32%
\linewidth,label={f)}]{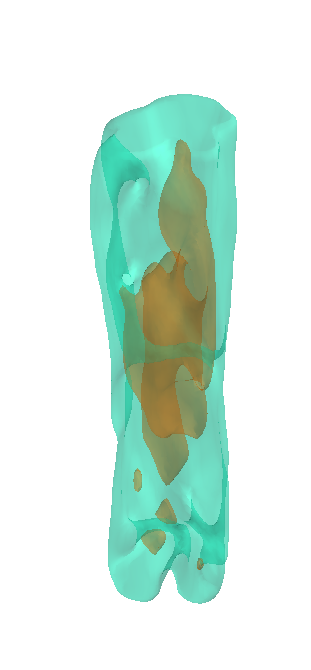} \xincludegraphics[width=0.32%
\linewidth,label={g)}]{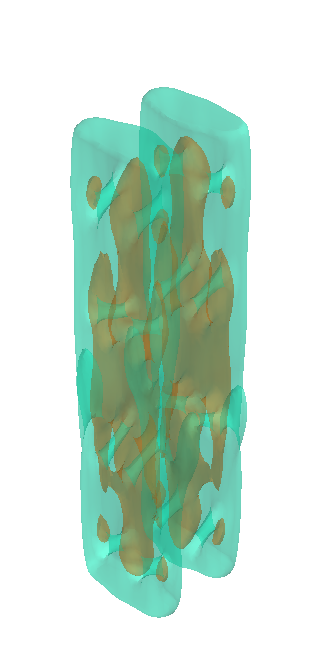} \xincludegraphics[width=0.32%
\linewidth,label={h)}]{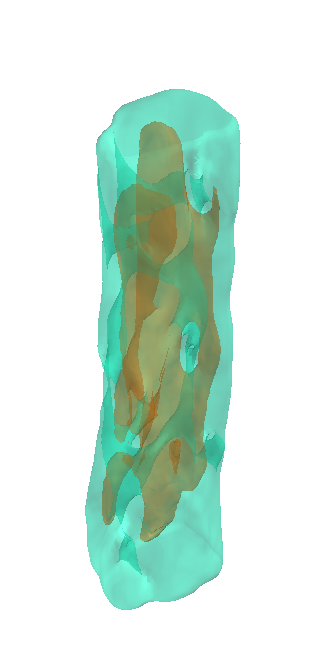} \xincludegraphics[width=0.32%
\linewidth,label={i)}]{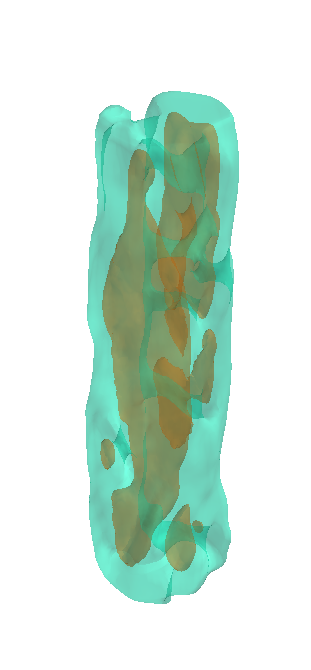}
\caption{ {The evolution of unstable multipole QDs perturbed by random noise
at the low $0.5\%$ level, for $N=5000$, the DDI and contact-interaction
scattering lengths $a_{\mathrm{dd}}=1000a_{0}$ and $a=100a_{0}$, and the LHY
strength $\protect\gamma_{\mathrm{LHY}}=1.6792\times 10^{9}a_{0}^{5/2}$.
Panels (a - c) show the evolution of the QD, produced by the ITP with the
phase-imprinting index $S=1 $ in the input, at $t=3.61$ ms, $6.15$ ms, and $%
9.02$ ms, respectively. Panels (d - f) and (g - i) represent the QDs created
by the ITP input with $S=2$ and $S=3$, respectively, displayed for the same
times. In all cases, the multipole configurations develop instability even
under the action of very weak perturbations, confirming that they initiate
genuine instability.}}
\label{rl_tme_0.01}
\end{figure}

\begin{figure}[tbp]
\xincludegraphics[width=0.2\linewidth,label={a)$\quad
t=1.02$ms}]{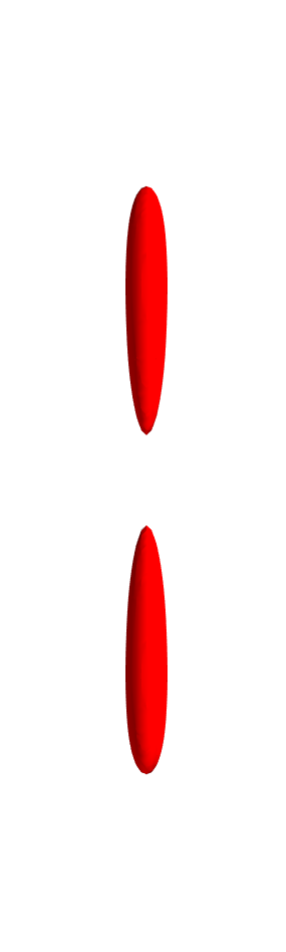} \xincludegraphics[width=0.2\linewidth,label={b)$%
\quad t=2.66$ms}]{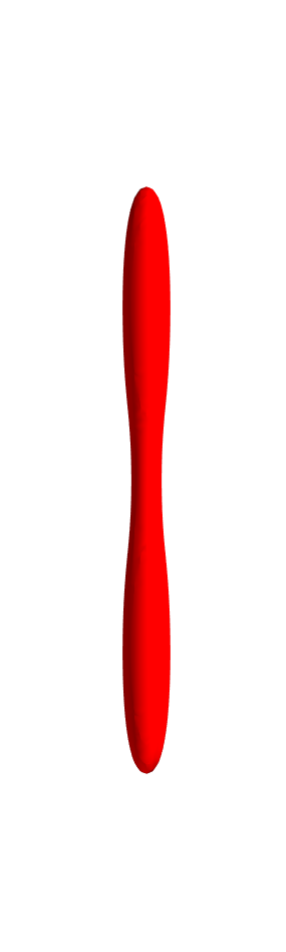}
\xincludegraphics[width=0.2\linewidth,label={c)$\quad
t=2.97$ms}]{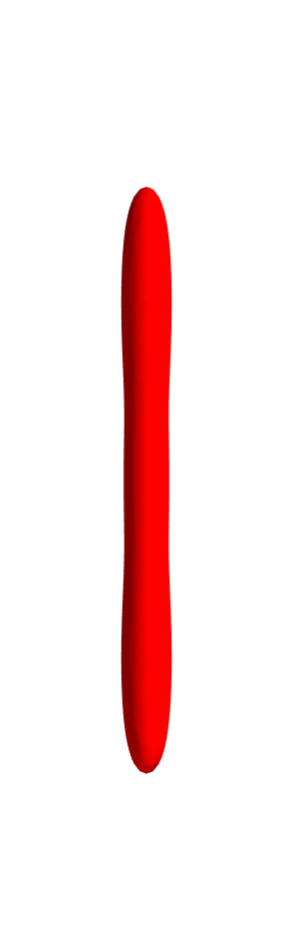} \xincludegraphics[width=0.2\linewidth,label={d)$%
\quad t=4.61$ms}]{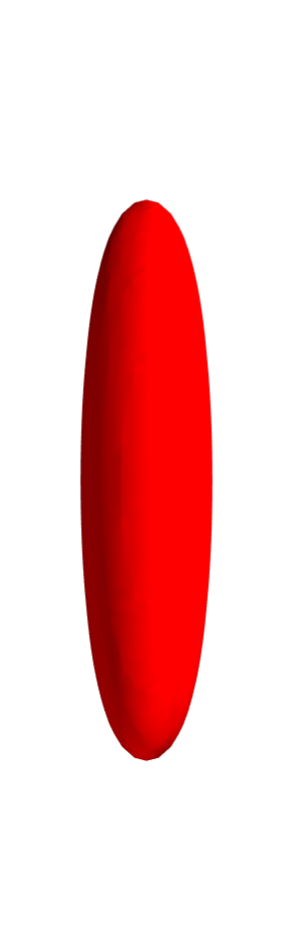}
\caption{ {A typical example of the simulated collision between identical
QDs with $S=0$ and $N=200$, which were set in motion along the $z$ axis by
kicks $\protect\eta =\pm 0.05$, starting from the initial separation $%
2z_{0}=38~\mathrm{\protect\mu}$m between the QDs. The DDI and
contact-interaction scattering lengths being $a_{\mathrm{dd}}=1000a_{0}$ and
$a=100a_{0}$. At these (and other) values of parameters, the collision leads
to the merger of the colliding droplets into a single one.} }
\label{deg_collision}
\end{figure}
Proceeding to characteristics of QD\ families, Fig.~\ref{n_mole}(a) presents
chemical potential $\mu $ as a function of the molecule number $N$ for
distinct vorticity values of ITP input. Obviously, the dependence shows $%
d\mu /dN<0$ for small $N$ and $d\mu /dN>0$ for large $N$. In the case of the
dominant repulsive nonlinearity, $d\mu /dN>0$ indicates that the localized
states may be stable, according to the anti-Vakhitov-Kolokolov (anti-VK)
necessary stability criterion \cite{HS}, while $d\mu /dN<0$ clearly implies
instability (the usual VK criterion, $d\mu /dN<0$, is the necessary
stability condition in systems dominated by the self-attractive nonlinearity
\cite{VK,Berge}). Further, Fig. ~\ref{n_mole}(b) displays the energy per
particle (\ref{E}) as a function of $N$. Similar to the $\mu (N)$
dependence, the energy is a non-monotonous function of $N$, with $dE/dN<0$
and $dE/dN>0$ at smaller and larger values of $N$, respectively. The
non-monotonous dependence of $\mu $ and $E$ on $N$ reflects the competition
between the MF attraction and LHY repulsion in the condensate. At low values
of $N$, the LHY effect is weak, and the attraction forces dominate, leading
to decreasing $\mu $. In this regime, dominated by the attractive DDI, the
QDs are aligned along the polarization direction, resulting in a quasi-1D
state, as shown in Fig. \ref{datt}. Panel (a) in the figure shows the QD for
$S=0$ with $N=100$. The increase of the molecule number to, $N=300$ and $%
N=500$, within the attractive regime, leads to the increase of the axial
size of the droplets, as shown in Figs. \ref{datt}(b,c). Panels (d-f)
pertain to $S=1$, with $N=100,300$, and $N=500$, respectively. As $N$ grows
beyond the attractive regime, the LHY term scales as $N^{3/2}$,
strengthening the repulsive pressure and increasing $\mu $ and $E$. It is
seen that the inclusion of the LHY correction leads to the robust
stabilization of the system with the increase of $N$. Figure~\ref{n_mole}(c)
shows the amplitude of the QD state as a function of $N$ for different $S$
configurations. In contrast to the trends observed for the chemical
potential and energy, the amplitude decreases monotonously with the increase
of $N$, being nearly identical for all vorticities $S=0$, $1$, and $2$
considered here. To further characterize the 3D structure of the multipole
QDs, we computed their average radial extension $\langle \rho \rangle $ and
axial size $\langle |z|\rangle $ as functions of $N$, for different
vorticities, as per the following natural definitions:
\begin{subequations}
\label{rad_and_axi}
\begin{align}
{\langle \rho \rangle} &{=} {\iiint \sqrt{x^{2}+y^{2}}\,|\psi(x,y,z)|^{2}%
\,dx\,dy\,dz,}  \label{radial_width} \\
{\langle |z|\rangle} &{=} {\iiint |z|\,|\psi(x,y,z)|^{2}\,dx\,dy\,dz.}
\label{axial_width}
\end{align}
As shown in Fig.~\ref{thickness}(a), the radial size $\langle \rho \rangle $
for $S=0$ increases with $N$, while the axial width initially increases and
attains a nearly constant value, see Fig.~\ref{thickness}(b). For the states
with $S=1$ and $S=2$, the radial width initially decreases while the axial
width increases. In the low-$N$ regime, the effectively attractive DDI along
the $z$ axis dominates over the LHY repulsive term, causing elongation of
the droplet in the axial direction, while lobes of the multipole state are
drawn inward in the radial direction. As $N$ increases, the LHY repulsion
becomes comparable to and eventually stronger than the DDI attraction,
suppressing further axial elongation and leading to radial expansion of the
droplet. At large $N$, the axial size approaches a nearly constant value,
due to the balance with the repulsive LHY term. It is observed that the
radial width increases with the growth of $N$ and $S$, whereas amplitude $%
|\psi |_{\max }$ exhibits a decreasing trend. This behavior is a natural
consequence of the radial spreading due to the increased $N$ value. Figure~%
\ref{s_vs_E} shows the variation of the chemical potential $\mu $ in panel
(a), and energy per particle $E$ in (b), with respect to the
phase-imprinting index $S$ of the ITP input $S$, for molecule numbers $%
N=3000,5000,7000$. In Fig.~\ref{s_vs_E}b), it is observed that the ground
state (energy minimum) corresponds to $S=0$ for each value of $N$.
\begin{figure}[tbp]
\xincludegraphics[width=1.0\linewidth]{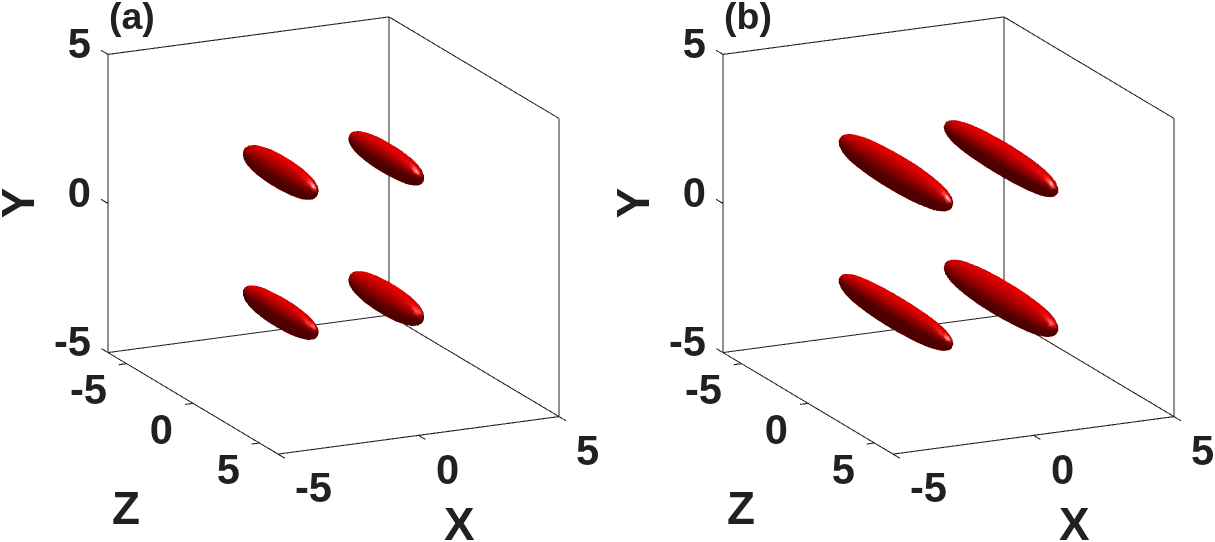} %
\xincludegraphics[width=1.0\linewidth]{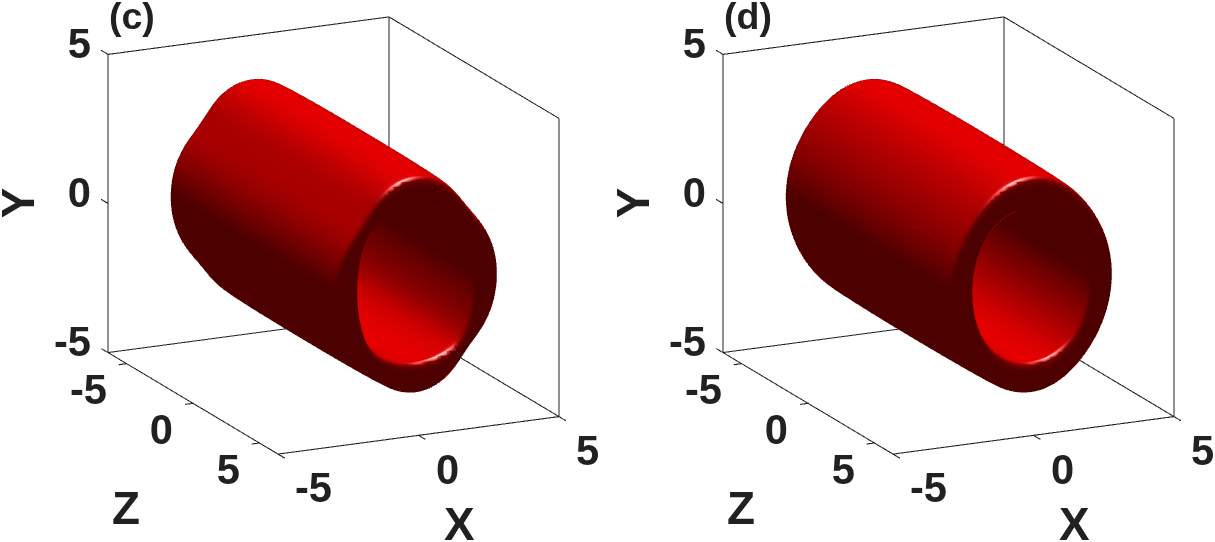}
\caption{ {Isosurface plots illustrating stationary density profiles of QDs
for $S=0$, under the action of the strong Gaussian confinement with strength
$A_{0}=30$ and HO trapping frequency $\protect\omega =1$ in Eq. (\protect\ref%
{elongated_trap}), satisfying condition~(\protect\ref{scal_pot}). Panels (a)
and (b) display four-droplet structures with molecule numbers $N=500$ and $%
N=1000$, respectively. Panels (c) and (d) show hollow-cylinder-shaped QDs
formed by higher molecule numbers, $N=6900$ and $N=7700$. All stationary
states are produced by the ITP method with the DDI and contact-inraction
scattering lengths $a_{\mathrm{dd}}=1000a_{0}$ and $a=100a_{0}$, and the LHY
strength $\protect\gamma_{\mathrm{LHY}}=1.6792\times 10^{9}a_{0}^{5/2}$.}}
\label{hollow}
\end{figure}
\begin{figure}[tbp]
\xincludegraphics[width=0.24\linewidth]{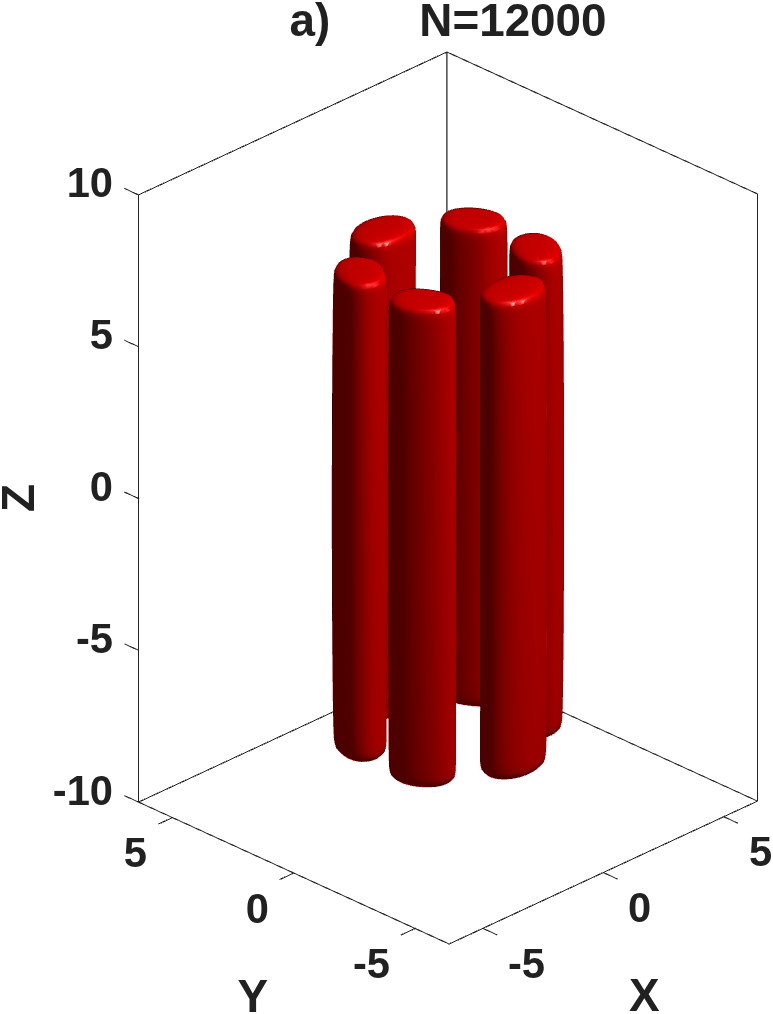} %
\xincludegraphics[width=0.24\linewidth]{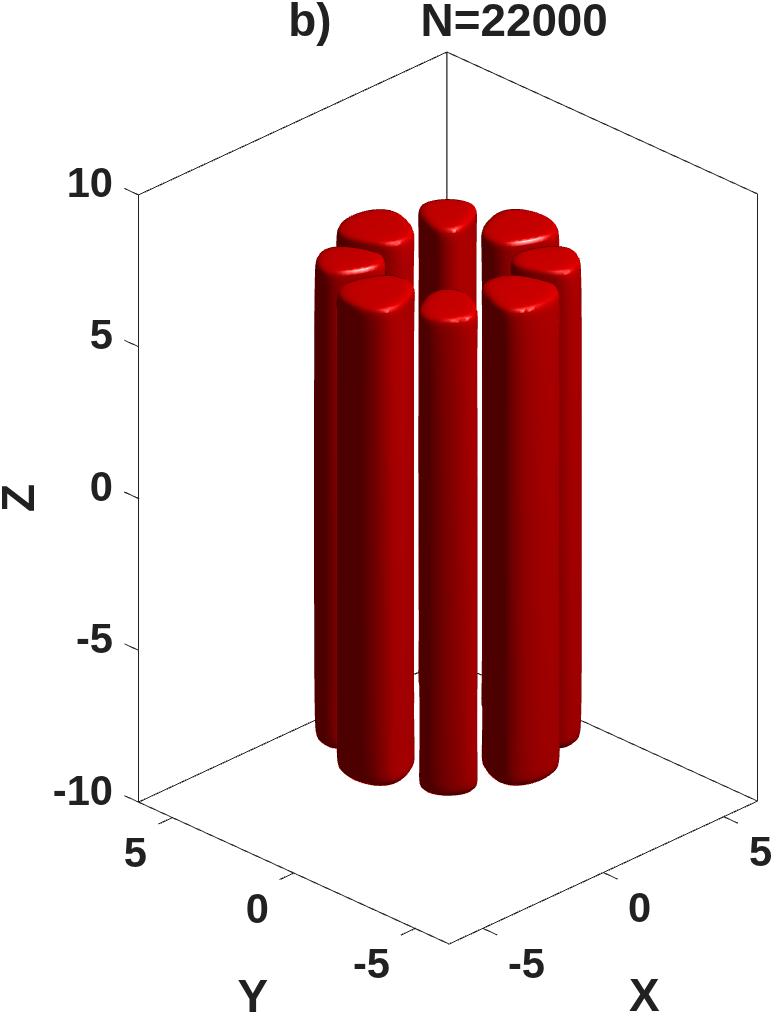} %
\xincludegraphics[width=0.24\linewidth]{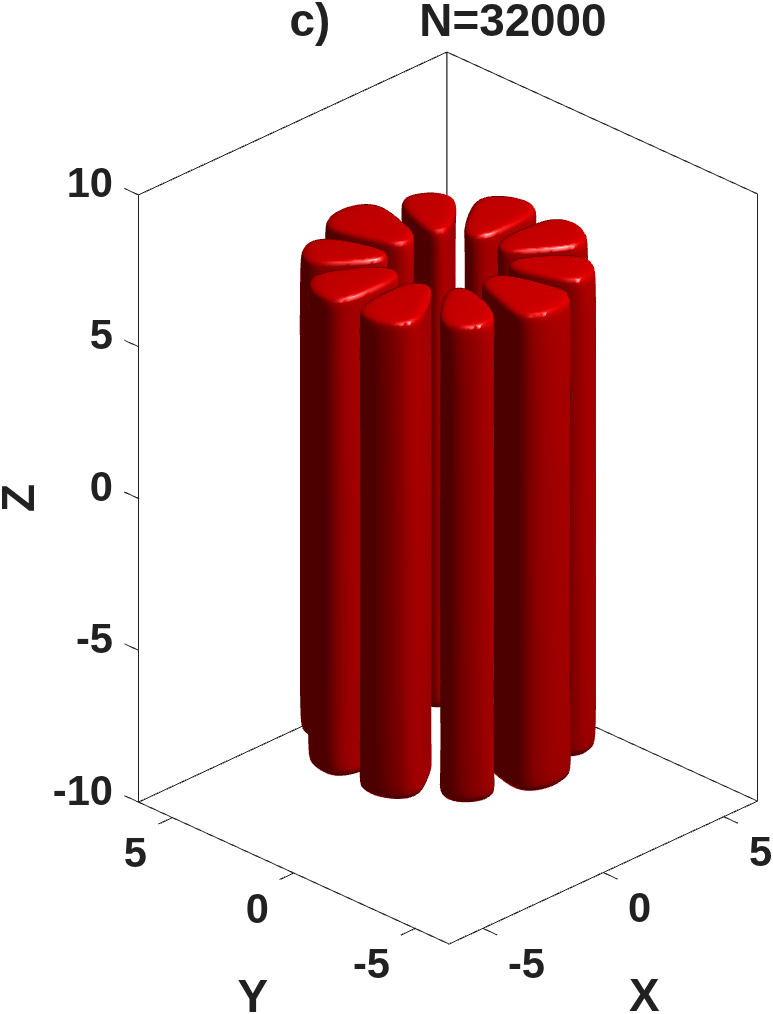} %
\xincludegraphics[width=0.24\linewidth]{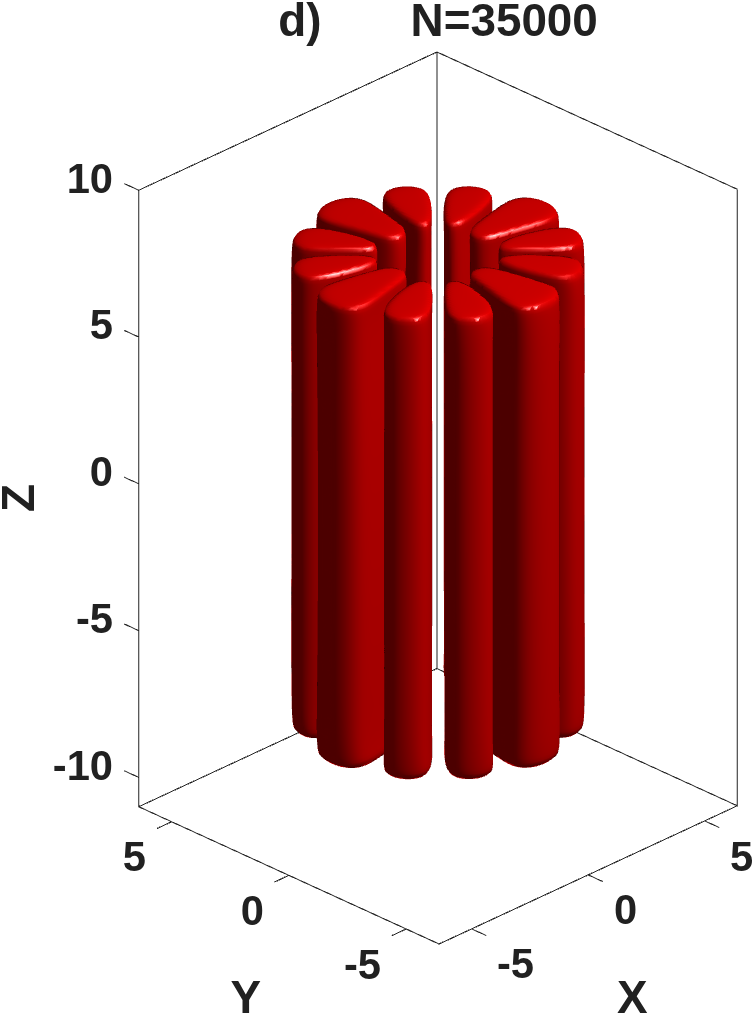}
\caption{Isosurface plots of stable stationary multipole QD density
profiles obtained under the action of the strong central Gaussian potential
hill with height $A_{0}>(\protect\alpha \protect\omega )^{2}/2$, for
different values of vorticity $S$ of the ITP input and molecule number $N$.
The DDI and contact-interaction scattering lengths are $a_{\mathrm{dd}%
}=1000a_{0}$ and $a=100a_{0}$, with LHY strength $\protect\gamma _{\mathrm{%
LHY}}=1.6792\times 10^{9}a_{0}^{5/2}$. Panel (a) shows a six-pole state with
$S=3$ and $N=12000$. Panel (b) displays an eight-pole state with $S=4$ and $%
N=22000$, while panels (c) and (d) show ten- and twelve-pole states, with $%
S=5$, $N=32000$ and $S=6$, $N=35000$, respectively. The trapping parameters
in Eq. (\protect\ref{elongated_trap}), taken for panels (a) - (d) are $%
A_{0}=30$ and $\protect\omega =1$, which satisfy condition~(\protect\ref%
{scal_pot}). The stationary states were produced by means of the ITP method.}
\label{hig_mul}
\end{figure}
\begin{figure}[tbp]
\xincludegraphics[width=0.49\linewidth,label={a)}]{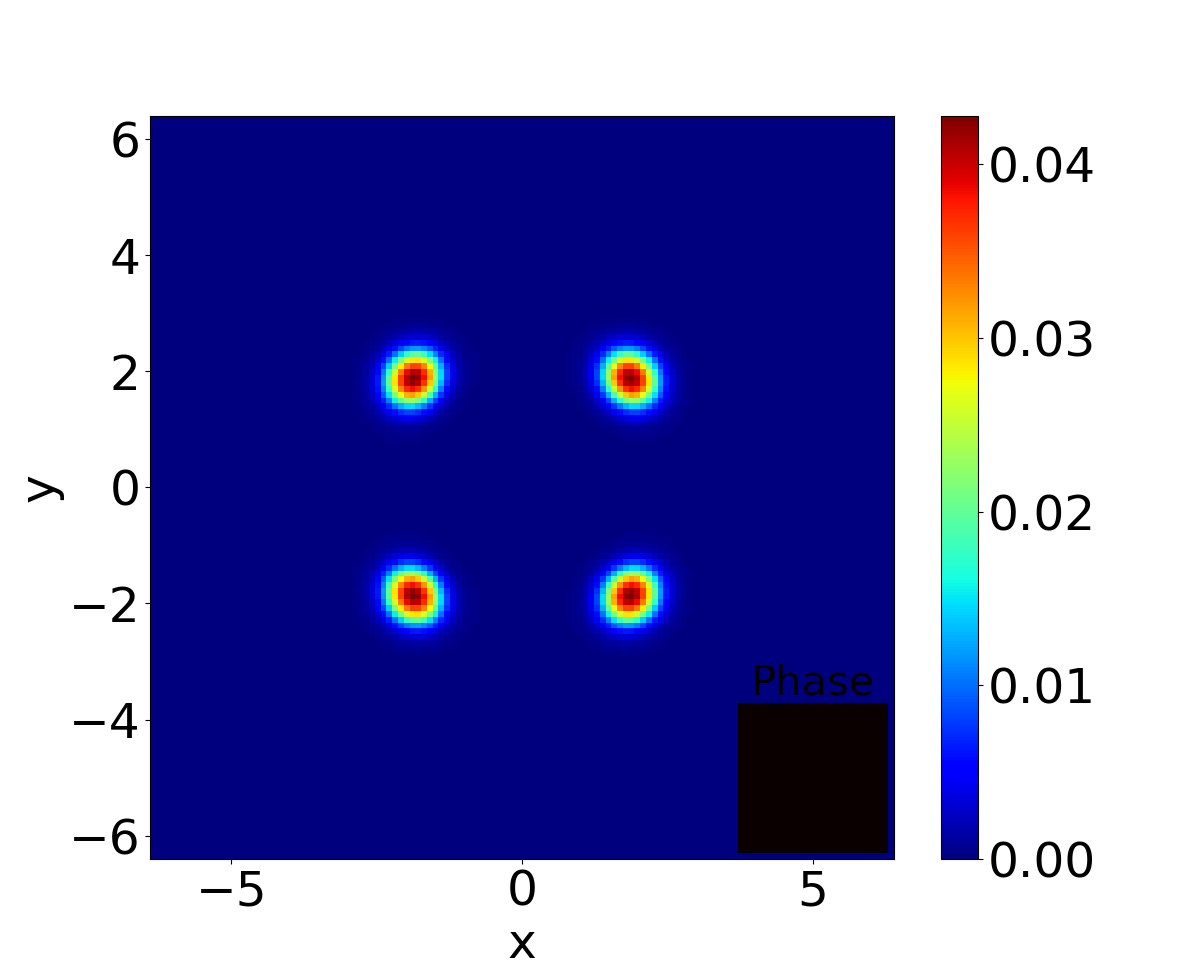} %
\xincludegraphics[width=0.49\linewidth,label={b)}]{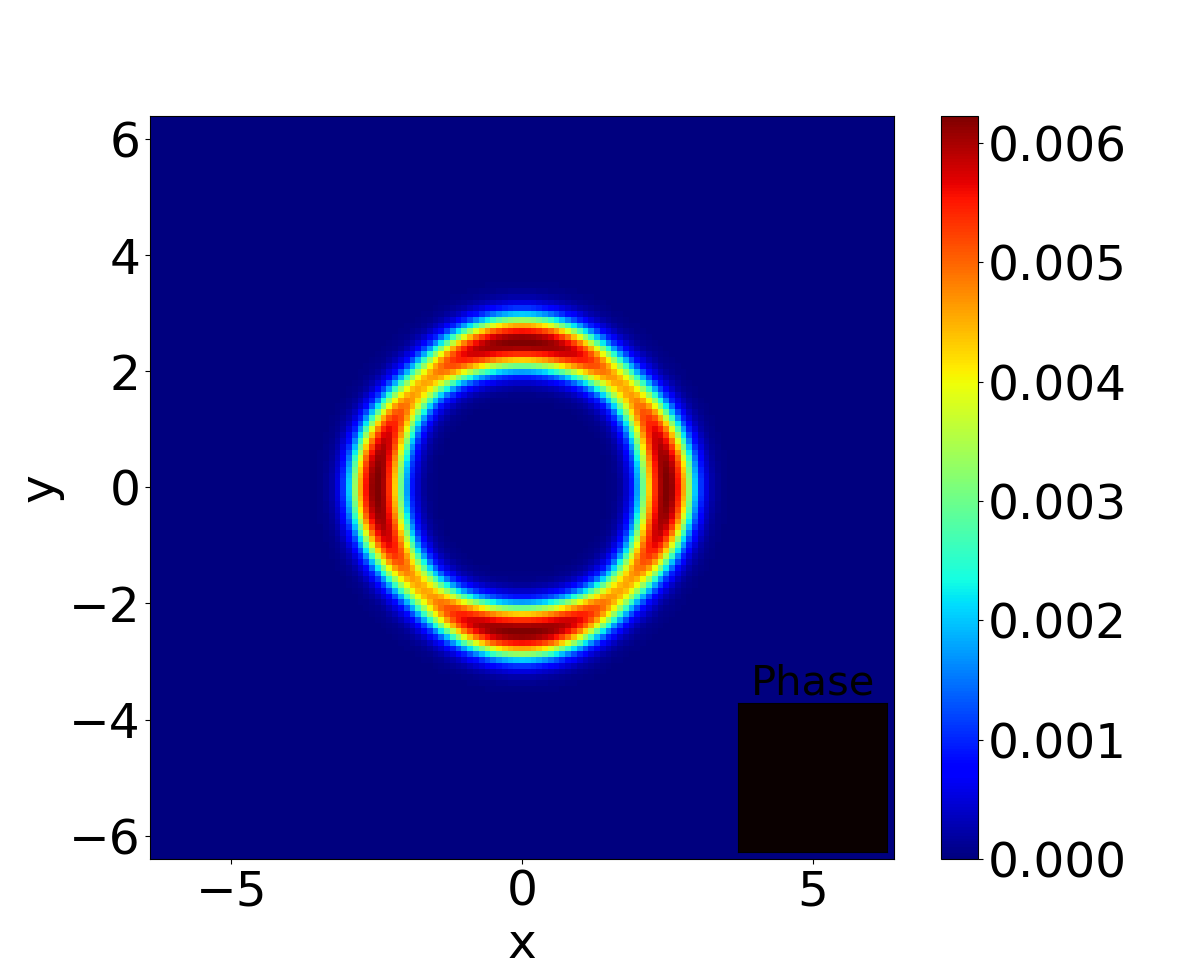} %
\xincludegraphics[width=0.49\linewidth,label={c)}]{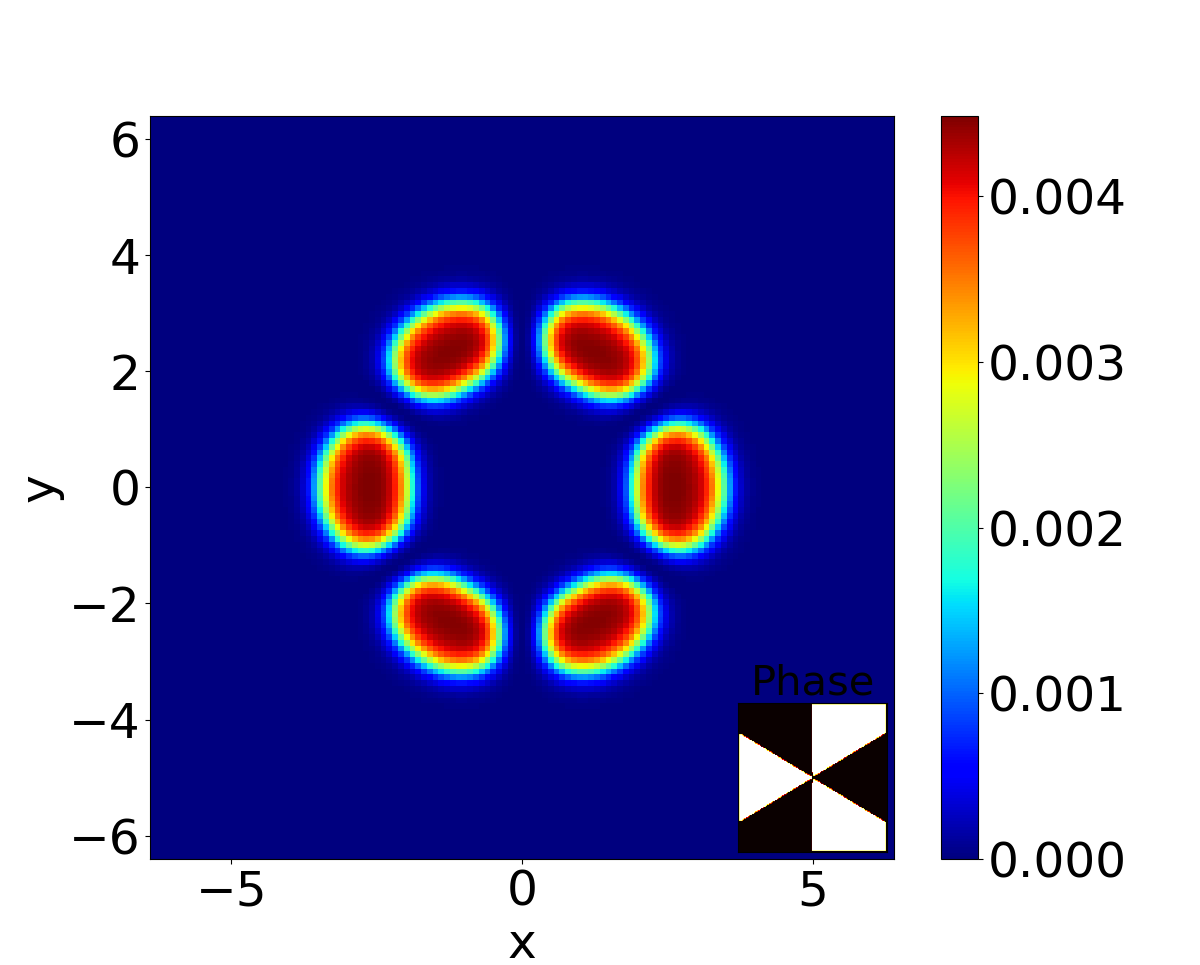} %
\xincludegraphics[width=0.49\linewidth,label={d)}]{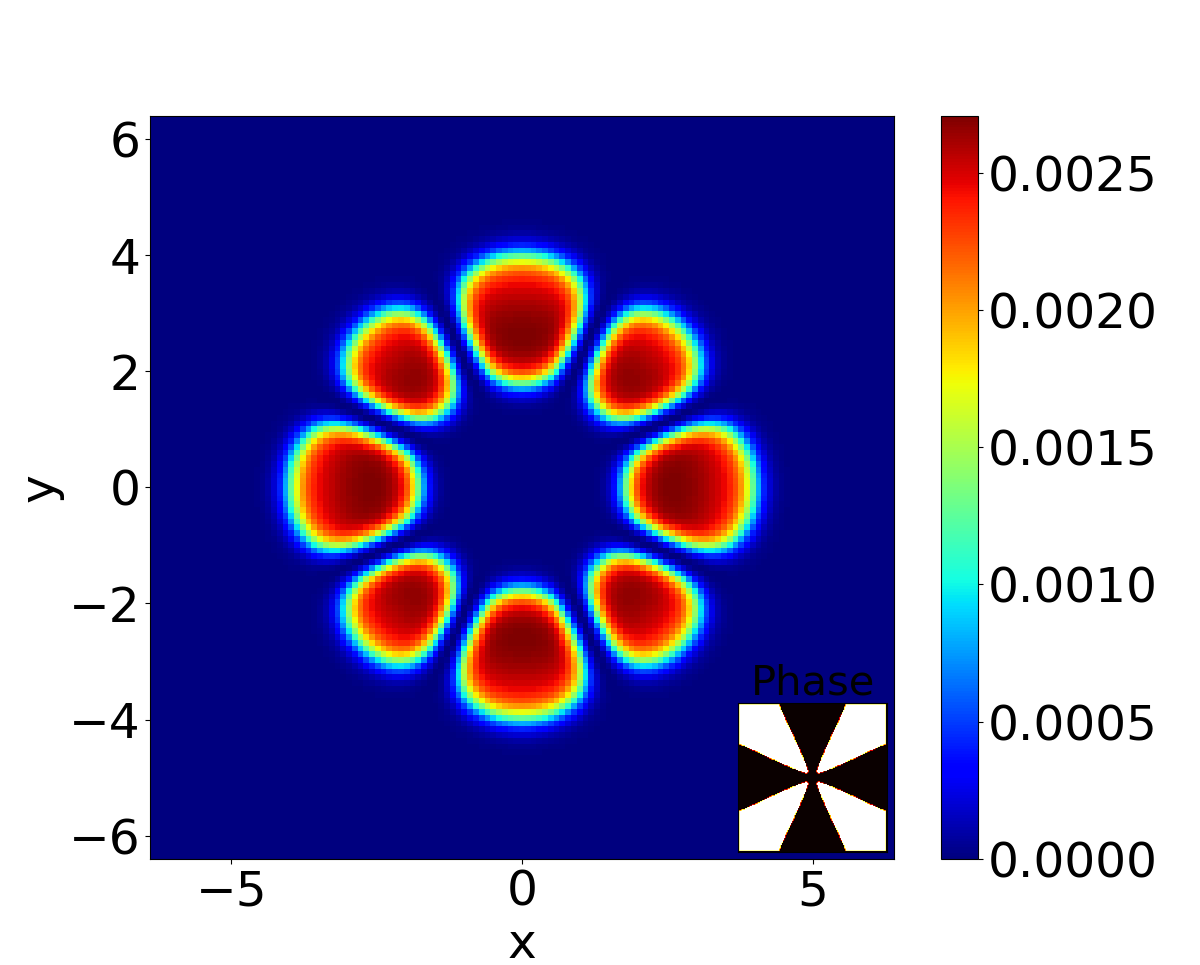} %
\xincludegraphics[width=0.49\linewidth,label={e)}]{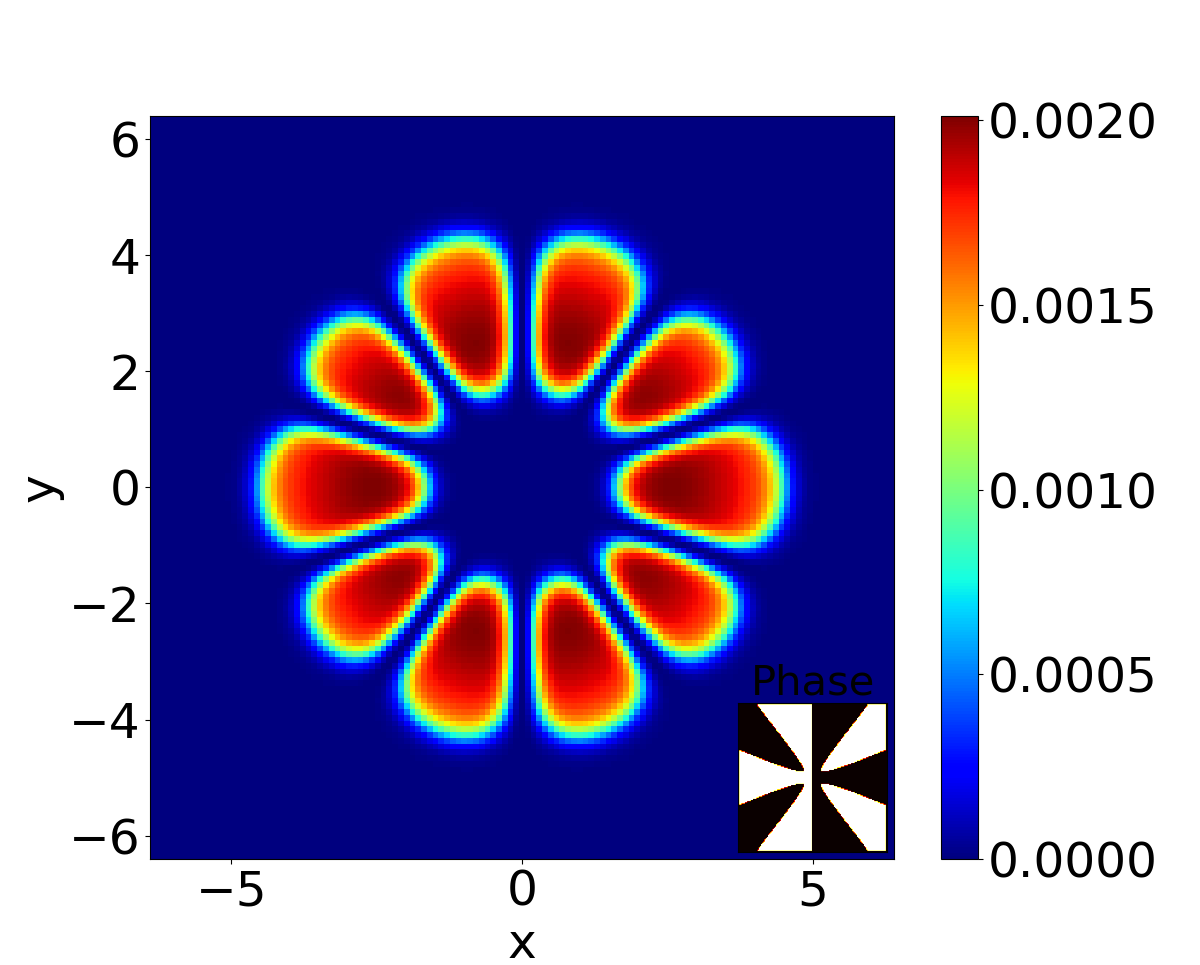} %
\xincludegraphics[width=0.49\linewidth,label={f)}]{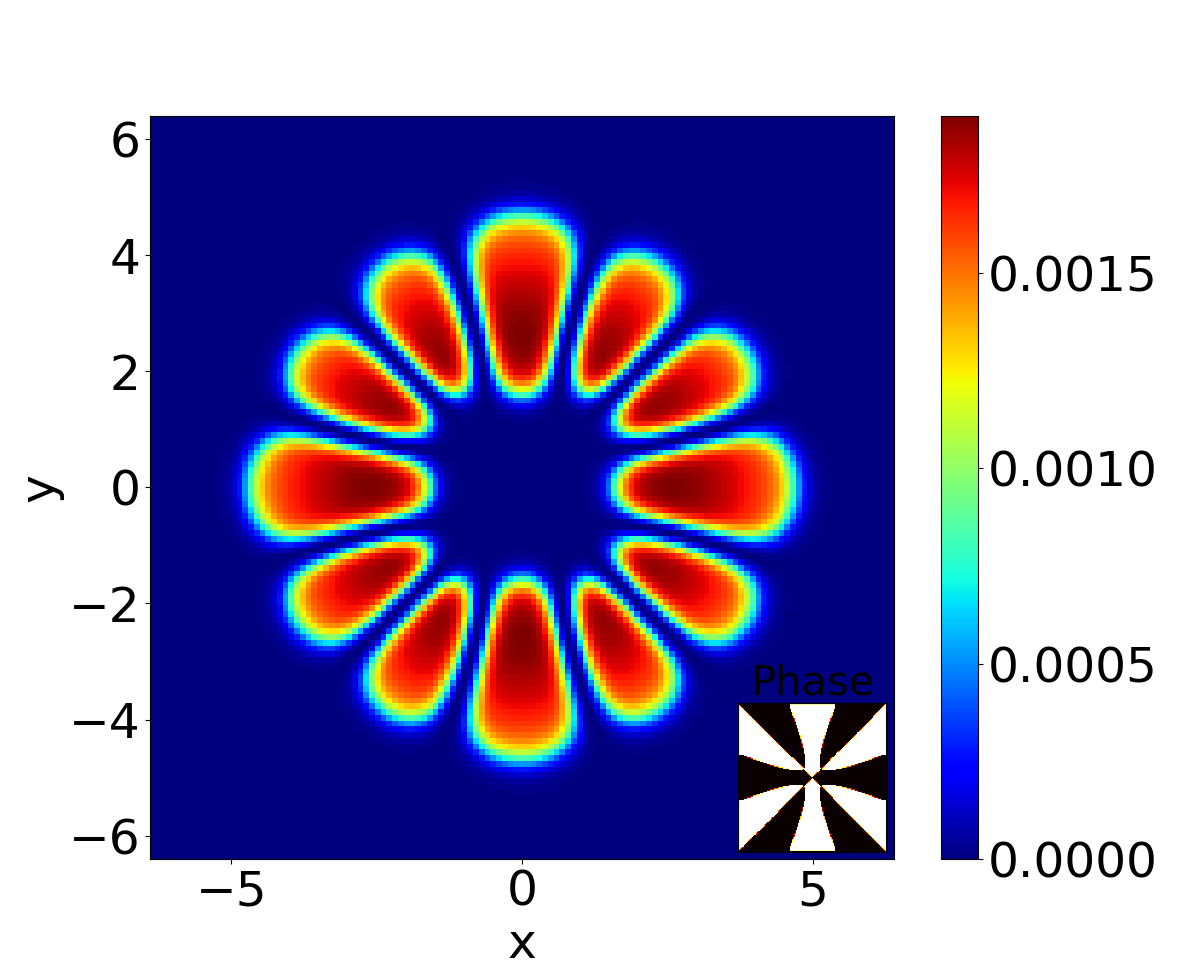}
\caption{ {Contour density plots in the $(x,y)$ plane for stable stationary
QDs. The DDI and contact-interaction lengths are $a_{\mathrm{dd}}=1000a_{0}$
and $a=100a_{0}$, respectively, and the LHY strength is $\protect\gamma_{%
\mathrm{LHY}}=1.6792\times 10^{9}a_{0}^{5/2}$. Panels (a) and (b) represent
a four-droplet configuration with $S=0$ and $N=1000$, while the increase of
the molecule number to $N=6900$ results in the merger of the separated
droplets into a single one with a hollow-cylinder shape. Panels (c - f)
display higher-order multipole QDs created by the ITP with the input
phase-imprinting indices $S=3$ - $6$ and $N=12000$, $N=22000$, $N=32000$,
and $N=35000$, respectively. The trapping parameters in Eq. (\protect\ref%
{elongated_trap}), satisfying condition~(\protect\ref{scal_pot}), are $%
A_{0}=30$ and $\protect\omega =1$, in panels (a - f). The inset in each
panel represents the corresponding phase profile.}}
\label{2d_hig}
\end{figure}

\subsubsection{The evolution of multipole droplets}

The above-mentioned anti-VK criterion, $d\mu /dN>0$, which holds in Fig.~\ref%
{n_mole}(a), is a necessary but not sufficient stability condition, in the
case of the dominant repulsive nonlinearity. To assess the full QD\
stability, we performed real-time simulations, initially perturbing the
stationary solutions by random noise and monitoring the subsequent
evolution. {The random perturbation was applied to the amplitude of the
established stationary solution, multiplying it by the perturbation factor
as follows:
\begin{equation}
\psi(\mathbf{r},0)=\left[1+\epsilon\,\delta(\mathbf{r})\right]\psi_{ITP}(%
\mathbf{r}),
\end{equation}
where $\psi_{ITP}$ is the stationary solution produced by the ITP method, $%
\epsilon$ specifies the perturbation strength ($\epsilon=0.01$ for the $1\%$
perturbation, and $\epsilon=0.005$ for the $0.5\%$ perturbation), and $%
\delta(\mathbf{r})$ is a binary random variable taking values $\pm 1$, with
equal probability, at each spatial grid point. The subsequent real-time
propagation conserves the norm of the wavefunction.}

{Characteristic examples of the perturbed QD dynamics, subjected to the
addition of the $1\%$ random noise, are presented in Fig.~\ref{rl_tme}. It
is observed that the fundamental QD, generated from the $S=0$ ITP input,
remains stable, as shown in Fig.~\ref{rl_tme}(a), at time $%
t=4.6415T_{0}\approx 4.1031$ ms, with $T_{0}=0.884$ ms. In contrast, the
multipolar QDs, corresponding to the ITP input with $S=1,2$, and $3$ (Figs.~%
\ref{rl_tme}(b-d)), exhibit clear signatures of instability.} The
instability develops more rapidly for larger $S$, leading to rapid
deformation and loss of the multipole structure. {To further probe the
persistence of the instability, we performed additional simulations with a
significantly smaller perturbation amplitude, $0.5\%$. The resulting
evolution is shown in Fig.~\ref{rl_tme_0.01}. Even under so weak
perturbation, the multipolar droplets with $S=1,2,3$ remain unstable. This
happens because, in the present degenerate confinement regime, the flat
potential minimum does not generate an annular potential that would support
stable multipole configurations. As a result, small perturbations initiate
inward motion of the density lobes, causing the multipolar structures to
merge toward the center of the trap. Panels (a-c) display the evolution of
the state corresponding to $S=1$ in the ITP input, at times $t=3.61$ ms, $%
6.15$ ms, and $9.02$ ms, respectively. Panels (d-f) and (g-i) show the
evolution of the states corresponding to $S=2$ and $S=3$, respectively, for
the same time intervals. The instability times reported here typically have
their values in the range of $3.6-9.0$ ms, which are relatively short but
still experimentally relevant time scales.} These results demonstrate that,
in the present case (the flat potential minimum), multipolar QDs with $S\geq
1$ are unstable, and do not exhibit long-lived transient quasi-stability
either, in the sharp contrast to the robust QD created by the input with $%
S=0 $.

Finally, we address head-on collisions between QDs in the present regime.
Stationary single-droplet solutions are initially obtained by ITP of the
extended GP equation Eq. (\ref{dgpe}) until full convergence. Collision
dynamics are then initiated by constructing an initial state composed of two
spatially separated QDs, generated by superposing two copies of the
single-QD wavefunction, displaced along the collision ($z$) direction. The
corresponding initial wavefunction is given by
\end{subequations}
\begin{gather}
\psi _{0}(x,y,z)=  \notag \\
\psi (x,y,z-z_{0})\,e^{-i\eta z}+\psi (x,y,z+z_{0})\,e^{+i\eta z},
\label{collision}
\end{gather}%
where $2z_{0}$ is the initial separation between the QDs and $\pm \eta $ are
opposite momenta (kicks) that set the QDs in motion toward each other. The
robustness of the $S=0$ state motivates its use in the simulations of the
collision dynamics. The resulting collision between the identical monopole
QDs with $S=0$ is \emph{inelastic}, resulting in their merger into a single
droplet, as shown in Fig. \ref{deg_collision}.

\subsection{The strong central Gaussian potential hill, \textbf{$A_{0}>(%
\protect\alpha \protect\omega )^{2}/2$}}

\label{strong}

\subsubsection{Stationary multipole QDs}

Next, we consider the regime with annular confinement, imposed by a
sufficiently tall central Gaussian potential hill, satisfying Eq.~(\ref{>}).
{Throughout this subsection, we fix $\alpha =\sqrt{2}$, $\omega =1$, and $%
A_{0}=30$ in Eq. (\ref{elongated_trap}).} The tall central Gaussian
suppresses the inward motion of the droplets toward the center, preventing
them from merging and enabling the formation of robust higher-order
multipole states.\ All stationary solutions are obtained using the input in
the form of the Gaussian with an imprinted vortical phase, as given by Eq.~(%
\ref{input}). Under the strong confinement, the stationary states generated
from the $S=0$ input exhibit distinct density morphologies, depending on the
molecule number $N$, as shown in Fig.~\ref{hollow}. {For relatively low
values of $N$, such as $N=500$, the DDI attraction dominates over the LHY
repulsion, splitting the QD with $S=0$ into a set of four elongated
droplets, as shown in Fig.~\ref{hollow}(a). Upon the increase of $N$ to $%
1000 $, while staying in the DDI-attraction-dominated regime, the QDs tend
to align along the polarization direction and exhibit enhanced axial
elongation, as shown in Fig.~\ref{hollow}(b). These multi-lobed
configurations are non-vortical split-density states. For larger molecule
numbers -- specifically, $N=6900$ and $N=7700$ -- the system enters the
LHY-dominated regime, where the repulsive beyond-MF contribution becomes
significant. In this regime, the four-droplet configuration gradually fuses
into a single hollow-cylinder droplet, as shown in Figs.~\ref{hollow}(c) and %
\ref{hollow}(d). A similar crossover from four-droplet states at low values
of $N$ to the hollow-cylinder structure at higher $N$ has been reported for
the dipolar $^{164}$Dy BEC confined in cylindrical geometries \cite%
{Young2023}.} Thus, the formation of higher-order multipole QDs is
determined not only by the trapping geometry, but also by the molecule
number. Even in the presence of a strong central Gaussian potential hill, a
relatively low value of $N$ for a given phase-imprinting configuration leads
to fewer poles or an uneven density distribution. Thus, an appropriate
balance between the confinement strength, phase-imprinting index $S$, and
molecule number is essential for formation of well-defined multipole
configurations.
\begin{figure*}[tbp]
\centering
\xincludegraphics[width=0.3\linewidth,label={\hspace{1cm}{a)}}]{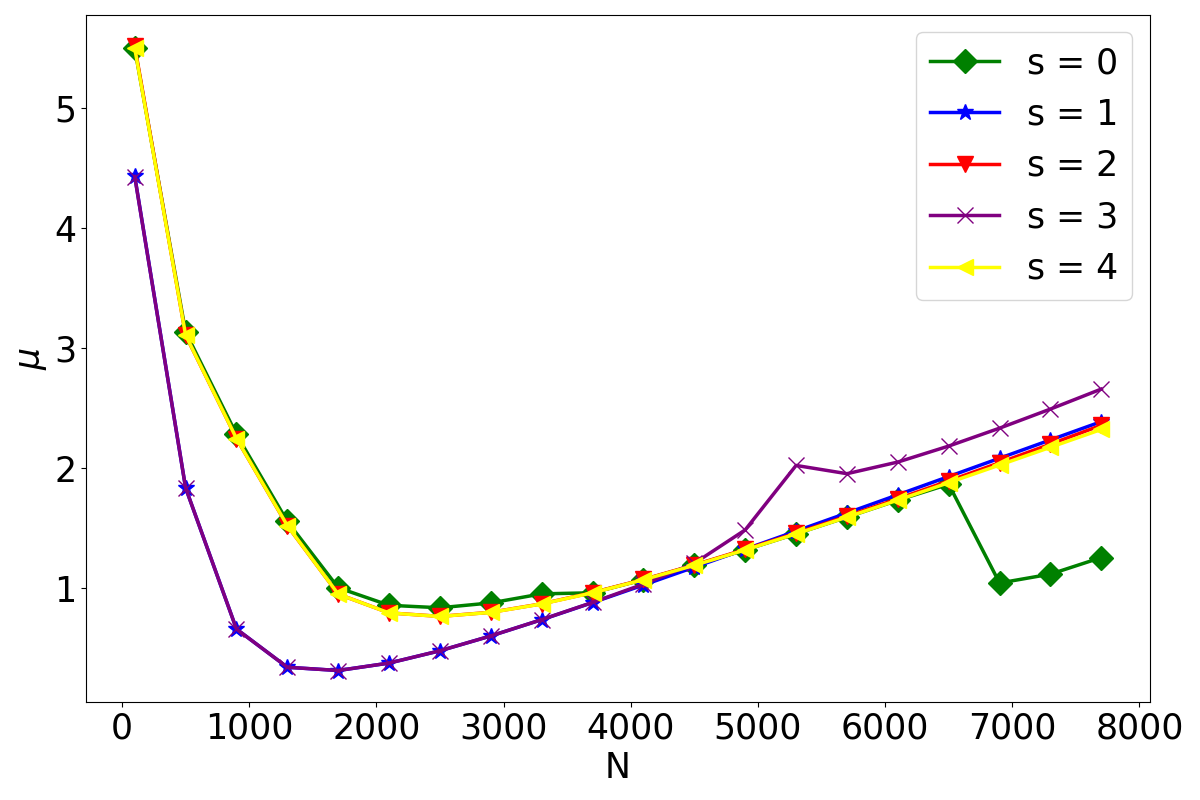} %
\xincludegraphics[width=0.3\linewidth,label={{b)}},pos={n}]{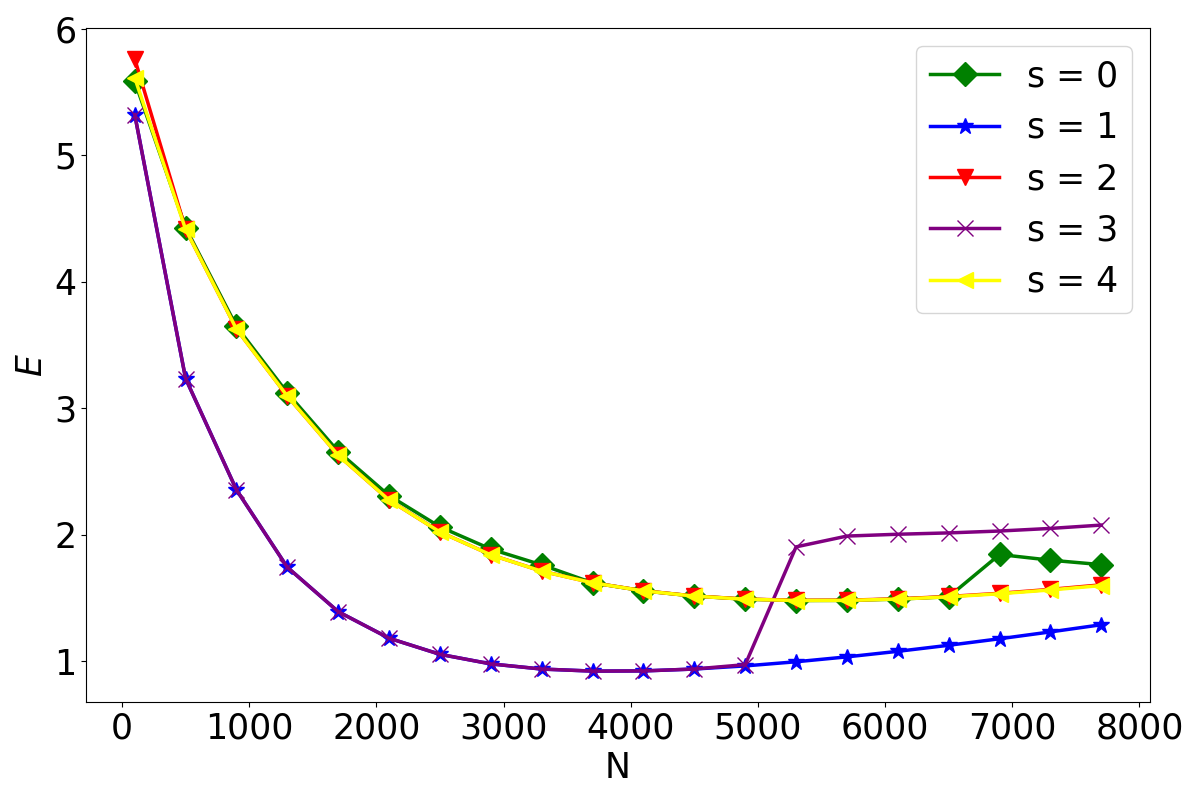} %
\xincludegraphics[width=0.3\linewidth,label={c)},pos={n}]{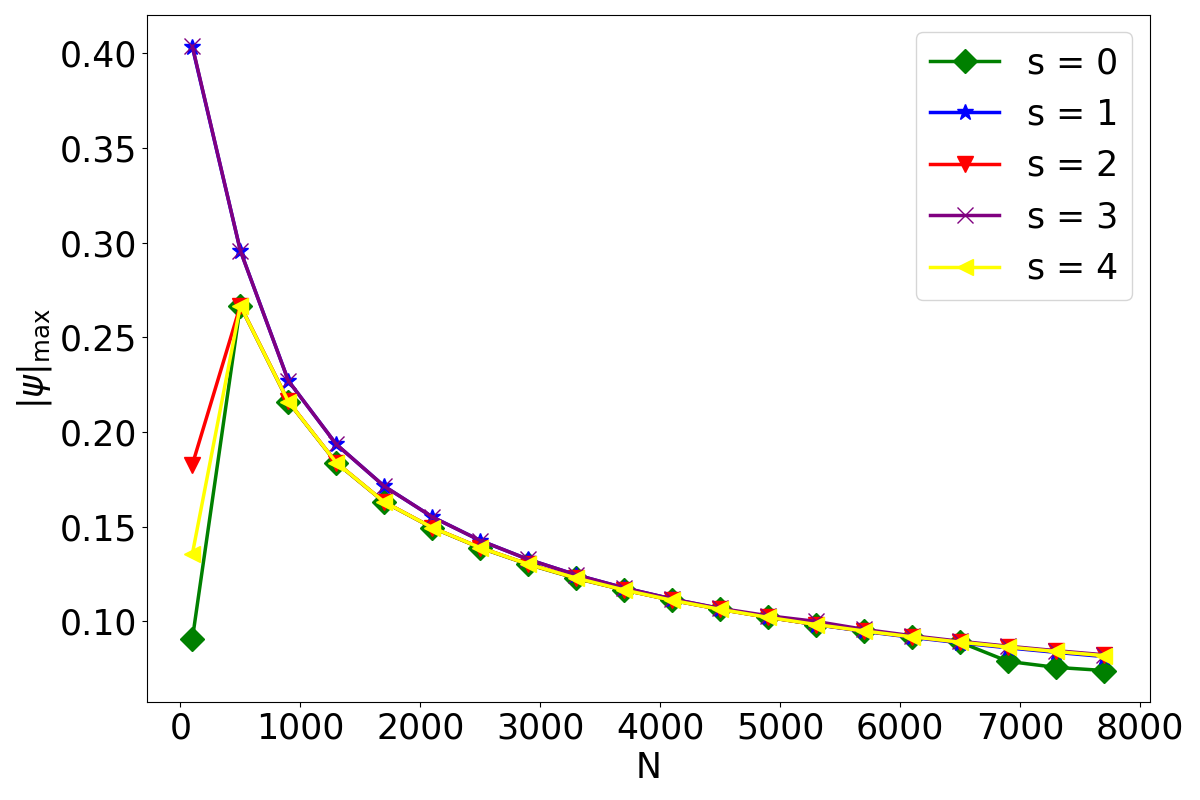}
\caption{ {(a) Chemical potential $\protect\mu$, (b) energy per particle $E$
(defined as per Eq. (\protect\ref{E})), and (c) amplitude $\protect\psi%
_{\max}$ vs. molecule number $N$ for stationary QDs generated by the ITP
inputs with phase-imprinting index (winding number) $S=0-4$, under the
action of the tall Gaussian with $A_{0}=30$ and $\protect\omega =1$, which
satisfy condition~(\protect\ref{scal_pot}). The DDI and contact-interaction
scattering lengths are $a_{\mathrm{dd}}=1000a_{0}$ and $a=100a_{0}$, the LHY
strength being $\protect\gamma _{\mathrm{LHY}}=1.6792\times
10^{9}a_{0}^{5/2} $. All stationary solutions were produced by means of the
ITP method.}}
\label{str_mu_vs_N}
\end{figure*}
{Representative examples of stationary multipole states are shown in Fig.~%
\ref{hig_mul}. Panel (a) displays a six-pole configuration corresponding to
vorticity $S=3$ of the ITP input, with molecule number $N=12000$, under the
action of the central Gaussian potential hill with strength $A_{0}=30$.
Increasing $S$ and $N$ leads to higher-order multipoles: an eight-pole state
is obtained for $S=4$ with $N=22000$, a ten-pole one appears for $S=5$ with $%
N=32000$, and a twelve-pole state is found for $S=6$ with $N=35000$, as
shown in panels (b), (c) and (d), respectively.} For phase-imprinting
indices exceeding $S=6$, the droplets merge, leading to\ a reduced number of
poles. Hence, under the strong Gaussian confinement, a maximum of $n=12$
poles of droplets, generated by the phase-imprinting index $S=6$, are
observed. Overall, the number $n$ of the lobes in the multipole QDs follows
the empirical relation $n=2S$ up to $S=6$, provided that $N$ and the
confinement strength are appropriately chosen.

{The spatial extent of the QDs is analyzed through their density structure
in the $(x,y)$ plane, as shown in Fig.~\ref{2d_hig}. Panels (a) and (b)
correspond to the modes with $S=0$ and $N=1000$ or $N=6900$, respectively.}
Panels (c-f) show the 2D cross sections of higher-order multipole states
with $S=3-6$ and $N=12000$, $22000$, $32000$ and $35000$ respectively. As $S$
increases, the QDs exhibit a systematic increase in their radial extent,
primarily driven by an increase in the molecule number required to form the
higher-order states. The radial expansion leads to pronounced azimuthal
density modulations, giving rise to petal-like configurations in the $(x,y)$
plane. The inset in each configuration refers to its corresponding phase
profiles.

Characteristics of the QD\ families are presented in Fig.~\ref{str_mu_vs_N}.
Figure~\ref{str_mu_vs_N}(a) shows the chemical potential $\mu $ as a
function of $N$ for different $S$ configurations. For all considered $S$
values, $\mu $ initially decreases, and and then increases with the increase
of $N$. The decreasing branches, with $d\mu /dN<0$, correspond to the regime
dominated by the effectively attractive DDI, while the increasing branches
with $d\mu /dN>0$ imply the domination of the LHY repulsion. In the latter
case, the QDs are expected to be stable according to the anti-VK criterion. {%
For $S=0$, the $\mu (N)$ curve exhibits a pronounced drop at a specific
molecule number $N=6900$. At this value of $N$, the stationary solution
undergoes a structural crossover from a four-droplet shape to a
hollow-cylinder one.}

\begin{figure}[tbp]
\xincludegraphics[width=0.48\linewidth,label={a)}]{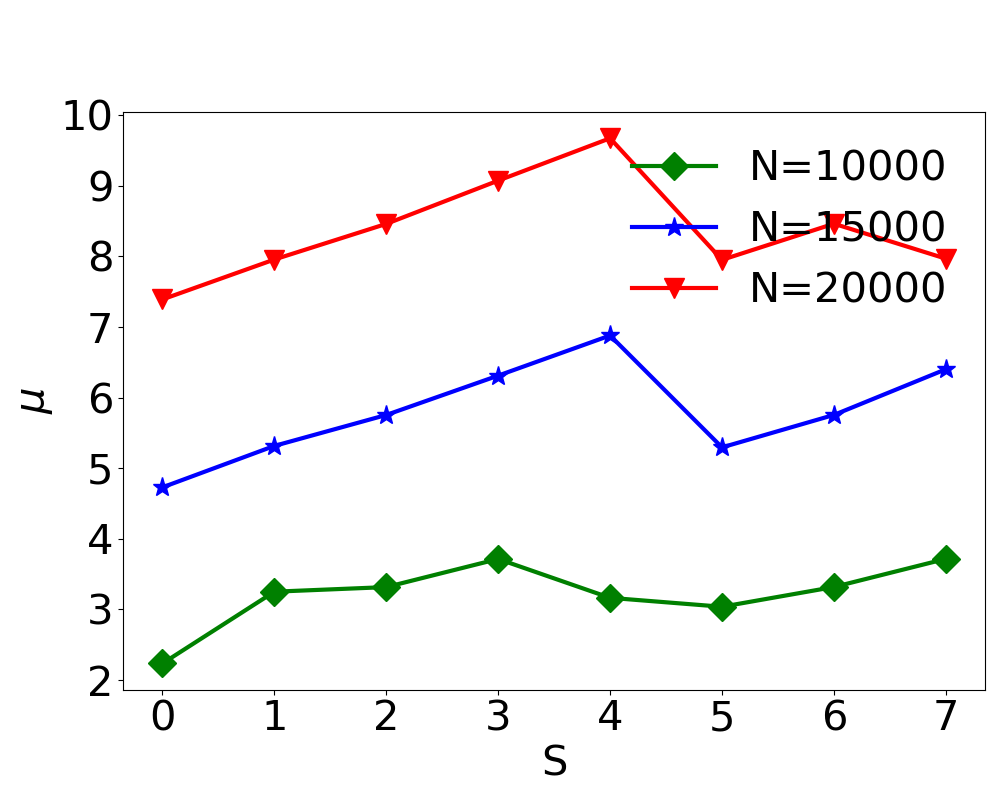} %
\xincludegraphics[width=0.48\linewidth,label={b)}]{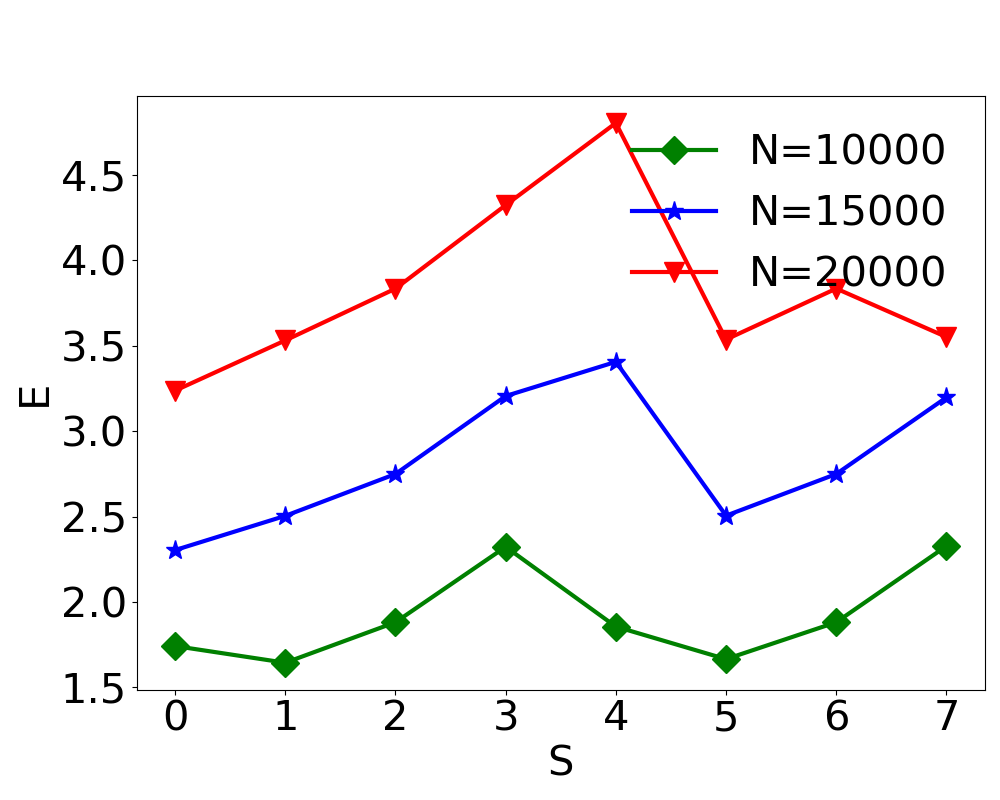}
\caption{{The variation of (a) chemical potential $\protect\mu$ and (b)
energy per particle $E$ vs. the phase-imprinting index $S$ of state's TP
input, as produced by the stationary solutions for different values of the
molecule number $N$, with the DDI and contact-interaction scattering lengths
$a_{\mathrm{dd}}=1000a_{0}$ and $a=100a_{0}$. Both the chemical potential in
(a), and the energy=per-particle in (b) exhibits non-monotonous behaviour.
With $N=10000$ and $S=1$, the configuration corresponds to the ground
(minimum-energy) state, while, with $N=15000$ and $N=20000$, the ground
state has $S=0$.}}
\label{s_vs_mu}
\end{figure}

{For the ITP-input's vorticity $S=3$, the $\mu (N)$ branch overlaps with
that of the $S=1$ configuration in Fig. \ref{str_mu_vs_N}(c), up to $N=4500$%
, beyond which the two branches separate. In the overlapping region, the $%
S=3 $ stationary solutions do not carry sufficient molecule number to
support well-developed higher-order QD structures, and relax toward
configurations corresponding to $S=1$ in the ITP input. Stable six-pole QD
structures emerge at $N$ exceeding the respective threshold value, $N
\approx 4500$, for $S=3$. The further increase in $N$ leads primarily to
broadening of the QD in the radial direction, due to the dominant effect of
the LHY-repulsive interaction. Similar behavior is observed for QDs
corresponding $S=4$ in the ITP input, when the stationary solutions overlap
with those for the $S=0$ and $S=2$ families. In that regime, the $S=4$ state
supports quadrupole QD configurations, rather than eight-pole ones. However,
as $N$ increases further, the $S=4$ branch gets separated from the ones with
$S=0$ and $S=2$. These observations indicate that the empirical relation $%
n=2S$ is not generally satisfied below the transition values of $N$. At the
transition points, higher-multipole states (with $S=4-6 $) only begin to
develop the corresponding $2S$-pole structure, which becomes fully
established at larger values of $N$.} The transition is determined primarily
by molecule number $N$, while the corresponding chemical potential $\mu $
serves as an indicator, rather than a determinant of the droplet morphology.
{Figure~\ref{str_mu_vs_N}(b) shows the energy per particle $E$ as a function
of $N$. In contrast to the chemical potential, the energy of the $S=0$ QD
shows a sudden jump at the transition point, and gradually decreases with
the increase of $N$. It is seen that the ITP input with $S=1$ realizes the
lowest energy; however, as $N$ increases further, the mode with $S=0$
becomes the ground state. } {This unusual energetic ordering at low molecule
numbers originates from the confinement-induced structural transition of the
stationary states. As shown by the qualitative analytical estimate presented
in Appendix A, a sufficiently strong central Gaussian barrier raises the
energy of the non-vortical $S=0$ configuration higher than that of the $S=1$%
. Consequently, the energy separation between the $S=0$ and $S=1$ branches
is reduced and may even be inverted in the weakly nonlinear regime. In the
present nonlinear system, the effect is further modified by the DDI and LHY
interactions, together with the confinement-induced structural transition.
At relatively small molecule numbers, the $S=0$ input produces a
four-droplet configuration due to the presence of the strong central
Gaussian barrier, whereas the $S=1$ input produces a simpler two-droplet
configuration, the latter one having a lower energy in this regime. As the
molecule number increases, the $S=0$ state undergoes a crossover to a
hollow-cylinder structure. Nevertheless, its energy remains higher than that
of the $S=1$ branch until the system enters the LHY-dominated regime, when
the hollow-cylinder configuration acquires the lowest energy among the
stationary states, restoring the conventional ordering in which the $S=0$
branch represents the ground state.}

Similar to the behavior of the chemical potential, below the transition
point, the energy curves for $S=3$ and $S=4$ coincide with those for $S=1$
and $S=0, 2$, respectively. Figure~\ref{str_mu_vs_N}(c) shows the droplet's
amplitude $\psi _{\max }$ as a function of $N$ for different vorticities. In
contrast to the chemical potential $\mu $ and energy $E$, $\psi _{\max }$
shows a decrease trend for all considered $S$ configurations. The
LHY-induced repulsion at larger values of $N$ supports the expansion,
thereby reducing the amplitude. {Figure~\ref{s_vs_mu} shows the variation of
the chemical potential $\mu $ in panel (a), and energy-per-particle $E$ in
(b), with respect to $S$, for different values of the molecule number, $%
N=10000,15000,20000$. Both $\mu$ and $E$ exhibit a non-monotonous dependence
on $S$, for the considered values of $N$. For $N=10000$, the $S=1$ state
corresponds to the energy minimum (the ground state). However, the
zero-vorticity mode becomes, as usual, the ground state for larger values of
the molecule number, \textit{viz}., $N=15000$ and $20000$.}

\subsubsection{Dynamical states of multipole droplets}

\label{strong_trap}
\begin{figure}[tbp]
\xincludegraphics[width=0.25\linewidth, label={a)}]{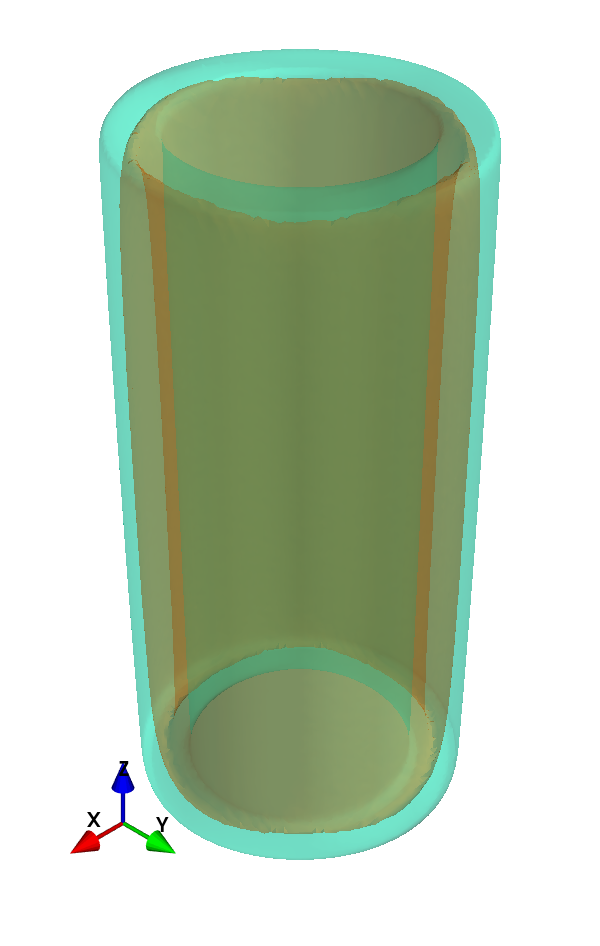} %
\xincludegraphics[width=0.25\linewidth,label={b)}]{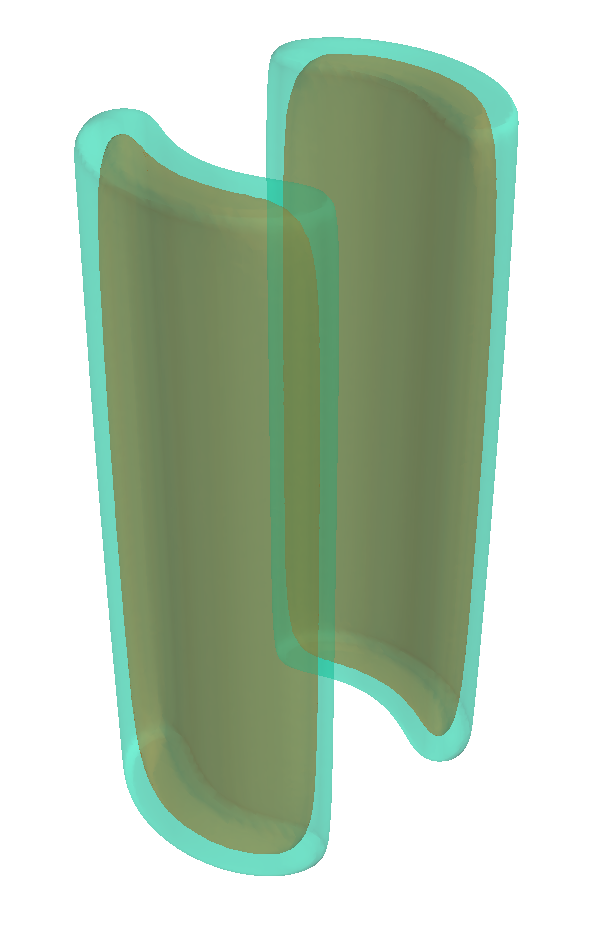} %
\xincludegraphics[width=0.25\linewidth,label={c)}]{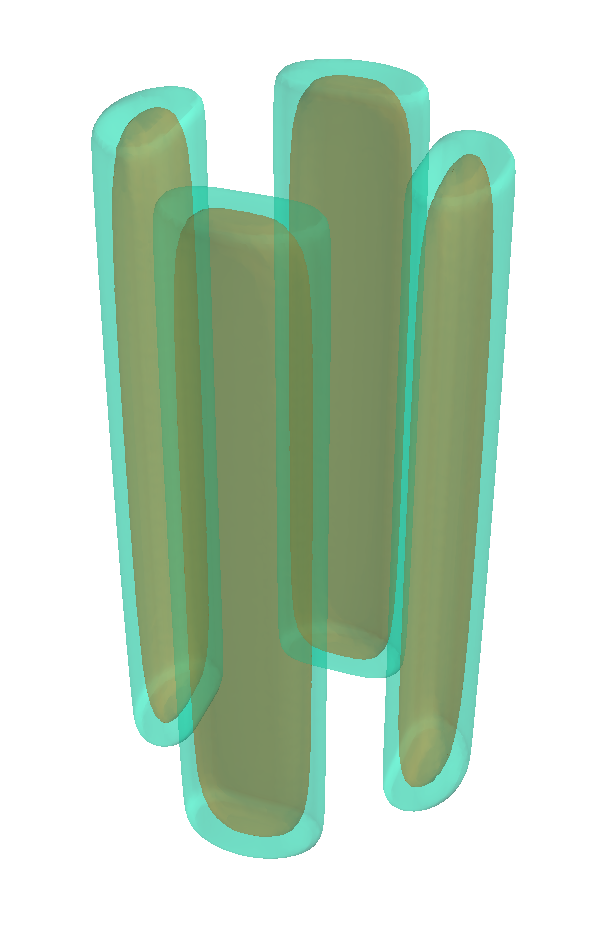} %
\xincludegraphics[width=0.25\linewidth, label={d)}]{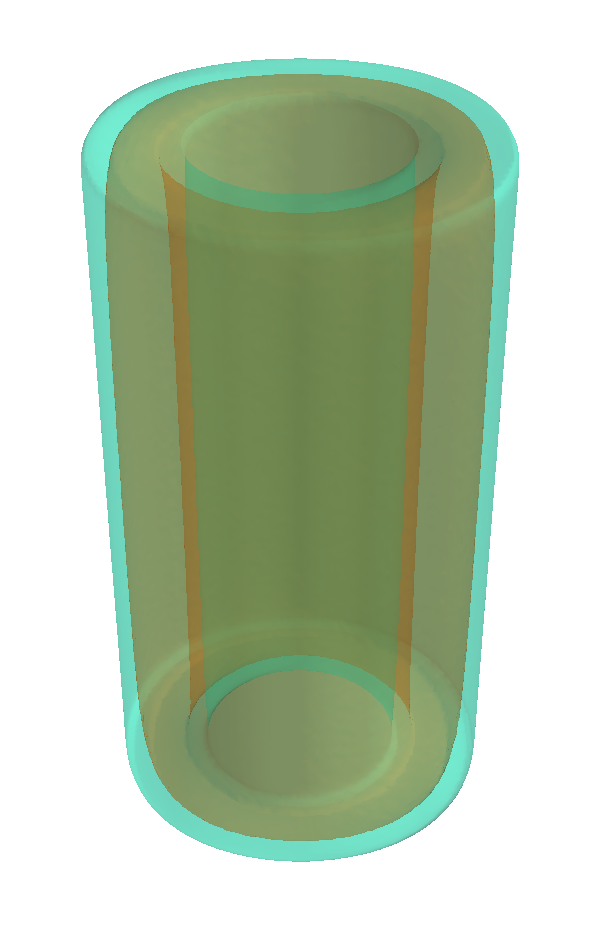} %
\xincludegraphics[width=0.25\linewidth,label={e)}]{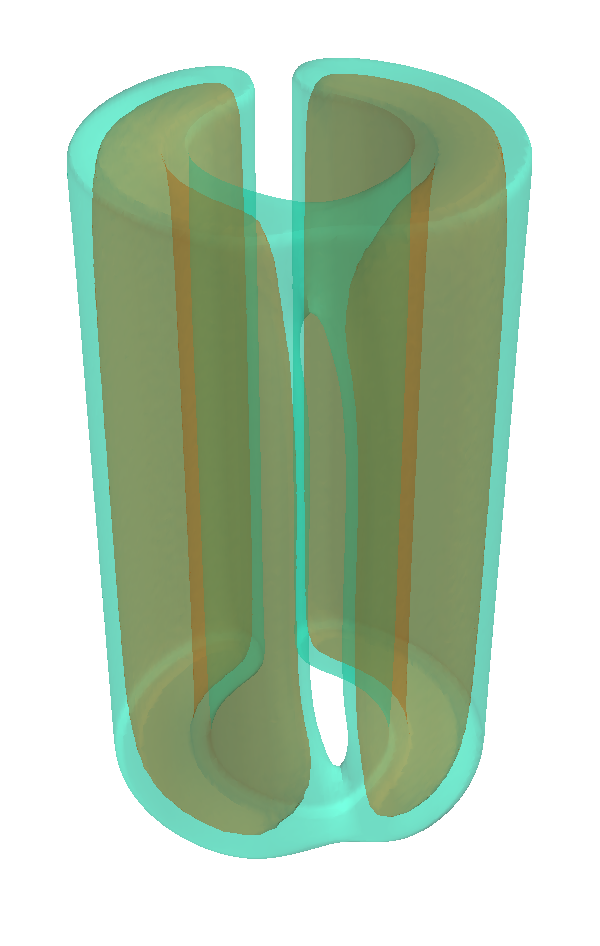} %
\xincludegraphics[width=0.25\linewidth,label={f)}]{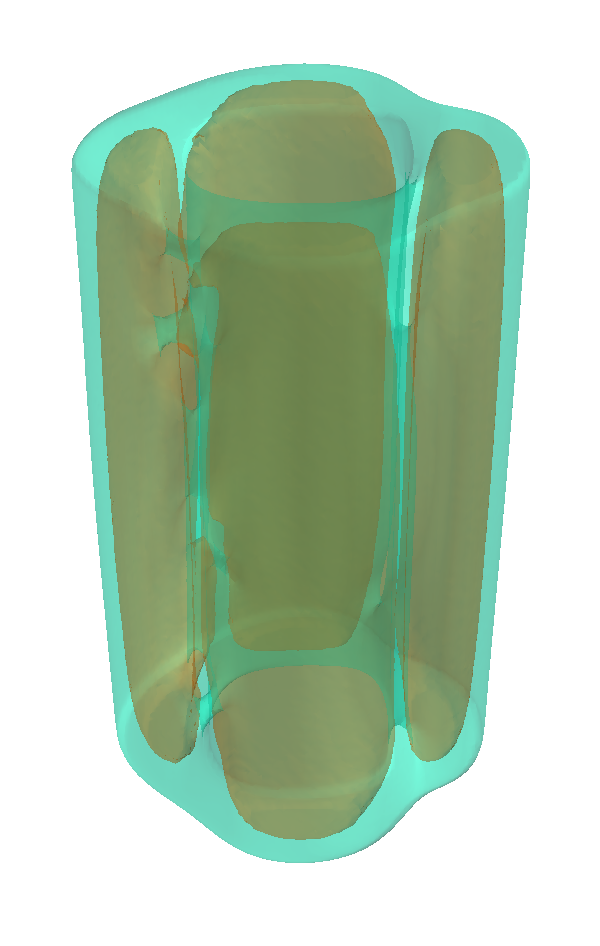}
\caption{ {Real-time dynamics of QDs perturbed by random noise at the $1\%$
amplitude level, for different phase-imprinting indices $S$, and different
molecule numbers $N$. Panels (a)--(c) display dynamically robust evolution
at time $t=4.103$~ms for the ITP input with $S=0$, $S=1$, and $S=2$,
respectively, for $N=10000$. Panels (d)--(f) correspond to a higher molecule
number $N=20000$, at the same time instant, for $S=0$, $S=1$, and $S=2$,
respectively, where the $S=0$ state remains dynamically stable at large
molecule numbers, while the ones with $S=1$ and $S=2$ develop instabilities.
All states were produced by the real-time propagation, initiated by the
converged stationary ITP solutions with added random perturbations. The
dipolar and contact interaction strengths are $a_{\mathrm{dd}}=1000a_{0}$
and $a=100a_{0}$, respectively.}}
\label{s_0_pert}
\end{figure}
\begin{figure}[tbp]
\xincludegraphics[width=0.27\linewidth,label={a)
 $t=3.69 $ms }]{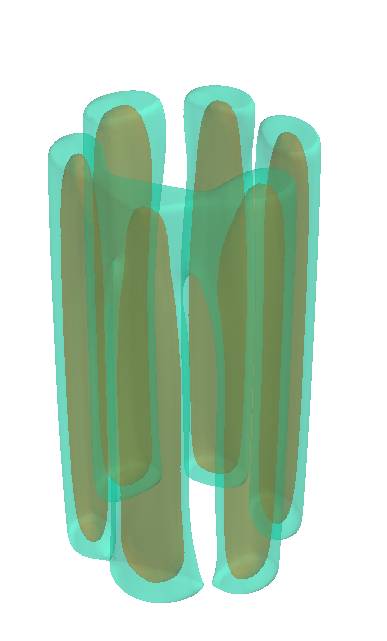} \xincludegraphics[width=0.27%
\linewidth,label={b) $t=6.15 $ms}]{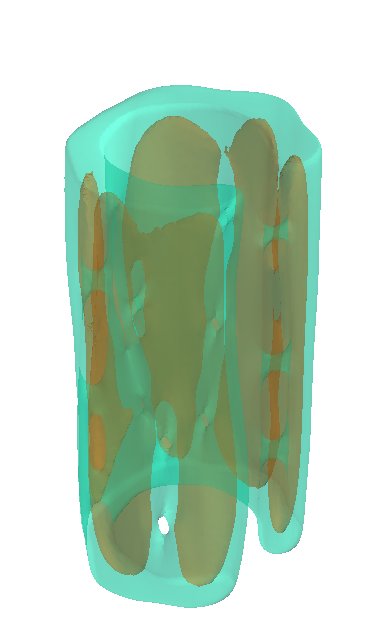} \xincludegraphics[width=0.27%
\linewidth,label={c) $t=9.02 $ms}]{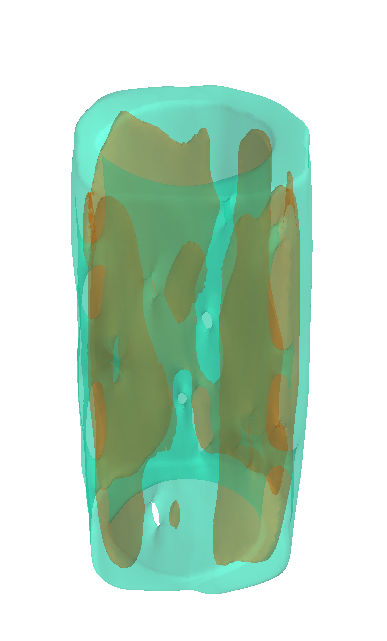} \xincludegraphics[width=0.27%
\linewidth,label={d)}]{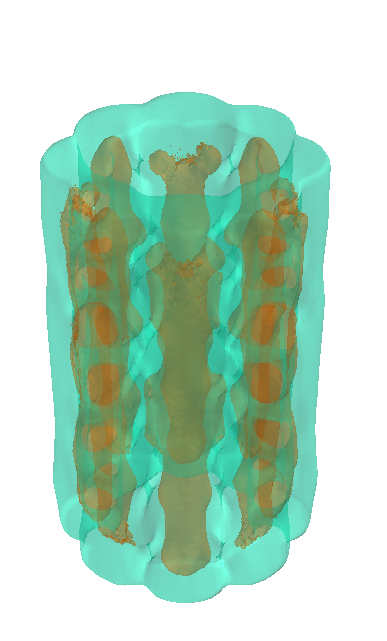} \xincludegraphics[width=0.27%
\linewidth,label={e)}]{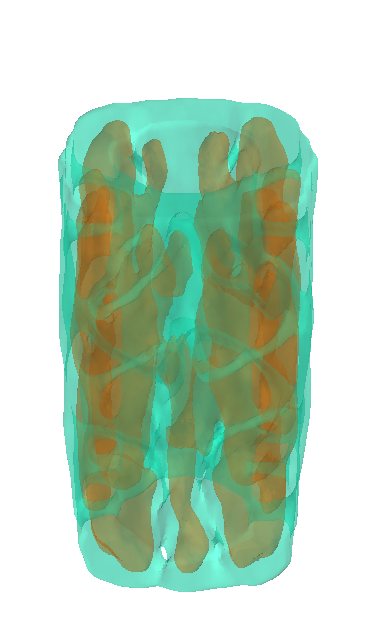} \xincludegraphics[width=0.27%
\linewidth,label={f)}]{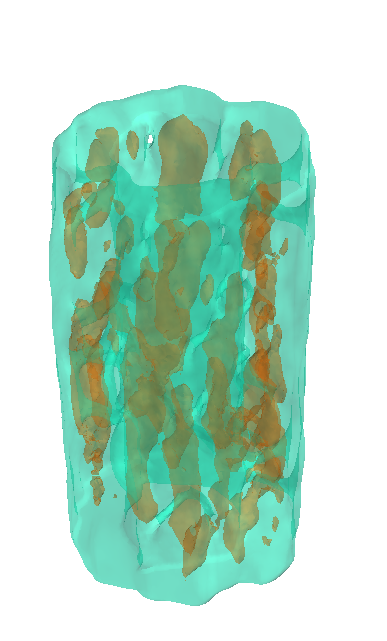}
\caption{ {The perturbed real-time evolution of unstable multipole droplets
created by the ITP input with phase-imprinting indices $S=3$ and $S=4$.
Panels (a)--(c) display the evolution of the $S=3$ state with molecule
number $N=12000$ at times $t=3.69$~ms, $6.15$~ms, and $9.02$~ms,
respectively. Panels (d)--(f) display the configuration generated by $S=4$,
with $N=22000$ at the same time instants. As time increases, instabilities
develop in the configurations. The DDI and contact-interaction scattering
lengths are fixed as $a_{dd}=1000a_{0}$ and $a=100a_{0}$. All results are
produced by the real-time propagation, initiated by the converged stationary
solutions with added random perturbations at the $1\%$ amplitude level.} }
\label{hig_mul_pert}
\end{figure}
\begin{figure}[tbp]
\includegraphics[width=1.0\linewidth]{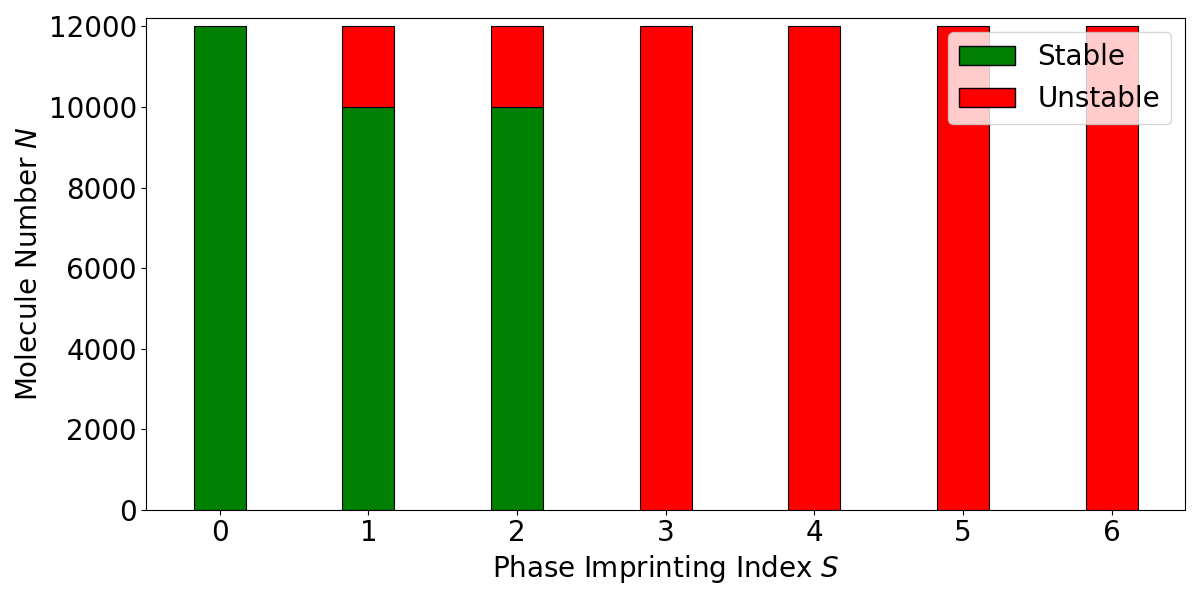}
\caption{Stability areas of multipole QDs in the $\left( N,S\right) $ plane
under the action of the strong Gaussian confinement, with parameters $%
A_{0}=30$ with $\protect\omega =1$, satisfying condition (\protect\ref{>}).
Green regions designate the completely stable QD with $S=0$, while the ones
with $S=1,2$ remain stable up to $N=10000$, being unstable above it (in red
regions). The QDs with $S=3-6$ are completely unstable.}
\label{str_stab}
\end{figure}

{The coexistence of stationary states does not determine their dynamical
stability, which was tested through the perturbed real-time dynamics.
Typical examples of stable and unstable evolution of QDs, subjected to the
addition of random noise at the $1\%$ amplitude level, are presented in Fig.~%
\ref{s_0_pert} at time $t=4.103$~ms. Panels (a)--(c) illustrate the
evolution of multipole QDs corresponding to $S=0$, $1$, and $2$,
respectively, and molecule number $N=10000$. All three configurations remain
dynamically stable under the the action of the random perturbation. For a
higher molecule number, $N=20000$, only the configuration with $S=0$, shown
in panel (d), is stable, while the states with $S=1$ and $S=2$, displayed in
panels (e) and (f), develops instabilities. This behavior correlates with
the energetic ordering of the stationary solutions shown in Fig. (\ref%
{s_vs_mu}). At $N=10000$, the $S = 1$ branch corresponds to the
lowest-energy stationary state, while the $S = 0$ and $S = 2$ ones have
higher energies. As mentioned above, a remarkable property of the system is
that it can make the multipole state with $S = 1$ a lower-energy one, in
comparison to its zero-vorticity counterpart. However, at $N = 20000$, the
QD with $S = 0$ becomes, as usual, the ground state, whereas the $S = 1$ and
$S = 2$ branches correspond to excited states with higher energies, being
therefore unstable with respect to small perturbations.}

{Somewhat similar instability of QDs at large atom numbers has been reported
previously~\cite{cidrim2018vortices} for 3D dipolar QDs of $^{164}$Dy atoms
in free space. Furthermore, Ref.~\cite{sanjay2026toroidal} reports unstable
evolution of the $S=1$ and $S=2$ dipolar multipole configurations at $N=5000$%
, under the action of the toroidal confinement. In the present system,
however, both the hollow-cylinder configuration corresponding to the $S=0$
state and the multipole states with $S=1$ and $S=2$, up to $N=10000$, remain
dynamically stable, owing to the combined stabilizing effects of the
trapping potential in Eq.~(2) and the repulsive LHY correction. Further, at $%
N=20000$, only the configuration with $S=0$ remains stable, both
energetically and dynamically. Further, we addressed the instability of
higher-order multipole QDs perturbed by random noise at the $1\%$ level, as
shown in Fig.~\ref{hig_mul_pert}. Panels (a)--(c) display the evolution of
the QD with $S=3$ and $N=12000$, at times $t=3.69$, $6.15$, and $9.02$~ms,
respectively. Panels (d)--(f) display the corresponding evolution of the QD
with $S=4$ and $N=22000$ at the same time instants. In both cases, the
instability growth becomes increasingly pronounced in the course of the
evolution. Figure~(\ref{str_stab}) summarizes the dynamical stability of the
multipole states produced by different phase-imprinting indices $S$, under
the action of the strong Gaussian confinement, where the green and red
regions correspond to dynamically stable and unstable states, respectively.}

\begin{figure}[tbp]
\xincludegraphics[width=0.3\linewidth,label={a)$\quad
t=2.05$ms}]{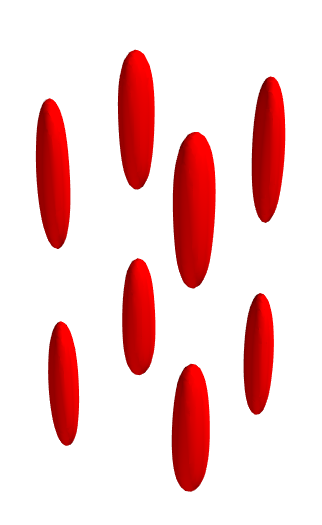} \xincludegraphics[width=0.3\linewidth,label={b)$%
\quad t=3.28$ms}]{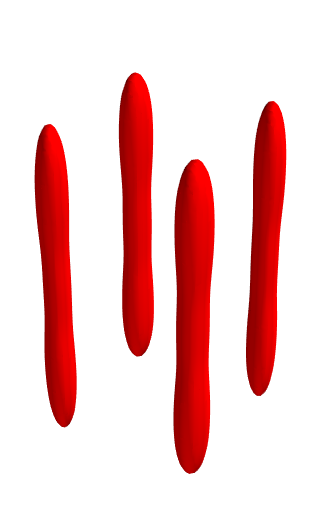}
\xincludegraphics[width=0.3\linewidth,label={c)$\quad
t=6.15$ms}]{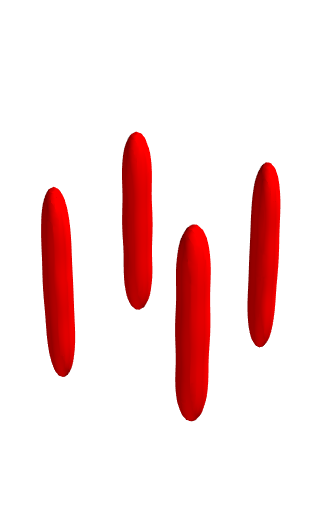}
\caption{ {(a)--(c) Collisions of two identical QDs with $S=0$ and $N=500$,
initially separated by distance $2z_{0}=36\,\mathrm{\protect\mu }$m and set
in motion along the $z$ axis by kicks $\protect\eta =\pm 0.1$ (see Eq. (%
\protect\ref{collision})). The collision is inelastic, leading to the merger
into a single four-lobe droplet. The DDI and contact-interaction scattering
lengths are $a_{\mathrm{dd}}=1000a_{0}$ and $a=100a_{0}$.}}
\label{s_0_col}
\end{figure}

\begin{figure}[tbp]
\xincludegraphics[width=0.3\linewidth,label={a)$\quad
t=2.05$ms}]{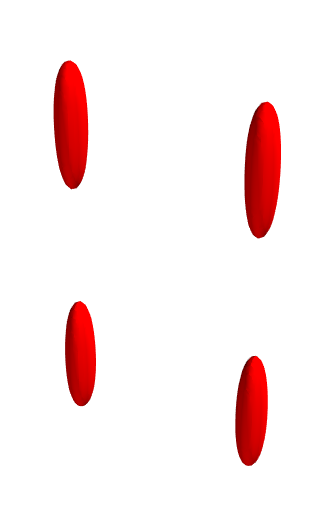} \xincludegraphics[width=0.3\linewidth,label={b)$%
\quad t=4.92$ms}]{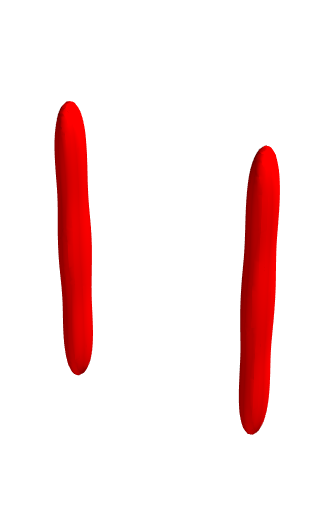}
\xincludegraphics[width=0.3\linewidth,label={c)$\quad
t=6.57$ms}]{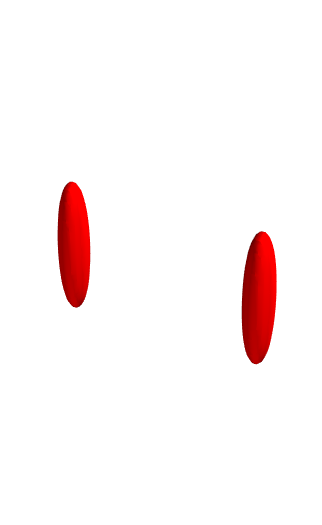}
\caption{ {(a)-(c) Collision of identical QDs with vorticity $S=1$ and
molecule number $N=200$, originally separated by distance $2z_{0}=36$ $%
\mathrm{\protect\mu }m$, which were set in motion by the kicks $\protect\eta%
=\pm 0.1$. The DDI and contact-interaction scattering lengths are $a_{%
\mathrm{dd}}=1000a_{0}$ and $a=100a_{0}$.}}
\label{s_1_col}
\end{figure}

{\ Additionally, we simulated collisions between the multipole QDs, in the
framework of Eq.~(\ref{collision}). Figures~\ref{s_0_col}(a-c) display the
collision dynamics of the QDs with $S=0$ and molecule number $N=500$.
Similar to the above figure~\ref{deg_collision}, inelastic collisions occur
in this case, resulting in merger of the colliding modes into a single
four-lobe one. Further, Fig.~\ref{s_1_col}(a-c) displays the collision
between QDs with $S=1$. In this case, the collision is again inelastic,
leading to the merger of the colliding droplets into a single two-lobe
configuration. Remind that the parameter $S$ represents the phase-imprinting
index in the initial state [Eq.~(\ref{input})], which determines the
resulting multipole configuration. Unlike stationary vortex states, the
multipole ones considered in this work (in the non-rotating frame) possess
zero angular momentum. As the collision is initiated by applying kicks along
the axial ($z$) direction, it introduces only the linear momentum along the
axial direction and does not generate angular momentum. The collision
dynamics are presented here for the states corresponding to $S=0$ and $S=1$.
Higher-order multipole states require larger molecule numbers to form
complete $2S$-pole configurations, and feature a larger spatial extent along
the axial direction. Therefore, the study of their collision dynamics
requires a substantially larger computational domain to resolve the
evolution and minimize boundary effects. A systematic study of collisions
involving higher-order multipole states and larger molecule numbers is
beyond the scope of the present work.}

\begin{figure}[tbp]
\xincludegraphics[width=0.49\linewidth,label={a)}]{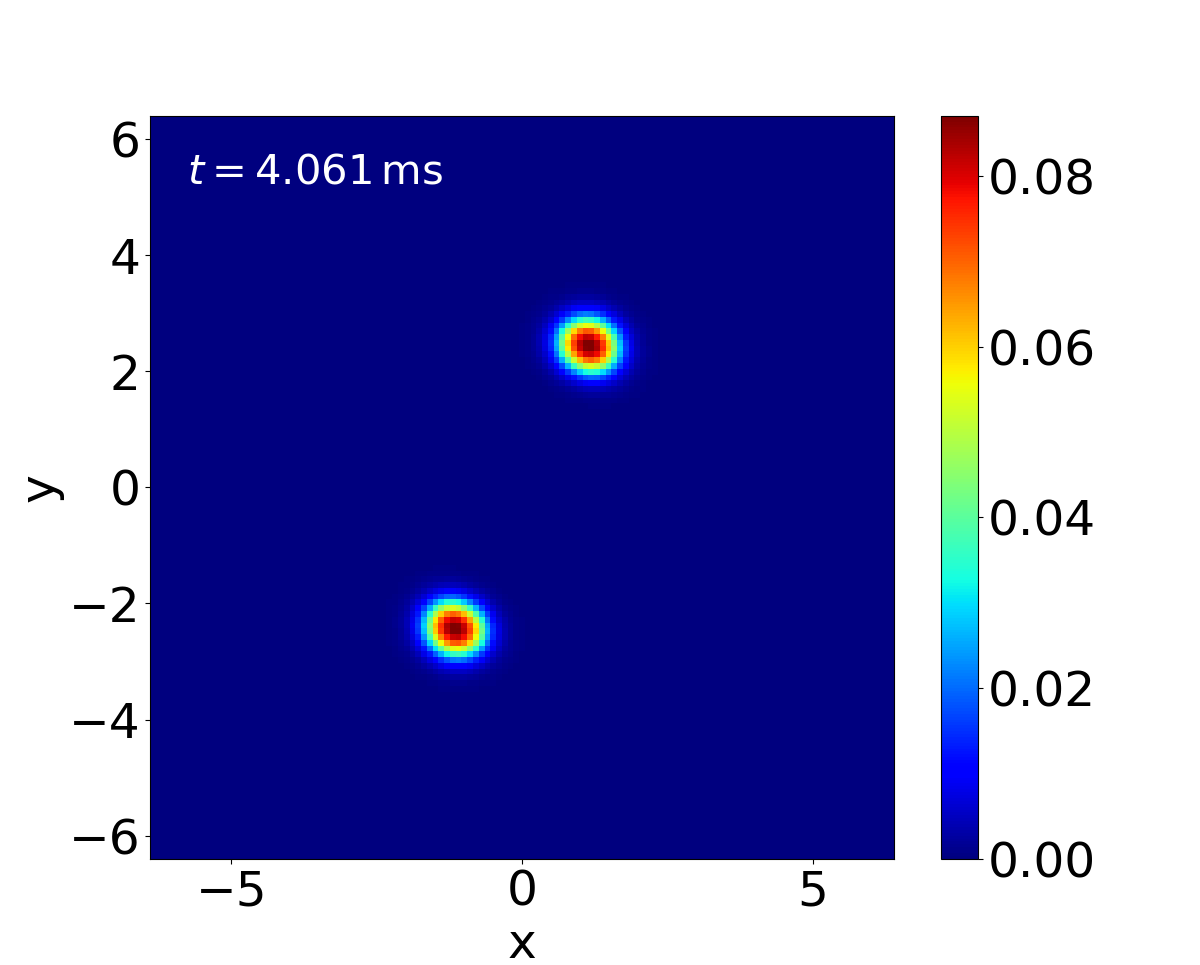} %
\xincludegraphics[width=0.49\linewidth,label={b)}]{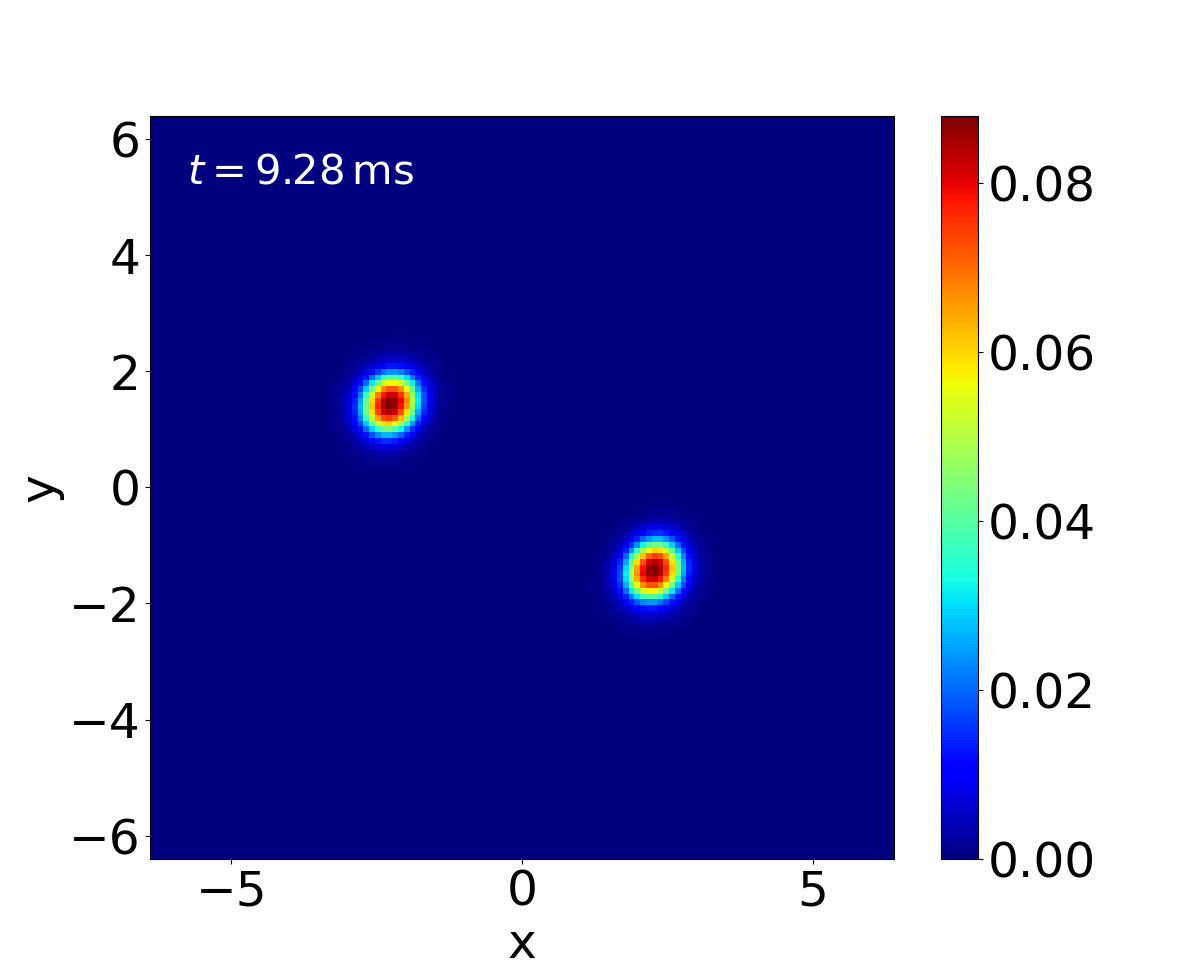} %
\xincludegraphics[width=0.49\linewidth,label={c)}]{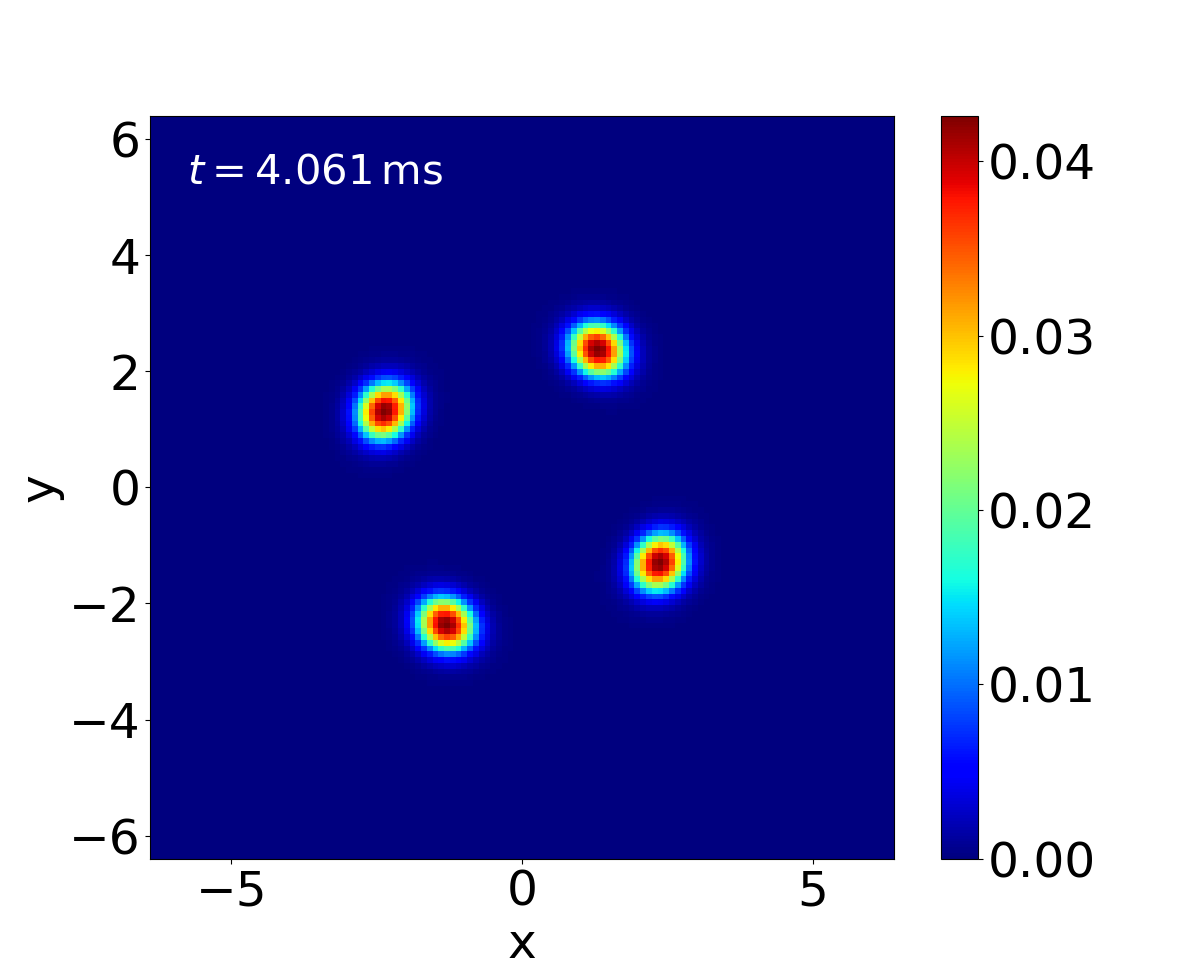} %
\xincludegraphics[width=0.49\linewidth,label={d)}]{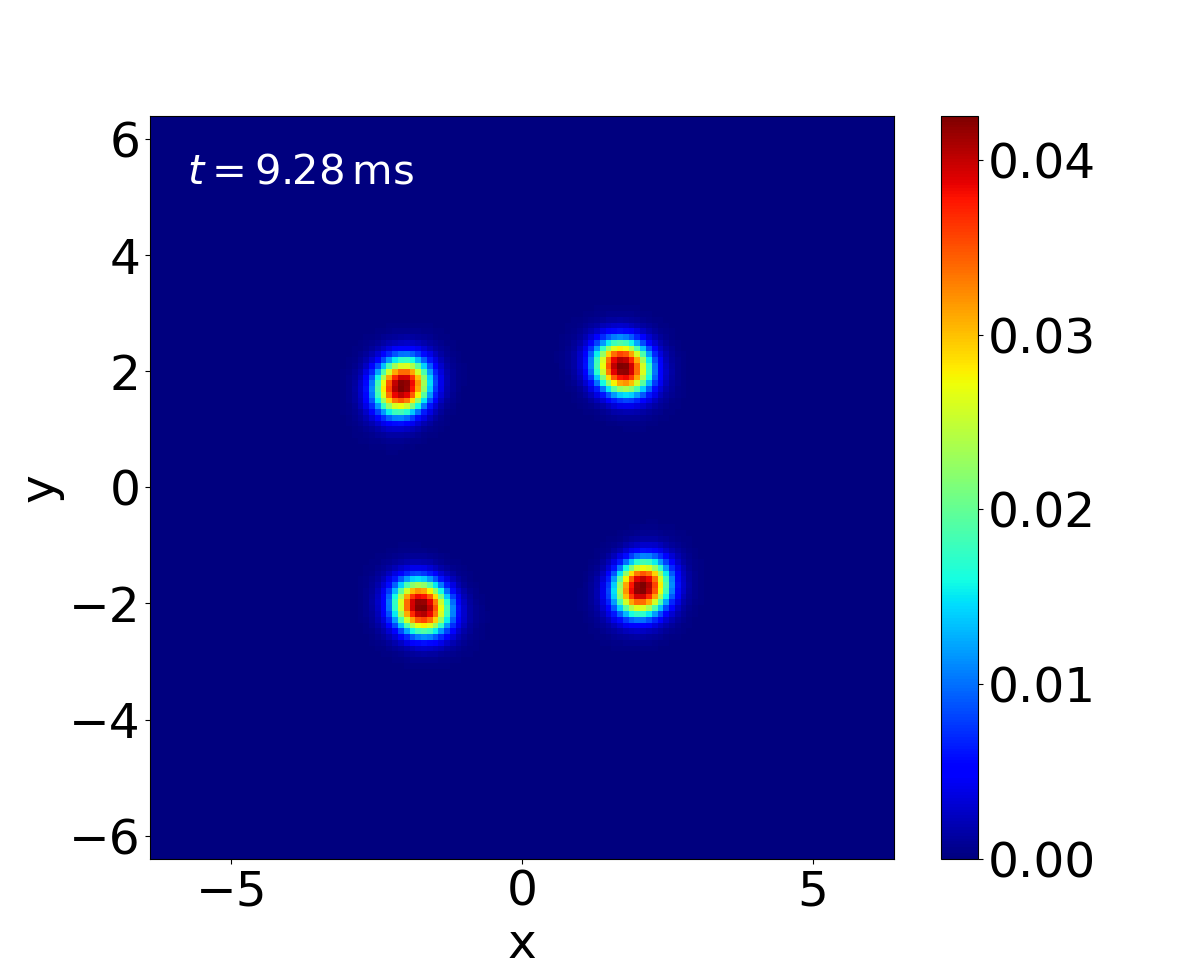}
\caption{ {The stable rotation of multipole QDs for different values of the
phase-imprinting index $S$. with $a_{\mathrm{dd}}=1000a_{0}$ and $a=100a_{0}$%
. The rotation is initiated by the application of the phase torque factor (%
\protect\ref{torque}) with $q=2$, and running real-time propagation. The
corresponding Density profiles in the $(x,y)$ plane are shown in panels
(a,b) for $S=1$ with $N=500$, and in (c,d) for $S=2$, with $N=1000$, at
times $t=4.08$ ms and $t=8.16$ ms.}}
\label{rotation}
\end{figure}

Finally, we tested the possibility to set the multipole QDs in rotational
motion, applying the torque to them,
\begin{equation}
\psi(x,y,z,t=0) = \psi(x,y,z)\exp(iq\theta)  \label{torque}
\end{equation}
where $q$ is the integer and $\theta$ is the angular coordinate in the $%
(x,y) $ plane. The stationary state, produced by means of ITP for each
vorticity, is used as the initial condition for the real-time evolution.
Figure. (\ref{rotation}) displays the resulting stable rotation of the
multipole QDs around the $z$ axis with $q=2$ for $S=1$ (a-b) and $S=2$
(c-d), at different moments of time periods. Naturally, larger integer
values of $q$ result in faster rotations. The angular velocity $\Omega$ is
evaluated numerically from the rotation period. In the limit of strong
confinement and circular symmetry, the rotating structure may be
approximated as a quasi-one-dimensional ring of radius $R$, which estimates $%
\Omega \sim q/R$. However, in the present case, the droplets are elongated
and lack circular symmetry, so that a unique radius cannot be defined. This
leads to systematic deviations from the estimate. For instance, in fig. \ref%
{rotation}(a)-(b), we obtain $\Omega \approx 0.515$, compared to the
estimate $q/R \approx 0.737$. For fig. \ref{rotation}(c)-(d), the numerical
value further decreases to $\Omega \approx 0.3185$, where $q/R \approx0.736$%
.

\section{Conclusion}

We have investigated the formation, characteristics, and dynamics of 3D
elongated multipole QDs (quantum droplets) in the BEC\ of the ultracold
molecular gas, confined by 2D HO (harmonic-trap) potential combined with the
central 2D Gaussian potential hill. The interplay between the DDI
(dipole--dipole interactions) and beyond-mean-field LHY (Lee--Huang--Yang)
correction with different phase imprint $S$ under the action of the
potential confinement gives rise to the variety of elongated multipole QD
states. In the case of the flat potential minimum at the center,
corresponding to the relatively low Gaussian potential hill), we observe the
formation of multipole (many-lobe) states featuring up to six density peak
(the latter pattern corresponds to $S=3$). A systematic numerical analysis
reveals the impact of the molecule number $N$ and vorticity $S$ of the ITP
input on the chemical potential, energy per particle, amplitude, and axial
and radial sizes of the QDs. In the regime dominated by the DDI attraction,
the droplets expand predominantly along the axial direction with the
increase of $N$, whereas in the case when the LHY repulsion is the dominant
factor, the radial expansion prevails. These regimes are distinguished by
the negative and positive slopes of the dependence of the chemical potential
on the molecule number ($d\mu /dN$), corresponding to the VK
(Vakhitov-Kolokolov) and anti-VK criteria, respectively. However, we find
that the compliance with the anti-VK criterion does not guarantee full
stability in the latter case. Simulations of the perturbed evolution show
that only the fundamental state with $S=0$ is truly stable, while the
multipole QDs with $S=1$, $2$, and $3$ decay, even under the action of weak
perturbations. Further, we explored collisions between stable QDs with $S=0$%
, exhibiting their merger into a single droplet.

Stable higher-order multipole QDs, with up to twelve density peaks, the
latter ones corresponding to $S=6$, are created in the strong-confinement
regime, which corresponds to the taller Gaussian hill. In this regime, the
stabilization is provided by the interplay of the geometric confinement,
DDI, and LHY repulsive term. In particular, the tall central potential hill
suppresses the inward-directed merger of the individual lobes, that
destabilizes multipole states in the weak-confinement regime. In the regime
dominated by the DDI attraction, which corresponds to low values of the
molecule number $N$, the state with $S=0$ is built as a set four elongated
sub-droplets, while in the regime dominated by the LHY repulsion, a
hollow-cylinder droplet emerges. The coexistence of stable stationary states
with different $S$ values is identified in the corresponding intervals of
values of $N$. When $N$ is not sufficiently large, higher-$S$
phase-imprinted inputs relax toward lower-$S$ ones. The stability analysis
reveals that the $S=0$ state remains stable for all $N$, while the ones with
$S=1$ and $S=2$ are stable below specific critical values of $N$, and
higher-order states with $S=3-6$ remain completely unstable. Simulations of
collisions in the strong-confinement regime again reveal merger of the
colliding QDs into a single droplet. Further, the multipole QDs exhibit
stable rotational dynamics when set in motion by the phase-imprinted torque
factor.

In summary, this work provides a comprehensive analysis of elongated
multipole QDs under the action of the weak and strong confinement, revealing
the crucial roles of the geometry, long-range DDI, and LHY term in
stabilizing the complex-structured modes. The demonstrated control over the
droplet morphology through the confinement strength and particle number
establish a straightforward route for the realization of multipole states in
dipolar molecular BEC. Future work may extend this approach to spin-coupled
and rotating systems, where additional constraints and symmetries may
further modify the stability and dynamical properties of QDs. \appendix

\section{An analytical estimate of the energy inversion of the $S=0$ and $%
S=1 $ states}

{The numerical results presented in Sec.\ref{strong_trap} demonstrate that,
under strong Gaussian confinement, the stationary state with $S=1$ may
possess a lower energy than the $S=0$ state at relatively small molecule
numbers. Although this phenomenon is obtained from the full nonlinear
extended Gross--Pitaevskii equation, its physical origin can be
qualitatively understood by considering a simplified analytical model. In
particular, the analysis presented below illustrates how the central
Gaussian barrier and the local nonlinear interactions modify the relative
energies of the two lowest states.}

\subsection{The linear limit}

{We consider the two-dimensional linear stationary Schr\"odinger equation
with the harmonic-oscillator trap and central Gaussian repulsive potential,
including the centrifugal term, proportional to the squared phase-imprinting
index $S$:}

\begin{equation}
{E\phi}{=} {-\frac{1}{2} \left( \frac{d^{2}}{dr^{2}} +\frac{1}{r}\frac{d}{dr}
-\frac{S^{2}}{r^{2}} \right)\phi + \left( \frac{\omega^{2}}{2}r^{2} + A_{0}
e^{-r^{2}/\alpha^{2}} \right)\phi .}  \label{Schr}
\end{equation}

{By rescaling the radial coordinate, we set $\alpha=1$. In the absence of
the Gaussian barrier ($A_{0}=0$), the harmonic-oscillator ground state ($S=0$%
) and first excited state ($S=1$) are}

\begin{eqnarray}
{E_{S=0}}&{=}&{\omega, \quad \phi_{0}(r)= \Phi_{0} e^{-\omega r^{2}/2},
N_{0} = \frac{\pi}{\omega}\Phi_{0}^{2},} \\
{E_{S=1}}&{=}&{2\omega, \quad \phi_{1}(r)= \Phi_{1} r e^{-\omega r^{2}/2},
N_{1} = \frac{\pi}{\omega^{2}}\Phi_{1}^{2},}
\end{eqnarray}
{$E_{S=0,S=1}$ and $N_{0,1}$ being the respective eigenvalues and norms.
Treating the Gaussian barrier as a weak perturbation, the corresponding
first-order corrections to the energy eigenvalues are}

\begin{eqnarray}
{(\Delta E_{S=0})_{A_{0}}} &{=}& {\frac{A_{0}}{N_{0}} \int_{0}^{\infty} 2\pi
e^{-r^{2}} \phi_{0}^{2}(r) r\,dr = \frac{\omega A_{0}} {\omega+1},} \\
{(\Delta E_{S=1})_{A_{0}}} &{=}& {\frac{A_{0}}{N_{1}} \int_{0}^{\infty} 2\pi
e^{-r^{2}} \phi_{1}^{2}(r) r\,dr = \frac{\omega^{2}A_{0}} {(\omega+1)^{2}}.}
\end{eqnarray}

{The energy ordering of the two lowest states is inverted when}

\begin{equation}
{E_{S=1}-E_{S=0} < (\Delta E_{S=0})_{A_{0}} - (\Delta E_{S=1})_{A_{0}},}
\end{equation}

{which yields the simple criterion}

\begin{equation}
{A_{0}> (\omega+1)^{2}.}  \label{A0}
\end{equation}

{Equation~(\ref{A0}) demonstrates that a sufficiently strong central
Gaussian barrier tends to reduce the relative energy of the $S=1$ state,
thereby allowing the inversion of the lowest two energy branches.}

\subsection{The contribution of the nonlinear terms}

{The above perturbative approach can be extended to include the local
mean-field and Lee--Huang--Yang (LHY) nonlinearities. The corresponding
two-dimensional stationary Gross--Pitaevskii equation is}

\begin{eqnarray}
{\mu\phi}&{=}&{-\frac12\left(\frac{d^2}{dr^2}+\frac1r\frac{d}{dr}-\frac{S^2}{%
r^2} \right)\phi+\left(\frac{\omega^2}{2}r^2+A_0e^{-r^2}\right)\phi}  \notag
\\
&&{+g\phi^3+\gamma\phi^4},  \label{GPE}
\end{eqnarray}
{where $g>0$ and $\gamma>0$ denote the strengths of the cubic and LHY
self-repulsive nonlinearities, respectively. The nonlinear contributions to
the Hamiltonian are}

\begin{eqnarray}
{H_g} &{=}& {\frac{g}{2} \int_0^\infty 2\pi \phi^4(r)\, r\,dr,} \\
{H_\gamma} &{=}& {\frac{2\gamma}{5} \int_0^\infty 2\pi \phi^5(r)\, r\,dr.}
\end{eqnarray}

{The corresponding corrections to the energy per particle are}

\begin{equation}
{\Delta E_g=\frac{H_g}{N}, \qquad \Delta E_\gamma=\frac{H_\gamma}{N}.}
\end{equation}

{Using the harmonic-oscillator wavefunctions given by Eqs. (A2) and (A3),
one obtains}

\begin{eqnarray}
{(\Delta E_g)_{S=0}} &{=}& {\frac{\omega N}{4\pi}g, \qquad (\Delta
E_g)_{S=1} = \frac{\omega N}{8\pi}g,} \\
{(\Delta E_\gamma)_{S=0}} &{=}& {\frac{4}{25} \left( \frac{\omega N}{\pi}
\right)^{3/2} \gamma,} \\
{(\Delta E_\gamma)_{S=1}} &{=}& {6 \sqrt{\frac{2\pi}{5^{7}}} \left( \frac{%
\omega N}{\pi} \right)^{3/2} \gamma.}
\end{eqnarray}

{Consequently,}

\begin{equation}
{\frac{(\Delta E_g)_{S=1}} {(\Delta E_g)_{S=0}} = \frac12,}
\end{equation}

{and}

\begin{equation}
{\frac{(\Delta E_\gamma)_{S=1}} {(\Delta E_\gamma)_{S=0}} = \frac32 \sqrt{%
\frac{2\pi}{125}} \approx 0.336.}
\end{equation}
{Since both ratios (A15) and (A16) are smaller than unity, this perturbative
analysis indicates that the local mean-field and LHY nonlinearities also
reduce the relative energy of the $S=1$ state with respect to the $S=0$
state, thereby reinforcing the trend toward the inversion of the energy
minimum. We emphasize that the above analysis is intended only as a
qualitative guide to the physical mechanism responsible for the inversion.
It is based on an axisymmetric perturbative model and therefore does not
account for the structural reorganization of the multipole quantum droplets,
the anisotropic dipole--dipole interaction, and the full nonlinear coupling
present in the extended Gross--Pitaevskii equation. Consequently, the above
analysis does not predict the critical molecule number at which the
inversion disappears. As demonstrated by the numerical solutions presented
in the main text, increasing the molecule number eventually restores the
conventional energetic ordering, with the $S=0$ state being the ground state.%
}

\end{document}